   \documentclass[reprint,aps]{revtex4-1}
                   \usepackage{dcolumn}

\usepackage{amsmath}
\usepackage{graphicx}
\usepackage{bm}
\usepackage{color}
 
\newcommand{\GeV}{\makebox{ GeV}}
\newcommand{\beq}{\begin{equation}}
\newcommand{\enq}{\end{equation}}
\newcommand{\beqa}{\begin{eqnarray}}
\newcommand{\beqast}{\begin{eqnarray*}}
\newcommand{\enqa}{\end{eqnarray}}
\newcommand{\enqast}{\end{eqnarray*}}

\def\GeV{\nobreak\,\mbox{GeV}}

\begin{document}

 \title {  Structure of Forward pp and $ {\rm p \bar p}$ Elastic Amplitudes at Low Energies}
\author{E. Ferreira $^{\rm a ~ \ast } $ 
\thanks{Corresponding author. Email: erasmo@if.ufrj.br\vspace{6pt}} }
\author{A. K. Kohara $^{\rm a}$   }
\author{J. Sesma $^{\rm b}$ }

\affiliation{ $ ^{\rm a}$ {\em Instituto de F\'{\i}sica, Universidade Federal do Rio de Janeiro \\
C.P. 68528, Rio de Janeiro 21945-970, RJ, Brazil }    }
\affiliation{ $ ^{\rm b}$ {\em
Departamento de F\'{\i}sica Te\'{o}rica, Facultad de Ciencias, 50009, Zaragoza, Spain}   }  

\begin{abstract}
 Exact analytical forms of solutions for Dispersion Relations for Amplitudes and Dispersion 
Relations for Slopes  are applied in the analysis of pp and $\rm {p \bar p}$  scattering 
data  in the forward range   at energies below $\sqrt(s)\approx 30 \GeV$. As inputs 
for the energy dependence of the imaginary part, use is made of  analytic  form for the 
total cross sections and for  parameters of the $t$ dependence of the imaginary parts, 
with exponential and linear factors.
   A  structure  for the $t$ dependence of the real amplitude  is written,  with slopes
 $B_R$    and a linear  factor  $\rho-\mu_R t$ that allows  compatibility of the  data 
with the predictions from dispersion relations for  the derivatives of the real amplitude  
at the origin. A very precise description is made of all $d\sigma/dt$ data, with  regular 
energy dependence of all quantities. It is shown that a revision of previous calculations 
of   total cross sections, slopes and  $\rho$ parameters in the literature  is necessary, 
and stressed that only determinations   based on $d\sigma/dt$ data covering 
sufficient $t$ range using appropriate forms of amplitudes can be considered as valid.  
 
 \end{abstract}

\pacs{13.85.Dz,13.60.Hb,13.85.Lg}

 \maketitle


\section {Introduction \label{introduction} }

In the scattering theory in quantum mechanics, 
the elastic differential cross sections are written in terms of a complex amplitude with 
independent imaginary and real parts, which are fiunctions of two variables $s,t$ (spin effects neglected). 
In the analysis of observables , besides the nuclear amplitude, account is 
taken      for the   contribution from the real Coulomb interaction. This is  
very basic and obvious, but we show in the present work that this structure is not 
usually obeyed in the    treatments of the   pp and p\=p systems, where   $d\sigma/dt$
is written without due account for the properties of the  amplitudes.
  We give a treatment of these elastic  processes 
using theoretical constraints and appropriate  forms  of the input quantities, arriving 
at realistic amplitudes to connect  measurements and theoretical dynamical models. 

Determinations of $\sigma$, $\rho$ and other parameters of  pp and p\=p forward elastic 
scattering are not direct experimental measurements. Rather, they result from
model-dependent analytical limiting procedures, performed with forms assumed for the 
imaginary and real parts of the complex elastic amplitude. The work done in the laboratory 
consists in measuring values of the number $\Delta N/\Delta t$ of event rates in intervals
$ t ~  -> ~ t+ \Delta t$. With attention given to fluxes and densities 
(we are only concerned with unpolarized beams and targets),  tables of 
$t$ distributions  in differential cross sections   $d\sigma/dt$  are produced. 
 We stress that the  identification of the 
amplitudes and their parameters   requires use of proper theoretical framework. 
  
   The differential cross section  is written as a sum  of absolute values
\begin{equation}
    \frac{d\sigma}{dt} =\frac{d\sigma _I}{dt}+ \frac{d\sigma _R}{dt}
   =\left(\hbar c\right)^{2} \big(   |T_I|^2   +|T_R|^2   \big) 
\label{parts}
\end{equation} 
and the disentanglement required for the   determination of the amplitudes $T_I(s,t)$ 
and $T_R(s,t)$ is not at all trivial. Help is brought from the interference of nuclear and 
Coulomb interactions and from dispersion relations connecting real and imaginary parts 
through  general principles of causality and analyticity. 

 Besides the entanglement to be resolved, we have that production rates are not obtainable 
directly at the origin  $t=0$, or even very  close to it,  but rather 
in sets of  points of an interval.  The determination of $\sigma$, $\rho$, 
slopes and other  quantities   requires  extrapolation of data in a $|t|$ range, using 
analytical expressions, and the results obviously  depend on   their forms.  
The mathematical structures of the 
amplitudes are mounted using parameters that must be found in confront with the 
observed $t$ distribution in $d\sigma/dt$.  Regularity in the behaviour of all 
quantities with the energy is important consideration   to obtain  sensible descriptions 
of the elastic processes. Experiments at different energies must be analysed globally, 
since  separate fitting procedures may lead to    values  that are useless 
as a step for the phenomenology of the area. 

The $t$ range of the data at  given energy must be   sufficient for  representation 
 through assumed   analytical forms. In the low energy range, up to  
$\sqrt{s} \approx 30 \GeV$,  often these conditions are not satisfied, even suffering  
insecure normalization in the  measurements of $d\sigma/dt$, and compilations of published   
values for typical parameters  result scattered in plots, without coherence and regularity.  
  We propose an investigation of this energy range, with  emphasis on the 
identification of the  amplitudes, searching to build a   bridge between measurements 
and mathematical description, necessary to guide models of the dynamics of 
the processes.  

 In the interval of $\sqrt{s} $ from 30 to 60 GeV, pp and p\=p   from ISR/CERN  
and Fermilab  data cover  large $t$ range with good precision,  showing fast 
increase in $\sigma$, a forward peak and a marked dip in $d\sigma/dt$. These 
measurements led  to the establishment of the successful Regge phenomenology 
\cite{Regge,DL}, based on the exchange of particles (pomerons, reggeons) in the $t$ 
channel. Several theoretical models were developed to describe  dynamically this 
region of data in the $s$ and $t$ variables \cite{models}. 
 
Above $\sqrt{s} = 60 \GeV$, experiments \cite{data_HE} have large energy gaps, passing 
fast  by SPS/CERN, Fermilab and reaching the TeV range of LHC \cite{data_LHC}. 
 Ingredients of QCD dynamics  enter with less or more detail in the interpretation 
of these  data \cite{SVM,us_LHC}. 
According to  QCD expectations, as the energy increases the response of the gluon 
density in the hadrons  increases and the hadronic interaction  becomes determined 
 by the  vacuum structure \cite{Tel-Aviv,CGC}. The interpretation of the forward 
scattering parameters in the LHC experiments at 7, 8 and 13 TeV is  not trivial, 
and ambiguities and possible discrepancies  are not clarified \cite{LHC_2017}. 
The potential of crucial information \cite{REAL_PART} in the real part of the forward 
amplitude at  high energies requires that doubts in the analysis of the data
be properly solved.

In this paper we  analyse forward pp, p\=p data with  $\sqrt{s}$ from $\approx 3$  to 
 $\approx 30 \GeV$, using  forms for real and imaginary scattering amplitudes restricted 
to the forward regime and exploring fully   the theoretical resources and constraints 
of dispersion relations treated exactly, in order to extract pure information 
on the forward quantities, as much as possible independently of peculiar 
microscopic models.
This is the most difficult range of data for the analysis, for  
both reasons of insufficiency  in the data and sophistication of the mathematical 
solutions of dispersion relations at low energies. Anyhow, we believe that in this  
 sector  we can learn about determination of amplitude parameters, and hope 
that this technical knowledge  may be useful in the present difficulties encountered in 
the analysis of the recent LHC experiments.

We  propose a treatment of  pp and  p\=p  forward elastic scattering  analysing all data 
of differential cross sections that seem qualified (namely covering necessary $t$ range 
with regularity) for the extraction of the real and imaginary
parts of the complex amplitude. We use the simplest and realistically possible analytical 
forms,   treating coherently  the Coulomb interference and the Coulomb phase, 
we use dispersion relations for the amplitudes (DRA) and for their slopes (DRS) 
 \cite{EF2007} with  exact solutions for the principal value integrals, 
obtaining coherent  energy dependence of all quantities. 
DRA and DRS predict  algebraic values for the real amplitudes and their derivatives at 
$t=0$, and our aim is to have sound proposals for the energy dependence of other parameters 
in order to reduce flexibility and  choices by fittings, and produce  a complete coherent 
description of all data in pp and  p\=p  unpolarized elastic scattering . To eliminate fluctuations that are not meaningful, we account for   normalization uncertainties 
(systematic errors), investigating a normalization factor  in each experiment that 
adjusts the  total cross section to the parametrized prediction. These factors are 
always very close to 1. 

    The  plots and numbers  presented  in the Review of Particle Properties \cite{PDG}
 of the Particle Data Group for forward scattering parameters  in  the low energy 
range, taken  from the   experimental  papers, are scattered and  misleading.
 The results of our work for all data that we analyse 
are regularly  distributed, presented in numbers and very precise  plots 
of $d\sigma/dt$  in Sec.\ref{data_analysis}, showing a  way for   
rationalization of the phenomenological knowledge.
 However, the proposed solutions are not meant to be conclusive, unique  
or fully convincing.  Alternatives    are possible and may be looked for. 

 In elastic pp  and   p\=p scattering in the forward direction, the $t$ 
dependences of the  amplitudes are mainly characterized by exponential forms, 
with slopes $B_I$ and $B_R$ that are essentially independent quantities, 
essentially not   equal to each other: the real and imaginary amplitudes 
do not run parallel along the $t$ axis. We take special care in the investigation 
of the behavior of the real part, which has structure deviating from 
a pure exponential form for $t$ values included in the forward range.   

Once the total cross sections for pp and p\=p   are parameterized in the energy, 
the usual Dispersion Relations for the Amplitudes (DRA) determine the real amplitudes 
at the  origin (namely the $\rho$ parameters). Similarly,  if the 
derivatives  of the imaginary parts of pp and p\=p  at $t=0$ are given as 
functions of the energy,   the derivatives of the real 
parts at $t=0$ are determined by the  Dispersion Relations for 
Slopes (DRS). At low energies it is essential that in both DRA and DRS calculations 
the exact solutions \cite{Exact} be used. The dispersion relations for the amplitudes
has been effectively used  in investigations of the energy dependence of total cross
section and $\rho$ parameter in pp and $\rm p\bar p$ scattering, being a very important 
tool of control in the analysis of the data \cite{Menon}.

The imaginary part is positive at  $t=0$ and decreases with an exponential form, 
which must be multiplied by a proper factor pointing to a zero, so that the well 
known dip may be  created in  $d\sigma/dt$. Actually, in our analysis the dip is 
located outside the examined $|t|$ range, but a linear factor pointing to a 
distant zero has influence in the shape of the imaginary  amplitude 
and its extrapolation for $t=0$ to use  the optical theorem. 
In the real part the effect of the  structure ($t$ dependence) beyond the 
exponential slope is present in the small $|t|$ region, and  is essential in  
DRA and DRS    for the determination of the parameter $\rho$, and here also a 
factor (linear, in our case) must be introduced. 
 
The solutions for the forward amplitudes  can be obtained with high accuracy, minimum 
freedom of parameters, and with remarkable simplicity and regularity in 
the  energy dependence of all quantities. 

Trusting to propose  a realistic assumption, we write  for the  pp or p\=p  elastic  
differential cross sections   
\begin{eqnarray}   
&&\frac{d\sigma}{dt} ( {\rm pp,p\bar p} ) (s,t) \nonumber \\
&&  =\pi\left(\hbar c\right)^{2}\Big\{\Big[\frac
{\sigma(\rho-\mu_R t) }{4\pi\left(  \hbar c\right)^{2}}~{{e}^{B_{R}t/2}
+F^{C}(t)\cos{(\alpha\Phi)}\Big]^{2}}\nonumber\\
&& +\Big[\frac{\sigma (1-\mu_I t) }{4\pi\left(\hbar c\right)^{2}}~{{e}^{B_{I}t/2}%
+F^{C}(t)\sin{(\alpha\Phi)}\Big]^{2}\Big\}~,}   
  \label{diffcross_eq}
 \end{eqnarray}
where $t\equiv-|t|$ and we call attention for the  different values expected 
for the slopes $B_{I}$
and $B_{R}$ of the imaginary and real amplitudes and introduce factors with 
linear $t$ dependence in each amplitude.  This expression is applied for 
pp an p\=p, and  the energy dependent quantities   $\sigma(s),
  B_I(s) , \mu_I(s),  \rho(s),  B_R(s), \mu_R(s)$ ,  are specific for each case.
 
In a given normalization we write for the real and  imaginary nuclear (upper label $N$)  amplitudes 
 \begin{equation}
    T^N_R (s,t)= \frac{1}{4 \sqrt{\pi} \left(\hbar c\right)^2} ~ \sigma (\rho-\mu_R t) ~ e^{B_R t/2}  
\label{real_TR}
\end{equation}
 and 
\begin{equation}
     T^N_I (s,t)=\frac{1}{4 \sqrt{\pi}\left(\hbar c\right)^2 } ~ \sigma (1-\mu_I t)~ e^{B_I t/2} ~. 
\label{imag_TI}
\end{equation}
The optical theorem is implicit in Eq.(\ref{imag_TI}). 
At $t=0$, we have  the usual definition of the $\rho$ parameter 
\begin{equation}
\rho = \frac{T^N_R(s,t=0)}{T^N_I(s,t=0)} ~ , 
\label{real_forw0}
\end{equation}
remarking  that  the value of $\rho$ obtained by fitting of data in 
a certain $t$  range depends on the analytical forms 
    (\ref{real_TR},\ref{imag_TI}) of the amplitudes.

In Eq.(\ref{diffcross_eq}), 
  $\alpha$ is the fine-structure constant, $\Phi (s,t)$ is the Coulomb
phase and $F^{C}(t)$ is related with the proton form factor 
\begin{equation}
F^{C}(t)~=(-/+)~\frac{2\alpha }{|t|}~F_{\mathrm{proton}}^{2}(t)~,
\label{coulomb}
\end{equation}%
for the pp$/$p$\mathrm{{\bar{p}}}$ collisions, where  
 \begin{equation}
F_{\mathrm{proton}}(t)=[\Lambda^2/(\Lambda^2+|t|)]^{2} ~, 
 \label{ff_proton}
\end{equation}%
with $\Lambda^2=0.71\ $GeV$^{2}$. 
 
In the present work we follow the usual  belief that the phase of the Coulomb-Nuclear 
interference  is based on the superposition of amplitudes in the eikonal formalism 
\cite{phase}. In Appendix {\ref{coulomb_phase}} we present  the calculation 
of  the Coulomb phase  adequate for  the amplitudes (\ref{real_TR}) and (\ref{imag_TI}).

The expressions
 for the  derivatives of the amplitudes are   
\begin{equation}
\frac{d}{dt}T^N_R(s,t) \Big|_{t=0}= 
   \frac{1}{4 \sqrt{\pi}\left(\hbar c\right)^{2}} ~ \sigma \big(\frac{\rho B_R}{2} -\mu_R \big)  ~   
\label{derivative_TR}    
\end{equation}
and 
\begin{equation}
\frac{d}{dt}T^N_I(s,t) \Big|_{t=0}= 
   \frac{1}{4 \sqrt{\pi}\left(\hbar c\right)^{2}} ~ \sigma \big(\frac{B_I}{2} -\mu_I \big)  ~.  
\label{derivative_TI}    
\end{equation}
The combinations of parameters
\begin{equation}
D_I  = \frac{B_I}{2} -\mu_I  ~   
\label{coefficient_I}    
\end{equation} 
and 
\begin{equation}
D_R  = \frac{\rho  B_R}{2} -\mu_R  ~,   
\label{coefficient_R}    
\end{equation} 
entering respectively as input and output in DRS, 
are directly related with data, and 
are crucial for the determination of  $\rho$, $\mu_R$, $B_R$.
 
It must be noted that the usual direct evaluation of the exponential  behavior in 
 $d\sigma(t)/dt=(d\sigma(t)/dt)(t=0)\times {\rm exp} (Bt)$ 
using a straight line for the measurements, actually 
informs the combined average 
\begin{equation}
B ~ = 2\frac{ D_I+\rho D_R}{1+\rho^2} = 
~ \frac{(B_I-2 \mu_I)+\rho (\rho B_R -2  \mu_R)}{1+\rho^2} ~. 
\label{average_B}
\end{equation}  
In a complete analysis of  data  all quantities in this expression, and not only
the average slope $B$ in   $d\sigma/dt$, must be determined.     

 The  forms written above for $T_I(t)$  and $T_R(t)$  are   representations valid 
for small $|t|$, of  amplitudes  for the full $t$ range, studied 
in several models \cite{models,us_LHC} at ISR/CERN and higher energies
that stress the peculiar properties of the real part of the elastic amplitude 
  with   common features of strong slope $B_R$ and a zero  for small $|t|$.  
In the low energy  region here studied, there are not sufficient data for large $|t|$, 
and the analysis is restricted to the forward forms of  Eqs.(\ref{real_TR},\ref{imag_TI}),  
showing that all quantities  (for pp and p\=p) in these expressions 
are necessary and sufficient  for the description of data obeying constraints 
from DRA and  DRS.

At very low energies, namely below ${\rm p_{LAB}}=4$ GeV the description  of 
elastic processes, are influenced  by details of  quark-quark and quark-antiquark 
interactions, with account for specific intermediate states, as for example in a 
framework of partial waves \cite{very_low_energies}. The measurement of polarized amplitudes 
\cite{polarization}, not considered here, depend on the precise 
values of  nonpolarized quantities, as we  obtain  in the present work.  
 In the energy range of our study, 
gluonic interactions are present, with global dynamics that is describable by 
simple analytical forms in the variables $s, t$.  
 
With total cross sections, imaginary slopes and  the linear terms $\mu_I$ 
written as  analytical forms  with powers and logarithms in the energy, 
 both  DRA and DRS require evaluation of principal value (PV)  integrals 
with the generic structure 
\begin{eqnarray}
I(n,\lambda,x)=
{\bf P}\int_{1}^{+\infty}\frac{x^{\prime\lambda} \, \log^n (x^\prime)}{ x^{\prime 2}-x^2}\,dx^\prime  ~. 
\label{PV_int_1}
\end{eqnarray} 
In recent studies, we have obtained the analytic exact  solution for these integrals 
in terms of the Lerch's transcendents \cite{Exact}, and these solutions are applied in the 
present work, with demonstration that they are of fundamental importance, 
particularly in the low energy range.

The mathematical formalism of our work is presented in Sec. \ref{formalism}
and the energy dependent inputs of the imaginary parts are written, with forms 
that are shown to be valid up to LHC energies  and also predict correctly   
  the integrated elastic cross sections. 
  
In Appendix  {\ref{coulomb_phase}}  we calculate  the phase of the Nuclear-Coulomb 
interference  for  real amplitude of the  form of  Eq.(\ref{real_TR}). 

In  Appendix \ref{DR_expanded}  we present in explicit form the calculation of dispersion 
relations for the amplitudes (DRA)  and for  their  derivatives (DRS) with the exact 
solutions in  terms of Lerch's transcendents.

In Appendix \ref{Landshoff} we present alternative equivalent formalism for 
the total cross section in the  language of   Pomeron and Reggeon trajectories. 

 With established energy dependent inputs $\sigma $(pp,p\=p), $B_I$(pp,p\=p), 
$\mu_I$ (pp,p\=p)  we give in Sec.\ref{data_analysis}    precise  description of the 
  forward range of elastic pp and $\rm {p \bar p} $ scattering, with 
the essential identification and separation  of the real and imaginary amplitudes, 
with coherence and regularity in the energy dependence  all quantities.

In Sec.\ref{conclusions-sec}  we present conclusions and summarize   achievements 
of our effort. 

\clearpage

\section{Formalism and Inputs for Dispersion Relations  \label{formalism}  } 

\subsection{ Inputs for Imaginary Part of Elastic Amplitude {\label{input} } }

 In this section we  introduce  the  forms of the imaginary part of the elastic amplitudes, 
and explain the determination of their parameters. 
 We stress that we only use qualified data on $d\sigma/dt$ that may be considered as able to 
allow reliable analysis in terms of amplitudes written in the analytical forms of
Eqs.(\ref{real_TR},\ref{imag_TI}). 
The method of construction of our proposal is interactive, with inputs and outputs nourishing 
each other.  In  a first free analysis, we obtain values of parameters for $\sigma$, $B_I$ 
and $\mu_I$, while the quantities of the real part are left  free.   
 The extracted values  are put in regular 
behaviour with the energy, leading to    analytical forms with   terms 
of powers and logarithms as described below. Adopting these representations for the imaginary 
amplitude, we use exact forms of dispersion relations for amplitudes (DRA) and for slopes (DRS) 
  to obtain the quantities of the real part, and then we review the imaginary  amplitude.   

In the reported experiments at  low energies the momentum in the lab system  $\rm p_{LAB}$ 
is more often used, while at high energies the use of the  the center of mass energy $\sqrt{s} $ 
is more common. For pp and $\rm p \bar p$ scattering the connection with the lab energy 
\begin{equation}
 E=\sqrt{\rm p_{LAB}^2+m^2} 
\label{LAB_energy}
\end{equation}
  is
\begin{equation}
s=2mE+2m^2 ~,
\label{cm_energy}
\end{equation}
where $m$ is the p/\=p mass.
To work with the dispersion relations, the most useful quantity is the dimensionless
ratio
\begin{equation}
x=E/m   
\label{x_variable}
\end{equation}
and then
\begin{equation}
\frac {s}{2m^2}= x+1 ~.
\label{s_x_relation}
\end{equation}
Approximate relations that are often used at high energies are obviously $s\approx 2mE$, 
$x \approx s/2m^2$,
$x \approx {\rm p_{LAB}}/m $. 
  
The usual parametrizations \cite{PDG}  for the total cross sections
of the pp and p\=p interactions  has  the   forms
\begin{equation}
\sigma^\mp(s)= P + H  \log^2\left(\frac{s}{s_0}\right)+R_1 \left(\frac{s}{s_0}\right)^{-\eta_1 }  
     \pm\, R_2  \left(\frac{s}{s_0}\right)^{-\eta_2 },   \label{sigma_s}
\end{equation}
with parameters $P,~ H, ~ R_1, ~ R_2  $ 
constants given in millibarns,
$s_0$ in $\GeV^2$, while $\eta_1, ~\eta_2 $  are  dimensionless.  The upper and lower indices 
$-$,$+$   refer to pp and  p\=p  scattering respectively. The representation is considered to be 
adequate for all energies $s \geq s_0$ 
and is based on a large number of values of $\sigma$ (pp, p\=p) found in experimental papers. 
We  have the radical claim that these values often are not of good precision, for several reasons, 
and in general because the optical theorem  is not applied to well  identified imaginary 
amplitudes.   

  This   form of amplitude is based on the Pomeron/Reggeon dynamics  assumed 
for the strong interactions \cite{DL}, and refers only to   purely elastic processes.
Contributions of diffractive nature, as first studied by Gribov and later formulated
by Good and Walker \cite{GW} are not included in this framework. Single diffractive, 
double diffractive and truly inelastic processes have not been measured in the energy 
range of our study, while  theoretical \cite{Tel-Aviv} work based on the gluonic 
dynamics of QCD and  measurements start at  energies of the ISR/CERN  and Fermilab 
experiments \cite{data_HE}, namely $\sqrt{s}\geq \approx 30 \GeV$. 
We show indeed that our treatment describes well the integrated elastic and total 
cross sections  at these higher energies.

 Dispersion relations are defined with
respect to the lab system energy, and, for low energies, terms like $\log^2(E+m)$ and
$(E+m)^{-\eta}$ appear  preventing to obtain  closed exact forms.
We then  obtain a representation for the  total cross sections in
terms of dimensionless variables
$x=E/m$, $x_0=E_0/m$,  with $x>1$, writing
\begin{equation}
\label{sigma_x}
\sigma^\mp(x) =  P + H  \log^2\left(\frac{x}{x_0}\right)+R_1 \left(\frac{x}{x_0}\right)^{-\eta_1}  
      \pm\, R_2 \left(\frac{x}{x_0} \right)^{-\eta_2 },   
\end{equation}
 and   analyse all   $d\sigma/dt$ data for  ${\rm p_{LAB}}$ from 4 to 500 $ \GeV/c$  
using the assumed structures of the amplitudes in  Eqs.(\ref{real_TR},\ref{imag_TI}). 

  We keep the   value $s_0=16~ \GeV^2$   
suggested \cite{PDG}  for Eq. (\ref{sigma_s}), now  appearing  as  
\begin{equation}
x_0 ~= ~s_0/(2 m^2) ~ = ~9.0741 ~,  
\end{equation} 
where $m=0.93827 \GeV$. 
With this choice, the parameters $P$, $H$, $R_1$, $R_2$ remain the same.
 and  numerical values are the same, given in Table \ref{table:inputs}. 
With 
$$ \left(\hbar c\right)^{2}  = 0.38938 ~ \GeV^2 ~ {\rm mb}  $$ 
 we also need   
  $$ \left(\hbar c\right)^{2}\frac{1}{m^2} = 0.4423 ~ {\rm mb} ~. $$

\begin{figure}[h]
      \includegraphics[width=8.0cm]{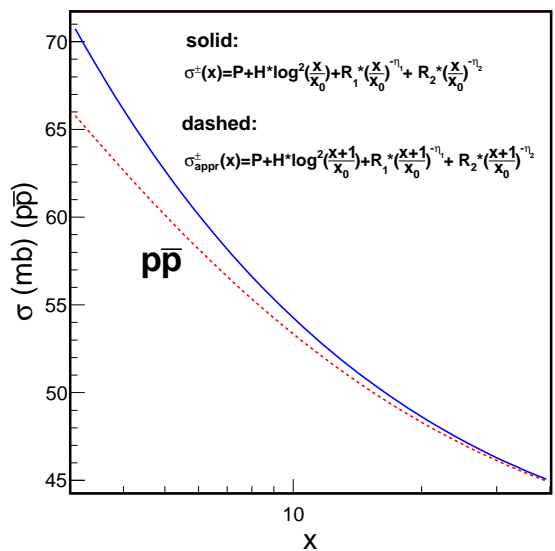}   
   \caption{\label{compare_sigma} Comparison between  calculations of $\rm {p \bar p} $ 
cross sections with   Eqs.(\ref{sigma_s})
 and (\ref{sigma_x}) using same parameter values, showing the deviation for low energies due to kinematics. } 

     \end{figure}

In Fig.\ref{compare_sigma} we show the comparison between Eqs. (\ref{sigma_s})
 and (\ref{sigma_x}) for p\=p  written with  the  same parameters. The difference between the curves 
represents the deviation of Eq.(\ref{s_x_relation}) to the approximated form $x \approx s/2m^2$.

For use  as inputs in  dispersion relations DRS we write the slopes  $B_I^{\rm pp}(x)$  
and $B_I^{\rm p\bar p}(x)$  in terms of the $x$ variable as 
    \begin{equation}
B_I^\mp(x) = b_0\! +\!  b_1 \log{x} + \!  b_2  \log^2{x} 
+\! b_3 ~ x^{-\eta_3}\! ~ \pm \!  b_4 ~ {x}^{-\eta_4} \! ,\hspace{5pt}
  \hspace{5pt}  \label{BI_x}
\end{equation}
 again with symmetry in the coefficients for    p\=p and pp.   
  As in  Eqs.(\ref{sigma_s}) and (\ref{sigma_x}),  for $s/2 m^2 >>1$
the slopes $B_I$ of Eq.(\ref{BI_x})    can be written with similar analytical 
forms in the variable $s$. 

In  Eq.(\ref{imag_TI}) we have in addition to the slope of the imaginary part, $B_{I}$, 
a term $\mu_I$ which is linear on $t$ dependence. The inputs $\mu_{I}(x)$ for both pp 
and p\=p are determined by a controlled   analysis. The result is that the difference 
between  the values of  $\mu_{I}(x)$ for pp and p\=p are not important, and we assume 
for both the form 
  \begin{equation}
\mu_I^\mp(x) = \mu_0 +\mu_1 \log{x} ~.
  \label{MUI_x}
\end{equation}

\begin{figure*}[h]
        \includegraphics[width=8.0cm]{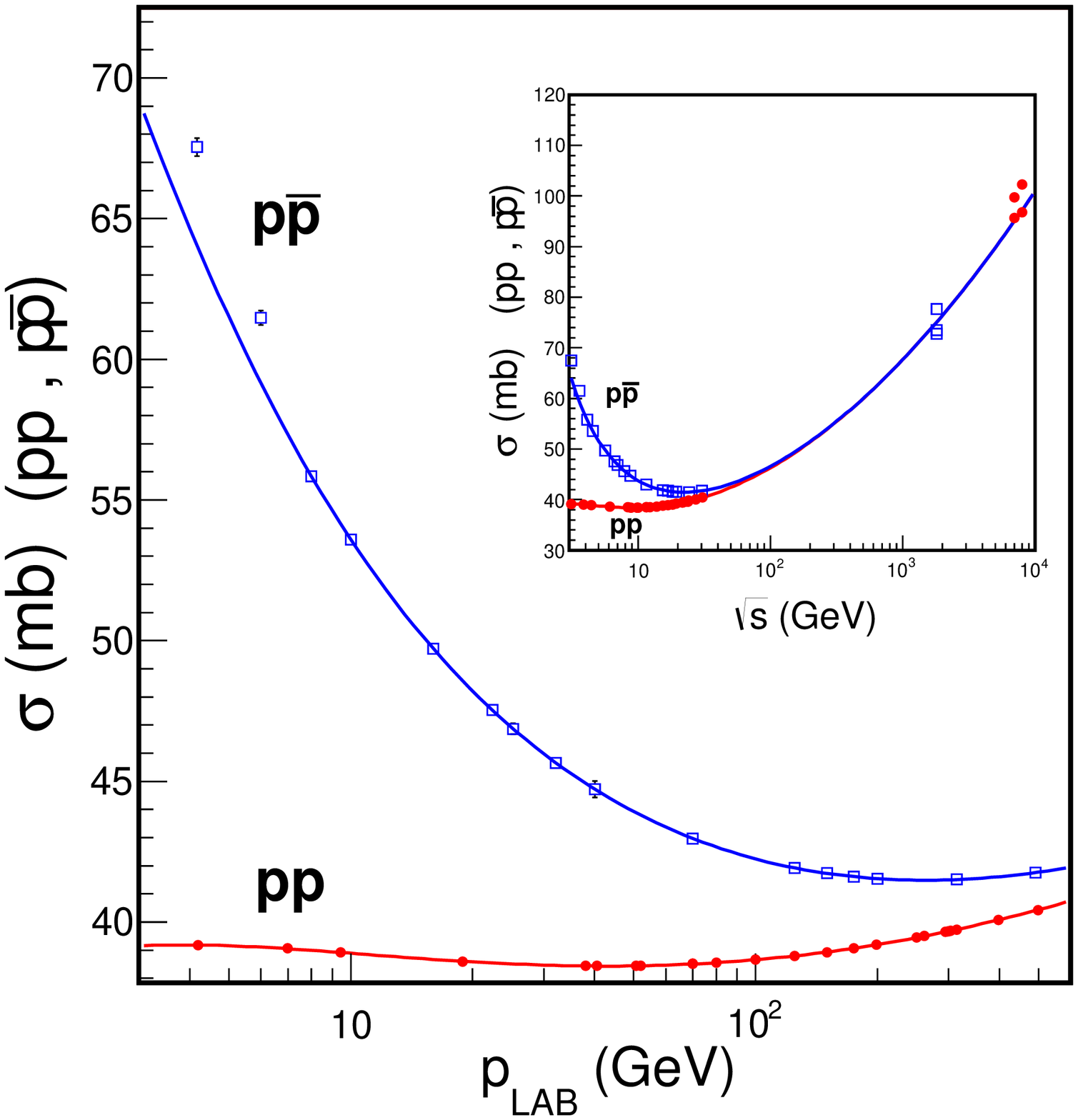}
    \includegraphics[width=8.0cm]{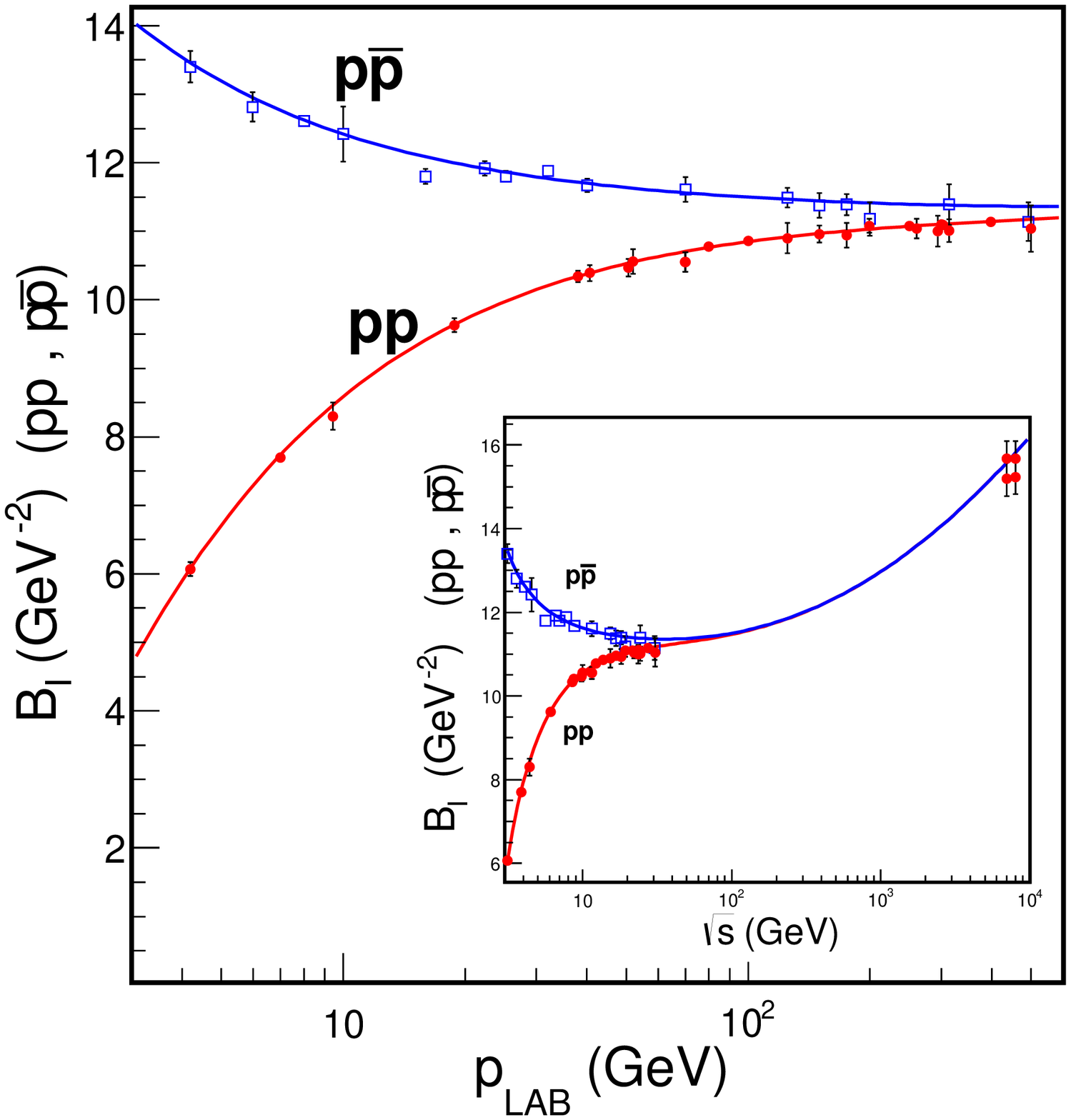}   
    \includegraphics[width=8.0cm]   {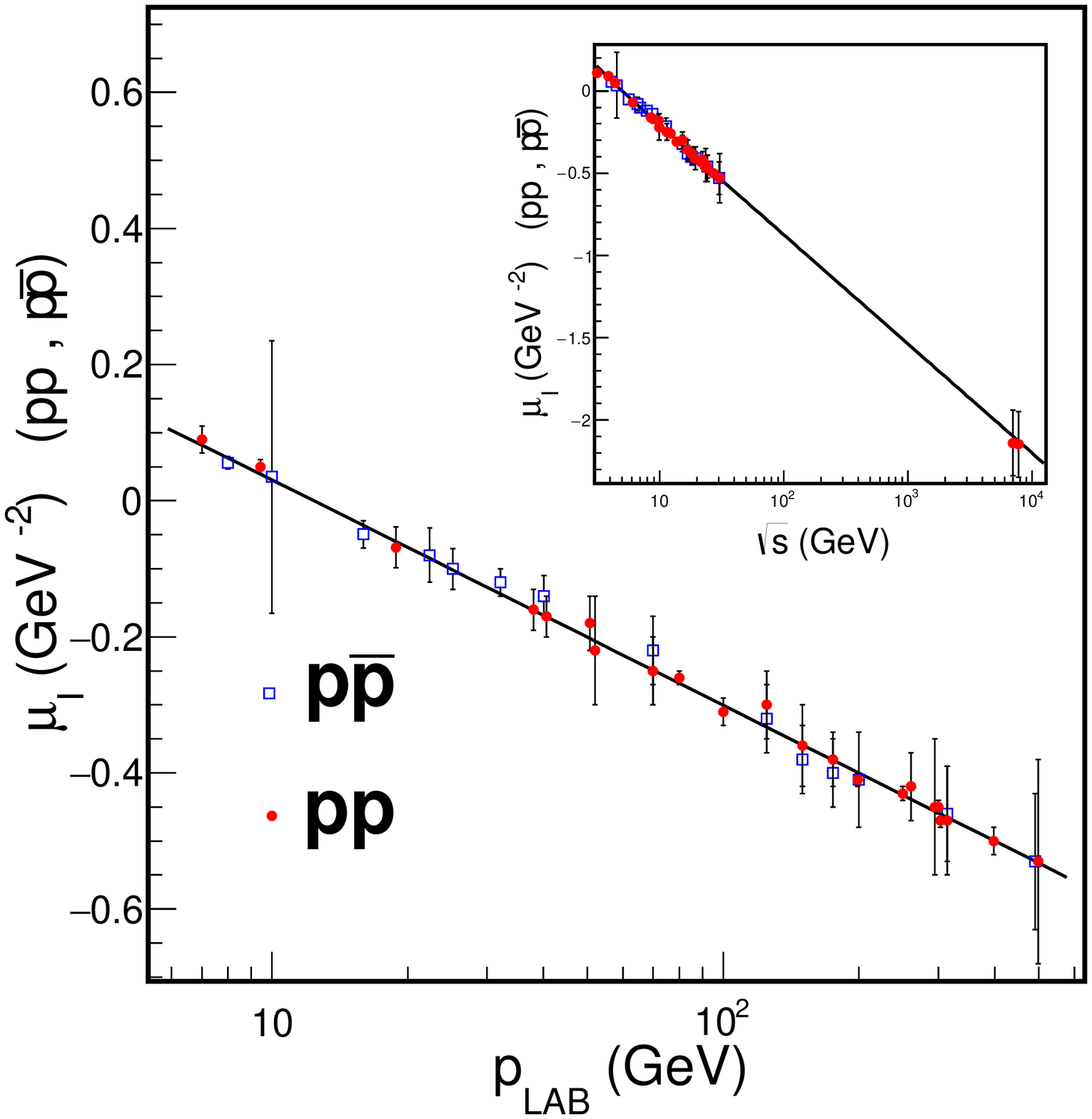}
   \caption{\label{inputs_fig} Input forms for $\sigma$, $B_I$ and ${\mu_I}$  of 
Eqs.(\ref{sigma_x},\ref{BI_x},\ref{MUI_x}), together with values obtained in the 
study of the data in Sec. \ref{data_analysis}. 
The connections of the analytical forms with our previously published calculations 
\cite{us_LHC} for very high energies (1.8 TeV and LHC) are  shown in insets in 
terms of $\sqrt{s}$. }
   \end{figure*}

The    numerical  values of the input parameters are given in Table \ref{table:inputs} 
and in Fig.\ref{inputs_fig} we show the quantities $\sigma(x)$, $B_I(x)$ 
and $\mu_I(x)$ of  Eqs.(\ref{sigma_x}),(\ref{BI_x}) and (\ref{MUI_x}) with the points obtained 
in the examination of the data  described  in Sec.\ref{data_analysis}. In the inset 
plots   with the variable $\sqrt{s}$  we show that the extrapolations  up to LHC 
energies (7 and 8 TeV) are compatible with the predicted results \cite{us_LHC,LHC_2017}. 
 \begin{table*} 
 \setlength{\tabcolsep}{3pt}
\caption{ Parameters of total cross section, slopes and linear terms of imaginaty parts
in Eqs.(\ref{sigma_x},\ref{BI_x},\ref{MUI_x}).}
\centering 
\begin{tabular}{c c c c c c c    } 
\hline\hline 
$ \sigma(x) $   &   &   &  &  &     &    \\
\hline 
$P({\rm mb}) $  & $H ({\rm mb}) $ & $R_1 ({\rm mb}) $ & $R_2({\rm mb}) $ & $\eta_1$ &  $ \eta_2$ &       \\
 \hline 
$34.37\pm 0.13$ &$ 0.272 \pm 0.00$  & $12.74 \pm 0.09$ & $ 8.143 \pm 0.180 $  & $ 0.4288 \pm 0.0100 $ &  $ 0.6144 \pm 0.0090$  &   \\
\hline \hline \\
$ B_I (x) $   &     &   &  &  &    &     \\
\hline 
$b_0 (\GeV^{-2})$ & $b_1(\GeV^{-2}) $      & $b_2(\GeV^{-2})$ & $b_3(\GeV^{-2})$ & $b_4(\GeV^{-2})$ &  $ \eta_3$&  $ \eta_4$    \\
 \hline 
$ 13.79\pm 0.12$  & $ - 0.625 \pm 0.070$ & $0.04255 \pm 0.01000 $ & $ -6.937 \pm 0.120 $ & $ 11.95 \pm 0.21 $ & $ 0.5154 \pm 0.0060$ &
              $ 0.772\pm 0.006$    \\
\hline \hline  \\
$ \mu_I (x) $   &     &   & $   $  &   &    &     \\
\hline 
$\mu_0 (\GeV^{-2})$ & $\mu_1(\GeV^{-2}) $      & $-$ & $-$ & $-$ &  $ -$&  $ -$    \\
 \hline 
$ 0.3724\pm 0.0096$  & $ -0.1441 \pm 0.0021$ & $- $ & $ - $ & $- $ & $ -$ &
               $ -$    \\
\hline \hline  
\end{tabular} 
\label{table:inputs} 
\end{table*}    

\begin{figure*}[h]
      \includegraphics[width=8.0cm]{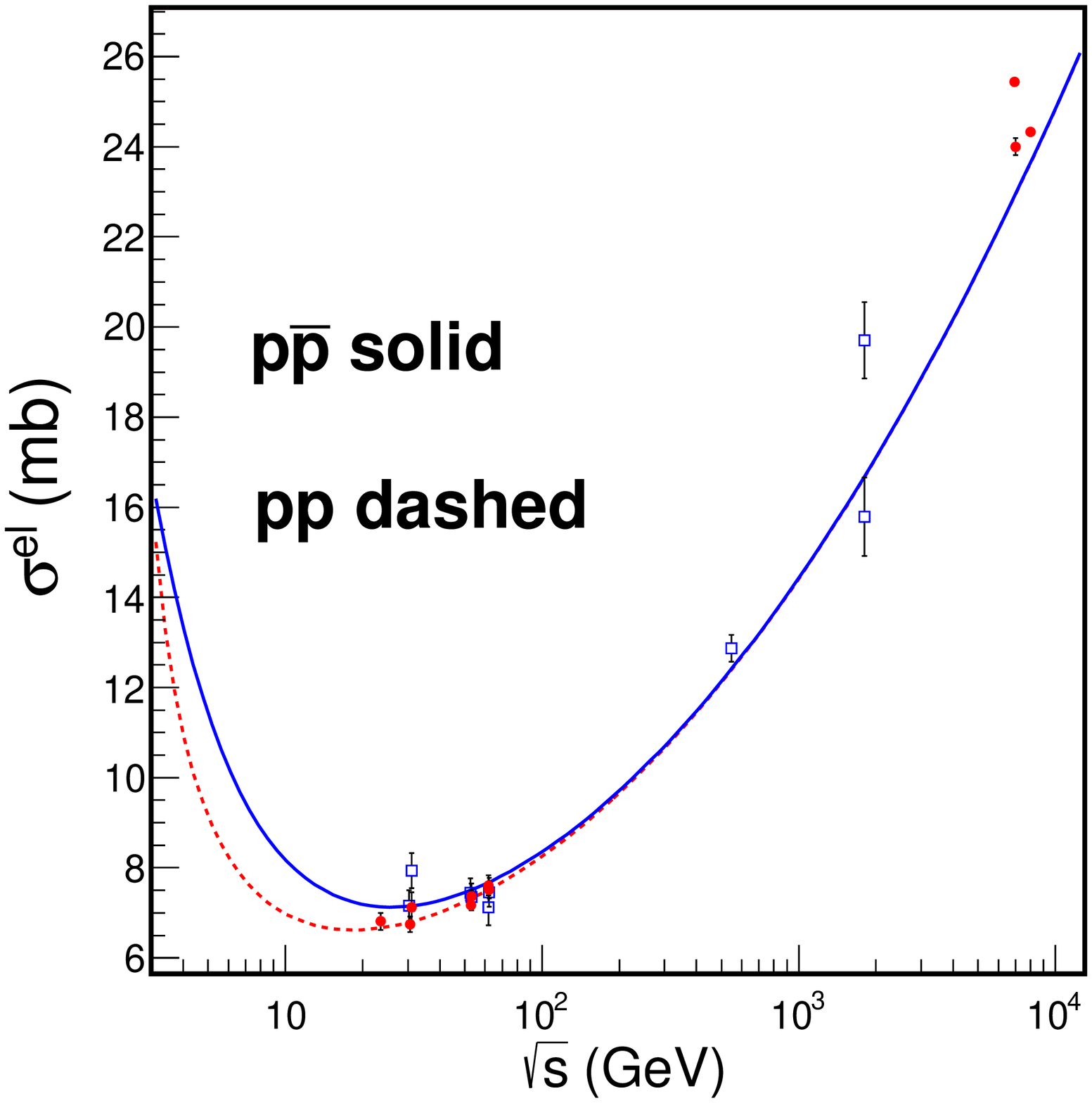}   
         \includegraphics[width=8.0cm]{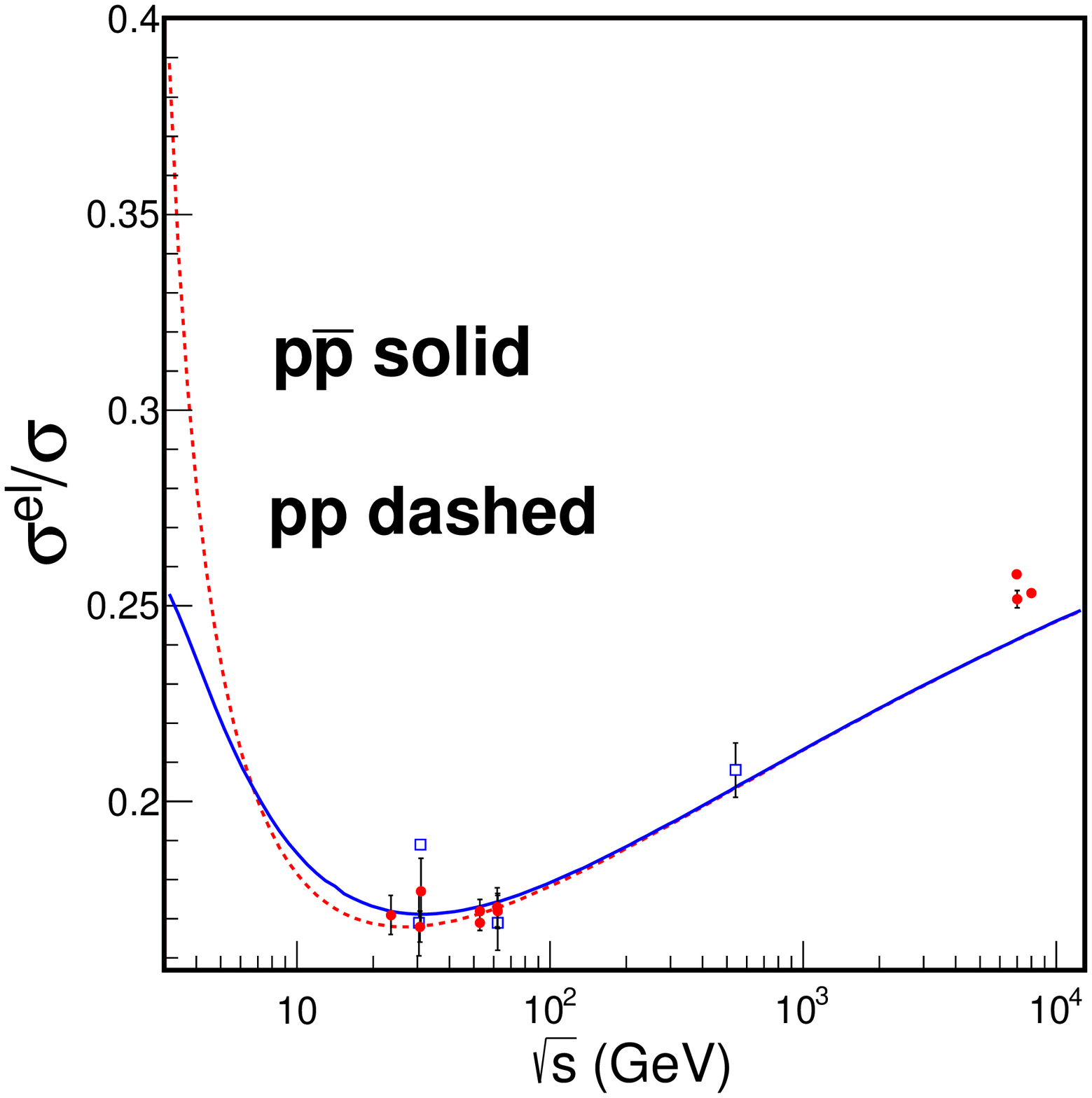}
   \caption{\label{sig_elastic}
(a) Integrated elastic cross section due to the  imaginary part of the amplitude. Recall that 
the difference $\sigma -\sigma_I^{\rm elastic}$ accounts for diffractive plus inelastic processes.
The large difference  indicates that diffractive processes are highly dominant 
at low energies.   Experimental points \cite{data_HE}  from ISR/CERN, Fermilab (1.8 TeV)  
and LHC (7 and 8 TeV)  \cite{data_LHC} are marked; 
(b) The same plot, for the ratio  $\sigma_I^{\rm elastic}/\sigma$ ~.
The agreement of the lines with the data points shown that our amplitudes are realistic.} 
     \end{figure*}

The integrated   elastic cross section of the imaginary part is given by 
\begin{eqnarray}
&& \frac{\sigma_I^{\rm el}}{\sigma}(x)=\frac{1}{\sigma}\int_0^{-\infty} \frac{d\sigma_I}{dt} \nonumber \\
&& = \frac{1}{16 \pi   \left(\hbar c\right)^{2} }\frac{\sigma }{B_I }
   \big[     \big(1+\frac{\mu_I}{B_I}\big)^2 +\frac{\mu_I^2}{B_I^2}  \big] ~ .   
\label{elastic_I}
\end{eqnarray} 
Plots of  $\sigma^{\rm elastic}_I$ and the ratio with $\sigma$ as function of the energy are 
given in Fig.\ref{sig_elastic}.  We recall  that this expression gives the ratio of the purely 
elastic processes.  The remaining part of the ratio $1- \sigma_I^{\rm el}/\sigma$ gives diffractive 
plus inelastic processes. We also mark in the figure the experimental values of  $\sigma^{\rm elastic}$ 
in the ISR range ($\sqrt{s} \approx$ 30 to 60 GeV) \cite{data_HE} and our published calculations for 1.8 GeV and LHC 
energies \cite{us_LHC}.  

 The dimensionless Fourier transforms of the amplitude $T(s,t)$ in 
Eqs.(\ref{real_TR},\ref{imag_TI}). 
  with respect to the momentum transfer
\begin{equation}
\tilde{T}(b;s)=\tilde{T}_{R}+i\tilde{T}_{I}
\label{Tbspace}
\end{equation}
are given by 
\begin{equation}
\tilde{T}_{R}\left(  b;s\right)  =\frac{\sigma}{2\pi B_{R}}\Bigg\{
\rho+\frac{\mu_{R}}{B_{R}}\left(  2-\frac{b^{2}}{B_{R}}\right) \Bigg\}  e^{-b^{2}/2B_{R}}  
\label{Tb_real}
\end{equation}
and
\begin{equation}
\tilde{T}_{I}\left(  b;s\right)     =\frac{\sigma}{2\pi B_{I}}\left\{
1+\frac{\mu_{I}}{B_{I}}\left(  2-\frac{b^{2}}{B_{I}}\right)  \right\}
e^{-b^{2}/2B_{I}}.
\label{Tb_imag}
\end{equation}
The profile corresponding to the imaginary forward  amplitude   
is  dominant over the real part for low $b$ values. However from $b\geq 14$ GeV ($ \simeq 2.8$ fm) 
the real part can be dominant and this effect is more pronounced due to the presence of the $\mu_{R}$ parameter in Eq.(\ref{Tb_real}). 

In Appendix \ref{Landshoff} alternative forms are written for the total cross section, $\sigma(x)$ or $\sigma(s)$,
 in terms of power instead of logarithm as in Donnachie-Landshoff formalism, with all accuracy.

\clearpage

\subsection{ \label{DR}  Dispersion Relations for Amplitudes and Slopes} 

 The well known dispersion relations for pp and p\=p  elastic scattering are
written in terms of even and odd dimensionless amplitudes,
\begin{equation}
  {\rm Re}\,F_+(E,t) =  K +
\frac{2E^2}{\pi} \,{\bf P} \int_{m}^{+\infty}dE^\prime \, \frac{ {\rm Im}\, F_+(E^\prime,t)}{E^\prime(E^{\prime 2}-E^2)} \, , 
\label{inteven}  
\end{equation}
\begin{equation}  
  {\rm Re}\,F_-(E,t)   =
\frac{2E}{\pi} \,{\bf P}\int_{m}^{+\infty}dE^\prime \, \frac{ {\rm Im}\, F_-(E^\prime,t)}{(E^{\prime 2}-E^2)} \, .
\label{intodd}
\end{equation}
With $x=E/m$, the even and odd combinations of amplitudes are related to the pp and p\=p systems through
\begin{eqnarray}
F_{\rm pp}(x,t)=F_+(x,t)-F_-(x,t),  \nonumber  \\ 
F_{\rm {p \bar p} }(x,t)=F_+(x,t)+F_-(x,t) ~.
\label{F_pp}
\end{eqnarray}
The optical theorem informs the normalization of the amplitudes  by
\begin{eqnarray}
\sigma_{\rm pp}(x)= \frac{{\rm Im} ~F_{\rm pp}(x,t=0)}{2m^2x} ~   
\label{teorema_otico}
\end{eqnarray}
and similarly for p\=p. 

With the  exponential and  linear  factors  in the imaginary parts,
we write   the inputs
\begin{eqnarray}
 && \frac{{\rm Im} ~F_{\rm pp}(x,t)}{2\,m^2\,x} = \sigma_{\rm pp}\, [1-\mu_I^{\rm pp}] \exp \left(B_I^{\rm pp}\,t/2\right)\,, 
 \label{F_pp2} \\
 && \frac{{\rm Im} ~F_{\rm p\bar p}(x,t)} { 2\,m^2\,x} = \sigma_{\rm p\bar p}\,[1-\mu_I^{\rm p \bar p}]\exp \left(B_I^{\rm p\bar p}\,t/2\right)\,,  \label{F_ppbar2}
\end{eqnarray}
with  functions $\sigma(x)$, $B_I(x)$ and  $\mu_I(x)$  for pp and p\=p  given in Eqs.(\ref{sigma_x}),
(\ref{BI_x}) and (\ref{MUI_x}).

As explained in  Sec. \ref{introduction}, the real parts are written  with  exponential  and linear factors 
  \begin{eqnarray}
&&  \frac{{\rm Re} ~F_{\rm pp}(x,t)}{2m^2x} \nonumber \\
&& = \sigma_{\rm pp}(x)[\rho_{\rm pp}(x) -\mu_R^{\rm pp}(x) t  ] ~ \exp[{B_R^{\rm pp}(x)t/2}] ~  
\label{real_forw}
\end{eqnarray}
and similarly for  p\=p.

The $\rho$ parameters are then obtained from 
\begin{equation}
    \frac{1}{2 m^2 x}\,{\rm Re}~ F_+(x,0) = 
  \frac{1}{2}\Big[(\sigma\rho) ({\rm p \bar p})\! +\! (\sigma\rho) ({\rm pp})\Big]  
 \end{equation}
and 
 \begin{equation} 
   \frac{1}{2 m^2x}{\rm Re}\,F_-(x,0)
   =\frac{1}{2}\Big[(\sigma\rho) ({\rm p\bar p})-(\sigma\rho) ({\rm p  p}) \Big]~  ~,  
\end{equation}
with the LHS given by dispersion relations (\ref{inteven}) and (\ref{intodd}).  

Thus the $\rho$ parameter of the real part is defined by 
 \begin{eqnarray}
\sigma_{\rm pp}(x)\rho_{\rm pp}(x)= \frac{{\rm Re} ~F_{\rm pp}(x,t=0)}{2m^2x}  \, ,    
\label{rho_defined}
\end{eqnarray}
and similarly for p\=p.

 The  derivatives of the real amplitude  at $|t|=0$ are written 
 \begin{equation}
  \frac{\partial{\rm Re}\,F_{\rm pp}(x,t)}{\partial t}\Big|_{t=0}   
  = {2\,m^2\,x }\,  \sigma_{\rm pp}(x) [\frac{\rho_{\rm pp}  B_R^{\rm pp}}{2}  -\mu_R^{\rm pp}](x) \, , 
 \label{derivative_real_pp}
 \end{equation}
 and similarly  for p\=p.
These quantities are determined by DRS, which give  the even and odd  combinations 
 \begin{equation}
  \frac{1}{2m^2 x}\frac{\partial{\rm Re}\,F_+(x,t)}{\partial t}\Big|_{t=0}   
  =  \frac{1}{2} \big[ \sigma_{\rm p \bar p} D_R^{\rm p \bar p}+\sigma_{\rm pp} D_R^{\rm pp} \big]
   \label{drdtplus}
\end{equation}
and  
 \begin{eqnarray}
  \frac{1}{2m^2 x}\frac{\partial{\rm Re}\,F_-(x,t)}{\partial t}\Big|_{t=0}  
  =\frac{1}{2} \big[ \sigma_{\rm p \bar p} D_R^{\rm p \bar p}-\sigma_{\rm pp} D_R^{\rm pp} \big] ~.   
\label{drdtminus}
\end{eqnarray}
The quantities $D_R^{\rm p \bar p} $ and $D_R^{\rm p p} $ are combinations of amplitude 
parameters as in Eq.(\ref {coefficient_R}). 

 Substituting these expressions into  Eqs.~(\ref{inteven}) and (\ref{intodd}), written in terms 
of the dimensionless variable $x$, we obtain
\begin{eqnarray}
&&{\rm Re}\,F_+(x,t) = K + \frac{2\,m^2\,x^2}{\pi}\, \nonumber  \\
&& {\bf P}\!\int_{1}^{+\infty}\frac{1}{x^{\prime 2}-x^2}\,\Big[\sigma_{\rm p\bar p}(x^{\prime})
       (1-\mu_I^{\rm p\bar p}(x^{\prime}))
\,\exp \left[B_I^{\rm p\bar p}(x^{\prime})\,t/2\right] \nonumber \\
 &&  +\,\sigma_{\rm pp}(x^{\prime})(1-\mu_I^{\rm p p}(x^{\prime}))
\,\exp \left[B_I^{\rm pp}(x^{\prime})\,t/2\right]\Big]\,dx^\prime\,       
\label{xinteven}
\end{eqnarray}
and  
\begin{eqnarray}
&&{\rm Re}\,F_-(x,t) = \frac{2\,m^2\,x}{\pi} \nonumber  \\ 
&&   {\bf P}\!\int_{1}^{+\infty}\frac{x^\prime}{x^{\prime 2}-x^2}\,\Big[\sigma_{\rm p\bar p}(x^{\prime})
(1-\mu_I^{\rm p\bar p}(x^{\prime}))
\,\exp \left[B_I^{\rm p\bar p}(x^{\prime})\,t/2\right]  \nonumber \\
 &&    -\,\sigma_{\rm pp}(x^{\prime})
(1-\mu_I^{\rm pp}(x^{\prime}))
\,\exp \left[B_I^{\rm pp}(x^{\prime})\,t/2\right]\Big]\,dx^\prime\,.   
\label{xintodd}
\end{eqnarray}
Taking t=0 we obtain the Dispersion Relations for the Amplitudes (DRA)
\begin{eqnarray}
&&{\rm Re}\,F_+(x,t=0)={2\,m^2\,x }\, [\sigma_{\rm p\bar p}\rho_{\rm p\bar p}+\sigma_{\rm pp}\rho_{\rm pp}](x) = K     \nonumber \\
&& +  \frac{2\,m^2\,x^2}{\pi}\,{\bf P}\!\int_{1}^{+\infty}\frac{1}{x^{\prime 2}-x^2}\,\Big[\sigma_{\rm p\bar p}   
  +\,\sigma_{\rm pp}  \Big](x^{\prime})\,dx^\prime\,     
\label{DR_even}
\end{eqnarray}
and 
\begin{eqnarray}
&&{\rm Re}\,F_-(x,t=0)  = {2\,m^2\,x }\, [\sigma_{\rm p\bar p}\rho_{\rm p  \bar p}-\sigma_{\rm pp}\rho_{\rm pp} ](x)      \\ 
&& = \frac{2\,m^2\,x}{\pi}\,{\bf P}\!\int_{1}^{+\infty}\frac{x^\prime}{x^{\prime 2}-x^2}\,\Big[\sigma_{\rm p\bar p}     
     -\,\sigma_{\rm pp}\Big](x^{\prime})\,dx^\prime \, . \nonumber 
\label{DR_odd}
\end{eqnarray}

The expressions in terms of PV integrals are given in Appendix \ref{DR_expanded}.

To obtain DRS, we take derivatives of Eqs.~(\ref{xinteven}) and (\ref{xintodd}) with respect to $t$,
 writing 
 \begin{eqnarray}
&& \frac{\partial{\rm Re}\,F_+(x,t)}{\partial t}  = \frac{m^2\,x^2}{\pi}{\bf P}\!\int_{1}^{+\infty}
\frac{1}{x^{\prime 2}-x^2}  \nonumber \\
&& \Big[\sigma_{\rm p\bar p}(x^{\prime})\,[B_I^{\rm p\bar p}-2\mu_I^{\rm p\bar p} ](x^{\prime})   \,\exp \left[B_I^{\rm p\bar p}(x^{\prime})\,t/2\right]   \\
 &&   +\,\sigma_{\rm pp}(x^{\prime})\,[B_I^{\rm pp}(x^{\prime})-2\mu_I^{\rm pp}(x^{\prime})]  \,\exp \left[B_I^{\rm pp}(x^{\prime})\,t/2\right]\Big]\,dx^\prime\,,  \nonumber      \label{A even}  \\
&& \frac{\partial {\rm Re}\,F_-(x,t)}{\partial t}= \frac{m^2\,x}{\pi} {\bf P}\!\int_{1}^{+\infty}
\frac{x^\prime}{x^{\prime 2}-x^2} \nonumber \\ 
&&  \Big[\sigma_{\rm p\bar p}(x^{\prime})\,[B_I^{\rm p\bar p}
-2\mu_I^{\rm p\bar p} ](x^{\prime}) 
\,\exp \left[B_I^{\rm p\bar p}(x^{\prime})\,t/2\right]   \\
 &&    -\,\sigma_{\rm pp}(x^{\prime})\,
[B_I^{\rm pp}(x^{\prime})-2\mu_I^{\rm pp}(x^{\prime})]
\,\exp \left[B_I^{\rm pp}(x^{\prime})\,t/2\right]\Big]\,dx^\prime\,. \nonumber 
\label{A odd}
\end{eqnarray}
With t=0, these equations  become the Dispersion Relations for Slopes (DRS) that we may write 
 \begin{eqnarray}
&&  4 \, \big[\sigma_{\rm p\bar p} D_R^{\rm p \bar p} + \sigma_{\rm p p}D_R^{\rm pp} \big](x) \\
&& = \frac{x}{\pi}\,{\bf P}\!\int_{1}^{+\infty}
\frac{2}{x^{\prime 2}-x^2}\Big[\sigma_{\rm p\bar p}D_I^{\rm p \bar p}           
   +\,\sigma_{\rm pp}
D_I^{\rm pp}(x^{\prime})
  \Big](x^{\prime})     \,dx^\prime\,,  \nonumber 
   \label{DRS_even}  \\
&&   4   \, \big[\sigma_{\rm p\bar p}D_R^{\rm p\bar p}-\sigma_{\rm p p}D_R^{\rm pp}\big](x) \\
&&  = \frac{1}{\pi}\,{\bf P}\!\int_{1}^{+\infty}
\frac{2x^\prime}{x^{\prime 2}-x^2}\Big[\sigma_{\rm p\bar p}D_I^{\rm p\bar p}             
   -\,\sigma_{\rm pp}
D_I^{\rm pp} 
  \Big] (x^{\prime}) \,dx^\prime\,.\nonumber
  \label{DRS_odd}
\end{eqnarray}
where   we have introduced the parameterization of the real amplitudes.

In  Eqs.(\ref{DRS_even}) and (\ref{DRS_odd})   the terms $D_I^{pp}$ and $D_I^{p\bar p}$ given by Eq.(\ref{coefficient_I}) keep analytical form similar to that of $B_I(x)$ given by Eq.(\ref{BI_x}), since 
the parametrization of $\mu_I$ is  linear in $\log (x)$.
The presence of the quantity $\mu_I(x)$ as an input in dispersion relations for slopes does not change the algebra of Eqs.(\ref{drdtplus_expanded}), (\ref{drdtminus_expanded}) (which were first written \cite{Exact} assuming $\mu_I=0$) and the contribution of $\mu_I$ can be given by the  change of the parameters 
\begin{eqnarray}
b_0\to b_0^\prime= b_0-2\mu_0 ~, \nonumber \\
b_1\to b_1^\prime = b_1-2\mu_1 ~.  
\label{change}
\end{eqnarray}

Introducing analytical expressions for the terms in the imaginary parts, we fall in principal value 
integrations of the form  (\ref{PV_int_2}) that we can solve exactly \cite{Exact}. 
The input forms in Eqs.(\ref{sigma_x}), (\ref{BI_x}) and (\ref{MUI_x}) taken into the expressions from 
DRA and DRS, with numbers given in  Sec.\ref{data_analysis} lead to values for $\rho$  and the 
coefficients of the 
derivatives   of the real parts at the origin $D_R$   for pp and p\=p. 
In the low energy end, namely with ${\rm p_{LAB}}$ up to 30 GeV, it is essential to use the exact solutions
for the PV integrals     that appear  in DRA and DRS 
in the calculations of $\rho$ and $D_R$.  
Illustrating plots are given in Appendix \ref{DR_expanded}.

  From comparison 
of the results with the   $\rho$ data,   the value of the separation constant   $K$ 
that appears in the expressions of DRA is determined. We obtain  the interval  
\begin{equation}
K  =   {\rm{from}} ~  (-310) ~ {\rm {to} }  ~  (-287)  ~.
\end{equation} 
In the examples and plots of the present paper we use the value $K=-310$ . 


Given the $\sigma(x)$, $B_I(x)$ and $\mu_I$  inputs, the quantities $\rho$ and $D_R$ are determined
by DRA and DRS. Since $D_R=\rho B_R/2-\mu_R$  is a combination of $\mu_R$ and  $B_R$, they must
be determined by the data. 
We obtain  that  $\mu_R$ presents very regular energy dependence   for both 
 pp and  p\=p  systems. We  introduce the forms
\begin{equation} 
 \mu_R({\rm pp})=  c_0+  c_1 ~ x^{-\nu_1}+c_2\log x    
\label{muRpp}
\end{equation}
and 
\begin{equation}
 \mu_R({\rm p\bar p})=  c_3+c_4 ~ x^{-\nu_4}+c_5\log x   ~.    
\label{muRppbar}
\end{equation}
Numerical values for the constants are given in Table \ref{table:muR}   
and plots are shown in  Fig.{\ref{muR_results}.

\begin{table} 
 \setlength{\tabcolsep}{3pt}
\caption{ Values of  parameters of   $\mu_R$ for  pp and p\=p    in Eqs.(\ref{muRpp},\ref{muRppbar}), 
obtained for  $K=-310$.   The corresponding 
lines are shown in Fig.\ref{muR_results}. 
 The central values of $c_0$ and $c_2$  are shown with 
high precision to put coincident the zeros of $\rho$  and $D_R/2+\mu_R$ for the choice $K=-310$. }  
\centering 
\begin{tabular}{c c c c} 
\hline\hline 
 pp  &    &      &    \\
\hline 
$ c_0 (\GeV^{-2}) $  & $ \nu_1$ & $c_1 (\GeV^{-2})$ & $c_2 (\GeV^{-2})$   \\
\hline 
$ 1.897\pm  0.160$ & $0.450 \pm 0.032  $  & $-17.87 \pm 0.92  $ & $-0.142\pm 0.001$   \\ 
\hline \\
{$\rm p\bar p$}   &   $   $ &       &        \\
\hline 
   $c_3 (\GeV^{-2})$ &  $ \nu_4$ & $ c_4(\GeV^{-2}) $ &  $ c_5(\GeV^{-2}) $  \\
\hline 
 $ 0.653 \pm 0.121 $ &  $0.385 \pm 0.101 $ & $-4.71 \pm 0.74$ & $ -0.075 \pm 0.001$   \\
\hline \\
\end{tabular} 
\label{table:muR} 
\end{table} 

\subsection {Output Quantities and Plots  \label{output figures}    }

We must  compare the results that are given  in Sec.\ref{data_analysis} with the predictions from 
dispersion relations for  the  quantities of the real part $\rho$, $D_R$ and $\mu_R$   for the 
 pp and p\=p systems.
    Figure \ref{DR_results}
   shows the energy dependence of the 
predictions from DRA and DRS  for   $\rho$ and for the  derivative coefficient 
$D_R=\rho B_R/2-\mu_R$.

In Fig.\ref{muR_results}(a)  we show the energy dependence obtained for the parameter $\mu_R$ for pp 
${\rm p \bar p}$ and in Fig.\ref{muR_results} (b) we form the quotient $B_R=2(D_R+\mu_R)/\rho$ 
using  as point values shown in the large table with fit results, and  the lines are 
calculated with Eqs.(\ref{muRpp}) and (\ref{muRppbar}) and the analytical results   for $\rho$ and $D_R$ 
(pp and p\=p) from DRA and DRS.  Note that  the points where $\rho$ pass by zero must coincide with 
the  zero of the sum  $D_R+\mu_R$. This condition results naturally in our solution.

 We thus  have a closed coherent determination of all quantities describing forward scattering, 
with no free local parameter.

\begin{figure*}[h]
      \includegraphics[width=8.0cm]{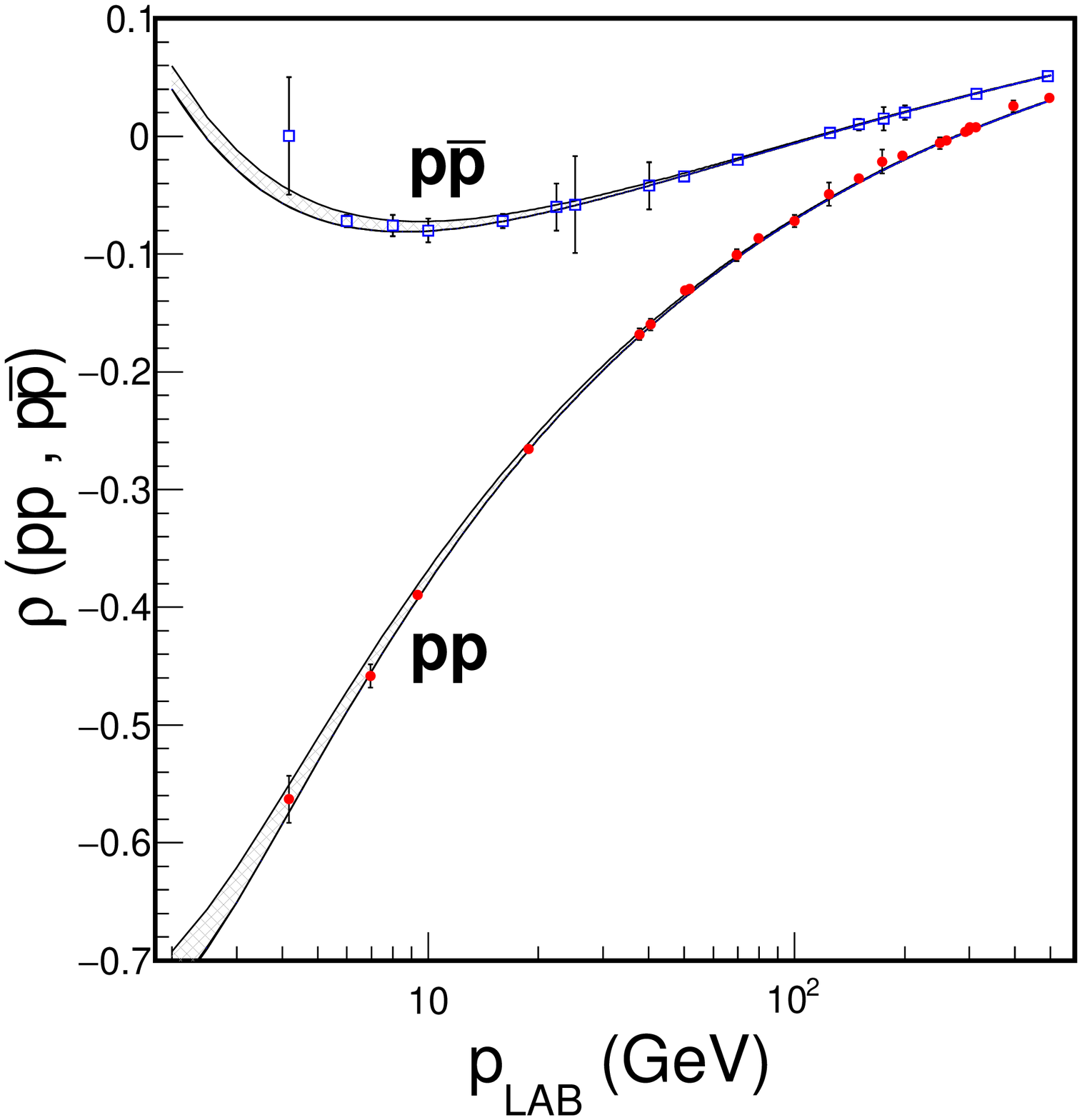}
            \includegraphics[width=8.0cm]{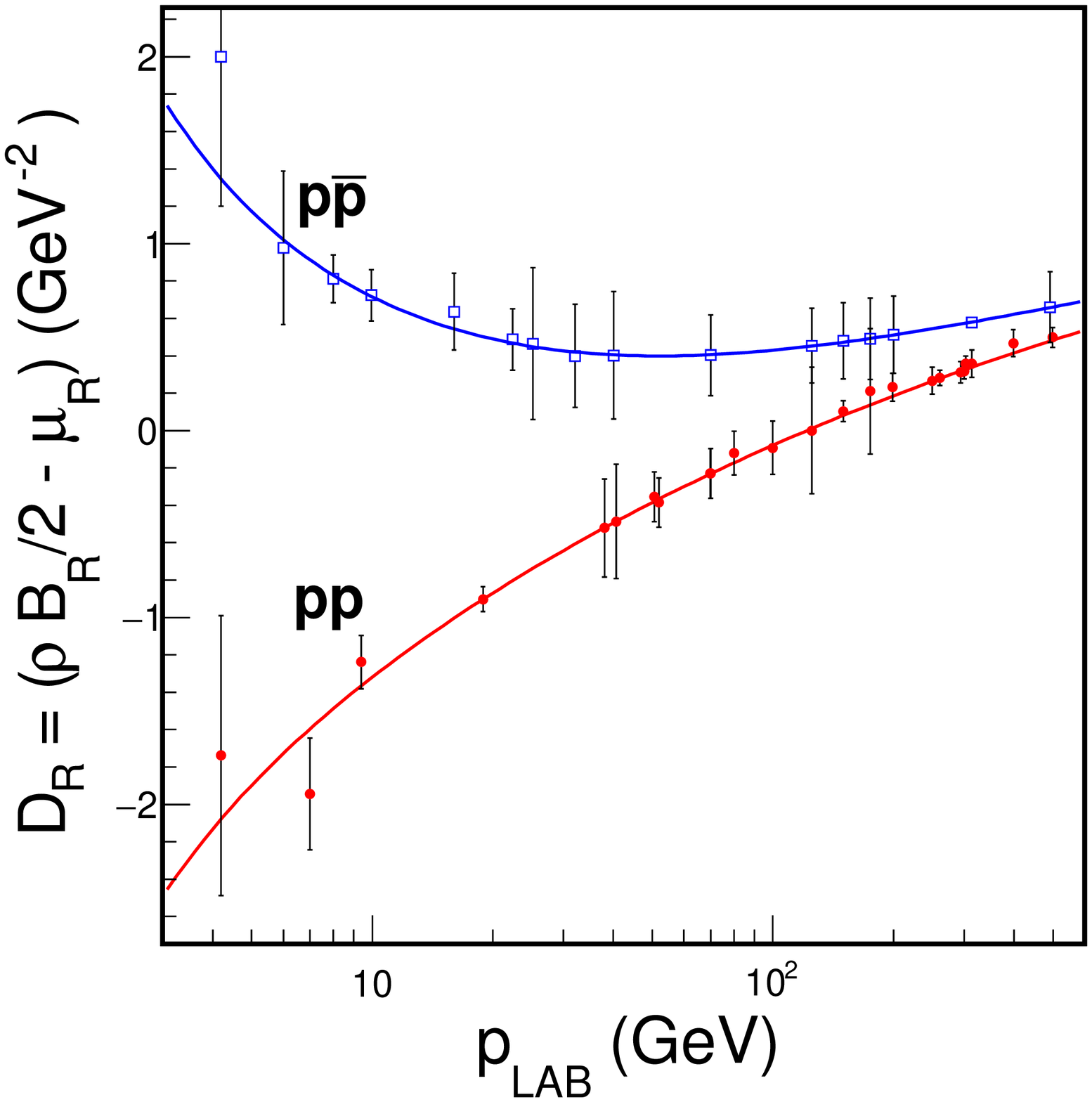}  
           \caption{\label{DR_results} (a) The  lines give the predictions for the quantity 
  $\rho$ predicted by dispersion relations for the amplitudes DRA, 
   showing the    band with variation of $K$ from -310 to -287. The  marked points 
     are obtained from the data. It is important to 
  remark that the positions of the $\rho$ zeros $|t|_0(pp)= 276.91 \GeV^2$  and 
  $|t|_0({\rm p \bar p})= 116.93 \GeV^2$ predicted by DRA are confirmed by the data; 
   (b) The lines show the predictions from the dispersion relations for slopes DRS 
   for the derivatives of the real amplitudes at the origin, which are represented by 
   the combinations $D_R=\rho B_R/2-\mu_R$  for pp and p\=p. The marked points represent these 
   combinations of parameters $\rho$, $\mu_R$, $B_R$  calculated with the values for the  
   experimental data presented in Sec.\ref{data_analysis}. } 
       \end{figure*}

\begin{figure*}[h]
            \includegraphics[width=8.0cm]{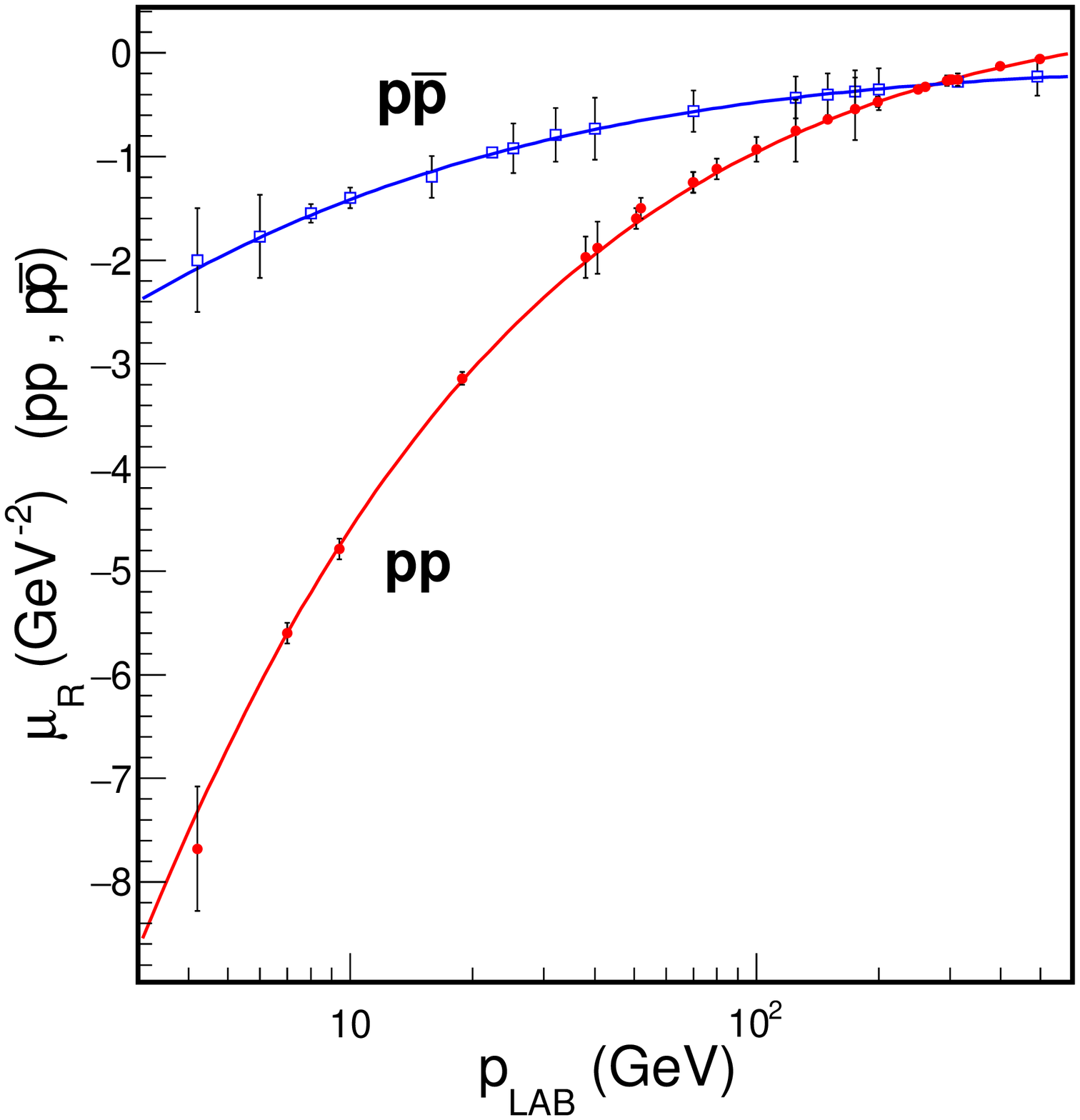}
   \includegraphics[width=8.0cm]{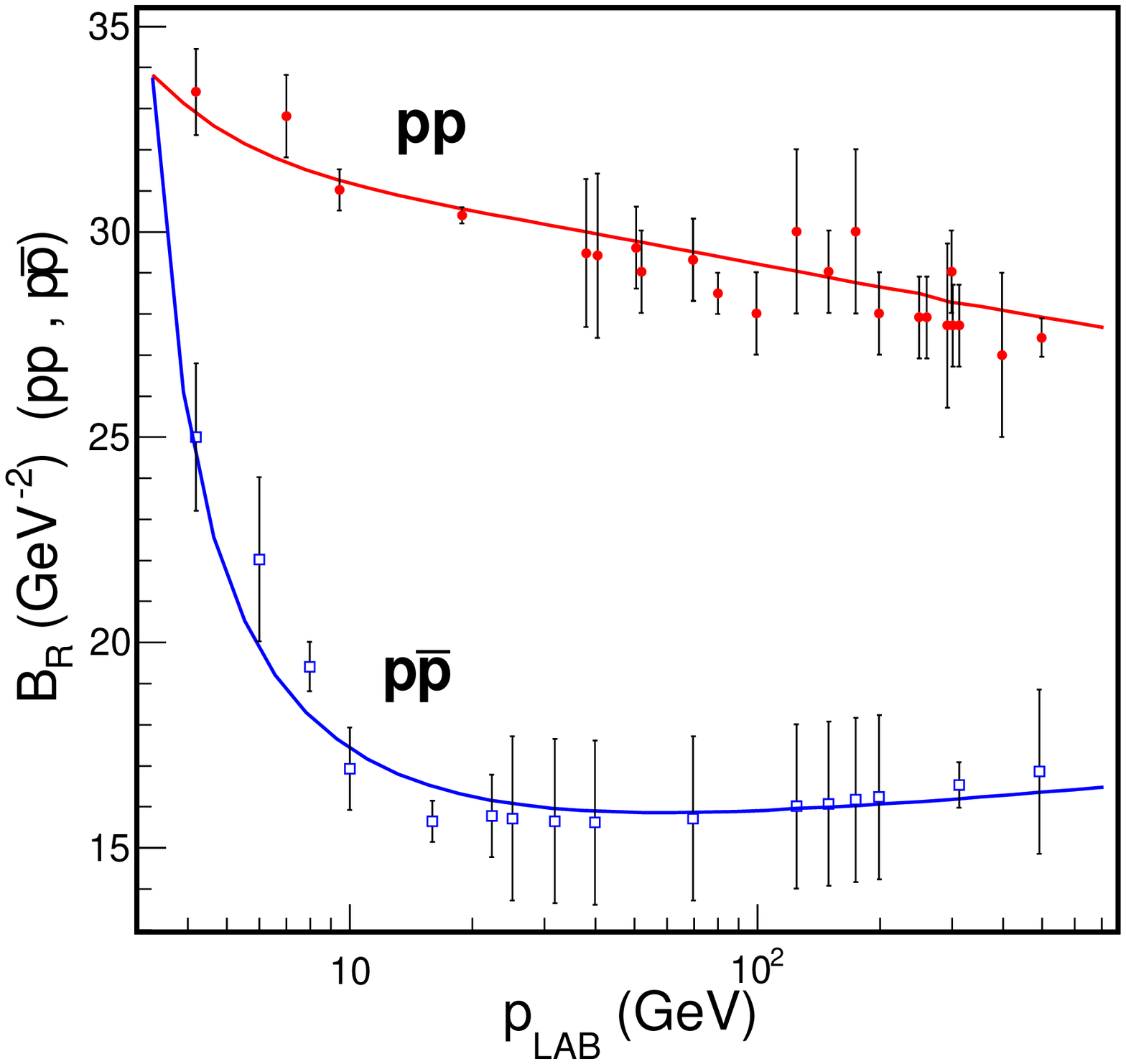}
        \caption{\label{muR_results}
(a) Representations of the values of the parameters $\mu_R^{\rm pp} $  and $\mu_R^{\rm p\bar p} $ 
   through Eqs. (\ref{muRpp},\ref{muRppbar}) and the points obtained from the data ;
(b) Energy dependence of the slopes $B_R ({\rm pp})$ and $B_R ({\rm p \bar p}) $ of 
the real amplitudes. The lines are fully predicted by the dispersion relations for slopes DRS
(that give $D_R=\rho B_R/2-\mu_R$ for pp and p\=p), dispersion relations for the amplitudes 
(that give $\rho$ for pp and p\=p) and by Eqs.(\ref{muRpp},\ref{muRppbar}) that give 
$\mu_R^{\rm pp}$  and  $\mu_R^{\rm p \bar p}$.     }
   \end{figure*}

 \clearpage

\section{Analysis of  data in forward direction \label{data_analysis} }

The data \cite{data_analysed}      presented in Tables {\ref{table:pp}}, {\ref{table:ppbar} 
  cover  $|t|$ ranges  accessible for the 
analysis,  that requires a     regular set of points with $|t| \leq 0.5 ~\GeV^2$. 
   Ideally, it would be nice to have good quality data from the very low  
$|t| \approx 0.001~ \GeV^2$ and going  up to 0.1 $\GeV^2$ , but this is not  always 
available in the low energy range.

The analysis   covers all data of   elastic  pp and p\=p scattering in the energy range of 
$\sqrt{s}$ from 3 to 30 GeV. 
These  data have been treated along the history with incomplete theoretical expressions 
for $d \sigma/dt$. 
In this  energy  range it is believed that the dynamics of forward scattering 
is mainly determined by gluonic interactions resulting in smooth energy dependence of all 
parameters. 
On the other hand, at  very low energies below ${\rm p_{LAB} = 4 \GeV} $ the direct quark-quark
and quark-antiquark  interactions may be more important. This seems to be particularly visible 
in the p\=p case. 

Above $\sqrt{s}= 30 \GeV$,   ISR/CERN and Fermilab data, covering very 
wide $t$ ranges, deserves to be studied with analytical forms including the whole  $t$ range. 
 There are many models \cite{models,us_LHC}  for this purpose,  and the forward scattering 
forms here studied are   part of these full-$|t|$ descriptions. 

 An important feature of our analysis  is the absorption of normalization errors that 
accompany the determinations of $d\sigma/dt$. Even when these normalization indeterminacies 
are small , their influence in the parameters is large. The experimental papers use 
criteria for normalization that we do not consider legal or correct, as ignoring existence of 
real part, ignoring realistic $t$ dependences in the amplitudes, and comparison with other 
experiments. Some experiments report $\sigma$ values using data that are not qualified 
for the analysis (as insufficient $t$ range). These difficulties lead to fluctuations 
in values of parameters  that do not represent physical effects and do not 
allow a regular  global description.

 After a  smooth description has been achieved  and parameters 
$P,H,R_1,R_2,\eta_1,\eta_2$ of Eq.(\ref{sigma_x}) are determined,  the values of $\sigma$ are 
imposed at each energy. Thus we introduce a constant normalization factor $f$ for 
each data set,  chosen so that the total cross section $\sigma$ equals the value 
determined  given by Eq. (\ref{sigma_x}). 
We thus write 
\begin{equation}
\frac{d\sigma}{dt}= f \times \frac{d\sigma}{dt}\big|_{\rm data} ~ .
\label{Normalization}
\end{equation}
The value of $f$ for the data set of  each experiment is given in 
Tables \ref{table:pp}, \ref{table:ppbar}. 
Thus the central values of $\sigma$ are assumed and $f$ is determined. The error bars 
in $\sigma$ and other quantities  represent sensitivities   of the fit  to each parameter individually, without
freedom for correlations.

 The resulting suggested parameter values from our analysis  are collected in Tables 
 \ref{table:pp} and  \ref{table:ppbar}. 
 The input energies for the data are written primarily in terms of ${\rm p_{LAB}}$, as has been more 
usual in the presentation of data  in this range, but the table also includes  $\sqrt{s}$ values.  
 In some  cases, mainly at very low energies, where data are not rich, 
 we combine information from different experiments of same or nearby energies, 
in the same numerical treatment  and in plots. We observe good matching of data sets.

We show   examples of the treatment of the  $d\sigma /dt$  data in many plots.   
More detailed information   is given in the figure captions.
  The log $|t|$ horizontal scale helps to expand and exhibit  the 
  small  $|t|$ behavior, and it is remarkable that often the descriptions work very well 
up to $|t|=0.2$, beyond the strictly forward range that determines parameters. We interpret 
that this is so thanks to appropriate  form assumed for the   amplitudes, and to the control 
established by DRA and DRS.

Ranges where $\rho$ passes  by zero are particularly delicate. 
The experiments at $\sqrt{s}$ =  23.542 and 23.882 GeV  ($p_{\rm LAB}$ = 294.4 and 303.1 GeV) 
give an example  in which there is  discrepancy in literature for the  $\rho$   sign.
 Our treatment   solves the discrepancy, 
 namely we show that $\rho$ passes by zero in this region, 
 and this is valid for both experiments. Parameters are in Table \ref{table:pp}. 

For p\=p in the range of our analysis $\rho$ is always small and the data are poor. 
We are then strongly dependent on the predictions from DRA and DRS.

We inspect and analyse the data  using a CERN Minuit  program   
for the determination of  $\chi^2$, nominally with six  parameters. We 
use the form for the total   pp and  ${\rm p \bar p} $  cross sections 
in Eq.(\ref{sigma_x}), and iteratively with observations of the behaviour of 
 $B_I$ and $\mu_I$, and use of dispersion relations DRA and DRS as guides for 
the real parts. Simultaneously we obtain a value for the subtraction constant $K$.
Determinations of $B_I$, $\mu_I$ and the normalization factor $f$ are made 
simultaneously. $B_I$, $\mu_I$  are well represented by the simple analytical forms
of Eqs.(\ref{BI_x},\ref{MUI_x}) and Table \ref{table:inputs}. 
Once the solution for each dataset is obtained, 
the error bars are obtained relaxing the value of each quantity in the fitting code, so 
that they represent the  sensitivity of the  $\chi^2$ value, but in general do not 
include correlations.    

Details of the data sets and of the calculation of the analytical representations are
given in the figure captions.

\clearpage

\subsection{pp data analysed  \label{pp_data} }

{\small 
 \begin{center}
\begin{table*} 
\centering
 \caption{ Values of the parameters for the amplitudes of  pp elastic scattering } 
\begin{tabular}{c c c  c c c c c c c c c c c  } 
\hline\hline 
$p_{\rm LAB}$  & $\sqrt{s}$  & N & $t_{\rm min} -  t_{\rm max}$& $f$ & $\sigma$ & $B_I-2\mu_I$ & $\rho$ &  $B_R$ & $\mu_R$ & $\mu_I$ & $\chi^2$  & ${\rm Re}$ \\
 \hline 
(GeV)  & (GeV)  &  & (GeV$^{-2}$) &  & (mb) & (GeV$^{-2}$) & & (GeV$^{-2}$) & (GeV$^{-2}$) & (GeV$^{-2}$) &  & \cite{data_analysed}  \\
\hline 
4.2 &3.14 &39 &0.00106-0.189 & 1.0520 &39.18$\pm$0.16 & 5.85$\pm$0.08 & -0.564$\pm$0.020 & 33$\pm$1 & -7.68$\pm$0.69 & 0.11$\pm$0.03 & 1.165 & aq
\\ \hline
7 &3.88 &59 &0.00141-0.31 & 1.0316 &39.05$\pm$0.08 & 7.52$\pm$0.04 & -0.460$\pm$0.010 & 33$\pm$1 &-5.60$\pm$0.10 &0.09$\pm$0.02 & 1.87 & aq
\\ \hline
9.43 &4.42 &34 & 0.00079-0.01283 & 0.9966 &38.92$\pm$0.09 & 8.2$\pm$0.2 & -0.390$\pm$0.002 & 31$\pm$1 & -4.8$\pm$0.1 &0.05$\pm$0.01 & 1.042 & h
\\ \hline
18.9 &6.11 &67 &0.0009-0.10883 & 0.9845 &38.6$\pm$0.04 &9.76$\pm$0.08 &-0.266$\pm$0.001 & 30$\pm$1 &-3.14$\pm$0.06 & -0.07$\pm$0.03 & 1.227 &  h
\\ \hline
38.01 &8.55 &65 &0.00086-0.11318 & 0.9792 &38.44$\pm$0.02 & 10.37$\pm$0.06 & -0.160$\pm$0.004 & 30$\pm$2 &-1.99$\pm$0.25 &-0.13$\pm$0.03 & 1.282 &h
\\ \hline
40.62 &8.83 &65 &0.00088-0.11379 & 0.9916 &38.44$\pm$0.04 & 10.73$\pm$0.1 & -0.161$\pm$0.005 & 29$\pm$2 & -1.88$\pm$0.25 &-0.17$\pm$0.03 & 1.55 &h
\\ \hline
42.5 &9.03 &19 &0.00193-0.03982 &1.0007 &38.43$\pm$0.25 &10.90$\pm$0.20 &-0.160$\pm$0.010 & 30$\pm$1 & -1.60$\pm$0.30 &-0.25$\pm$0.30 & 0.762 &j
\\ \hline
50.62 &9.84 &66 &0.00096-0.11508 & 0.9728 &38.44$\pm$0.05 & 10.83$\pm$0.1 & -0.132$\pm$0.004 & 30$\pm$1 &-1.6$\pm$0.1 &-0.18$\pm$0.04 & 1.651 &h
\\ \hline
52 &9.97 &72 &0.00063-0.0306 & 0.9800 &38.45$\pm$0.06 & 11$\pm$0.08 &-0.130$\pm$0.004 & 29$\pm$1 &-1.5$\pm$0.1 &-0.22$\pm$0.08 & 1.297 &i
\\ \hline
52.2 &9.99 &18 &0.00187-0.05041&0.9903  &38.45$\pm$0.20 &11.30$\pm$0.30 &-0.132$\pm$0.020 & 29$\pm$2 &-1.50$\pm$0.10 &-0.22$\pm$0.20& 1.729 &j
\\ \hline
69.84 &11.53 & 73 & 0.00111-0.10817 & 1.0094 &38.51$\pm$0.06 & 11.05$\pm$0.1 & -0.101$\pm$0.005 & 29$\pm$1 &-1.25$\pm$0.1 &-0.25$\pm$0.05 & 1.238 &h 
\\ \hline
70 &11.54 &124 &0.00185-0.08352 & 1.0038 &38.51$\pm$0.07 & 11.05$\pm$0.1 & -0.101$\pm$0.005 & 29$\pm$1 &-1.25$\pm$0.1 &-0.25$\pm$0.05 & 0.888 &n
\\ \hline
80 &12.32 &58 &0.00066-0.02928 & 0.9953 &38.55$\pm$0.06 & 11.3$\pm$0.04 & -0.087$\pm$0.004 & 29$\pm$1 &-1.12$\pm$0.1 &-0.26$\pm$0.01 & 0.960 &i
\\ \hline
100 &13.76 &140 &0.00170-0.15113 & 1.0053 &38.66$\pm$0.07 & 11.48$\pm$0.03 & -0.073$\pm$0.005 & 28$\pm$1 &-0.93$\pm$0.12 & -0.31$\pm$0.02 & 1.054 &n
\\ \hline
100 &13.76 &73 &0.0022-0.0388  & 1.0045 &38.66$\pm$0.07 &11.48$\pm$0.03 &-0.073$\pm$0.005 &28$\pm$1 &-0.93$\pm$0.12 &  -0.31$\pm$0.02 & 1.252 & r
\\ \hline
125 &15.37 &92 &0.00164-0.09828 & 1.0158 &38.8$\pm$0.05 &11.5$\pm$0.2 &-0.05$\pm$0.01 &30$\pm$2 &-0.75$\pm$0.3 &-0.3$\pm$0.05 & 0.845 &n
\\ \hline
150 &16.83 &92 &0.00164-0.09828 & 0.9943 &38.93$\pm$0.06 & 11.68$\pm$0.02 & -0.037$\pm$0.003 & 29$\pm$1 &-0.64$\pm$0.03 & -0.36$\pm$0.06 & 1.198 &n
\\ \hline
150 &16.83 &68 &0.0022-0.0392 & 0.9920 &38.93$\pm$0.06 & 11.68$\pm$0.02 & -0.037$\pm$0.003 & 29$\pm$1 &-0.64$\pm$0.03 & -0.36$\pm$0.06 & 1.192 &r
\\ \hline
175 &18.17 &55 &0.00181-0.09766 & 1.0103 &39.07$\pm$0.07 & 11.7$\pm$0.16 & -0.022$\pm$0.01 &30$\pm$2 &-0.54$\pm$0.3 &-0.38$\pm$0.04 & 1.152 &n
\\ \hline
199 &19.37 &69 &0.00066-0.0315 & 0.988 &39.2$\pm$0.06 &11.9$\pm$0.1 &-0.017$\pm$0.004 & 28$\pm$1 &-0.47$\pm$0.05 &-0.41$\pm$0.01 & 1.143 &i
\\ \hline
250 &21.7  & 64 & 0.0022-0.039 & 0.9880 &39.45$\pm$0.06 & 11.94$\pm$0.01 & -0.006$\pm$0.005 & 28$\pm$1 &-0.35$\pm$0.02 & -0.43$\pm$0.01 & 0.712 & r
\\ \hline
261 &22.17 &63 &0.0005-0.02978 & 0.994 &39.5$\pm$0.06 &11.88$\pm$0.1 &-0.0034$\pm$0.002 & 28$\pm$1 & -0.33$\pm$0.03 &-0.42$\pm$0.05 & 1.291 & i
\\ \hline
294.4&23.54&31 &0.00037-0.0102 & 1.0118 &39.65$\pm$0.08 & 11.9$\pm$0.1 &0.003$\pm$0.002 & 28$\pm$2 &-0.27$\pm$0.05 & -0.45$\pm$0.1 & 0.608 & k
\\ \hline
300 &23.76 & 60 & 0.0022-0.0388 & 0.9941 &39.67$\pm$0.08 & 12$\pm$0.03 &0.004$\pm$0.002 & 29$\pm$1 &-0.26$\pm$0.03 & -0.45$\pm$0.01 &1.078 & r
\\ \hline
303.1&23.88&66 &0.00066-0.0316 & 0.9823 &39.69$\pm$0.06 & 12.03$\pm$0.03 & 0.007$\pm$0.003 & 28$\pm$1 &-0.26$\pm$0.01 & -0.47$\pm$0.01 & 1.295 &i
\\ \hline
313.7&24.3 &31 &0.00108-0.01313 & 1.0054 &39.73$\pm$0.05 & 11.95$\pm$0.05 & 0.007$\pm$0.003 & 28$\pm$1 &-0.26$\pm$0.06 & -0.47$\pm$0.08 & 0.763 &l
\\ \hline
398 &27.36 &60 &0.00047-0.02579 & 0.979 &40.07$\pm$0.07 & 12.14$\pm$0.01 & 0.025$\pm$0.005 & 27$\pm$2 &-0.13$\pm$0.01 &-0.50$\pm$0.02 & 1.272 &i
\\ \hline
499.1&30.63 &32 &0.0005-0.0176 & 1.0080 &40.43$\pm$0.05 & 12.1$\pm$0.15 & 0.032$\pm$0.003 & 27$\pm$2 & -0.06$\pm$0.01 &-0.53$\pm$0.15 & 0.776 &k
\\ \hline   \hline
\end{tabular} 
\label{table:pp} 
 \end{table*}
      \end{center}
 }

\subsubsection{Piles of data in four experiments}

  Figure \ref{pilhas_pp} shows   data from four experiments \cite{data_analysed}(h),(i),(n),(r) that cover 
regularly large energy and $|t|$ ranges. In the plots these   are 
called Beznogikh (1973)\cite{data_analysed}(h), Kuznetsov (1981)\cite{data_analysed}(i), 
Fajardo (1981)\cite{data_analysed}(n) and Burq (1983) \cite{data_analysed}(r). 
  The parameters for the lines are given in Table \ref{table:pp}. 
We observe that the representations obtained  by Eqs.(\ref{diffcross_eq}),(\ref{real_TR}) and
(\ref{imag_TI}) are very faithful to the data  up to surprisingly large $|t|$  values, namely 
above $|t|=0.1  \GeV^2 $  and up to 0.3  in some cases. The compatibility of different experiments 
can   be noted.
   \begin{figure*}[h]
    \includegraphics[width=8.0cm]{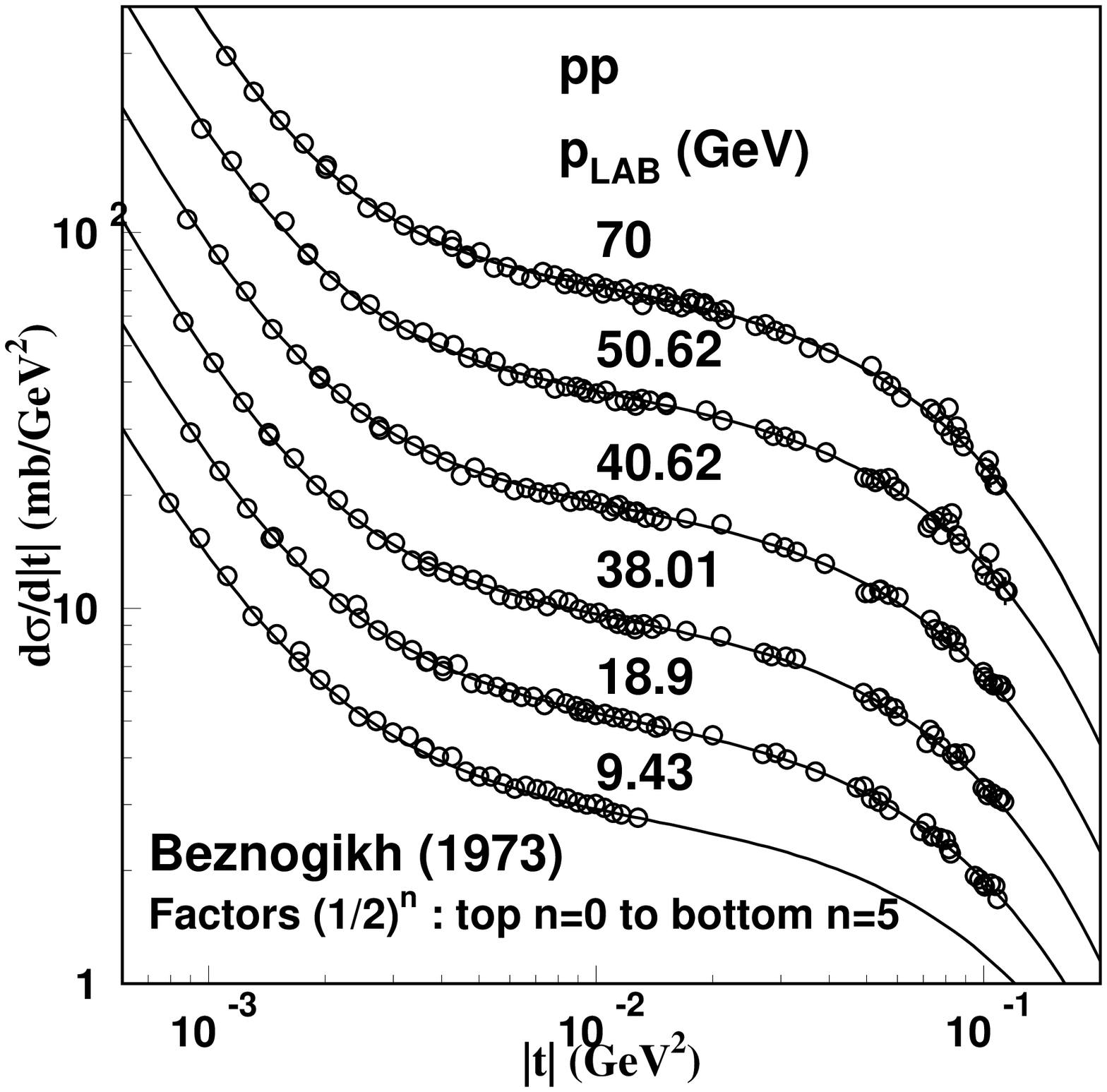}  
   \includegraphics[width=8.0cm]{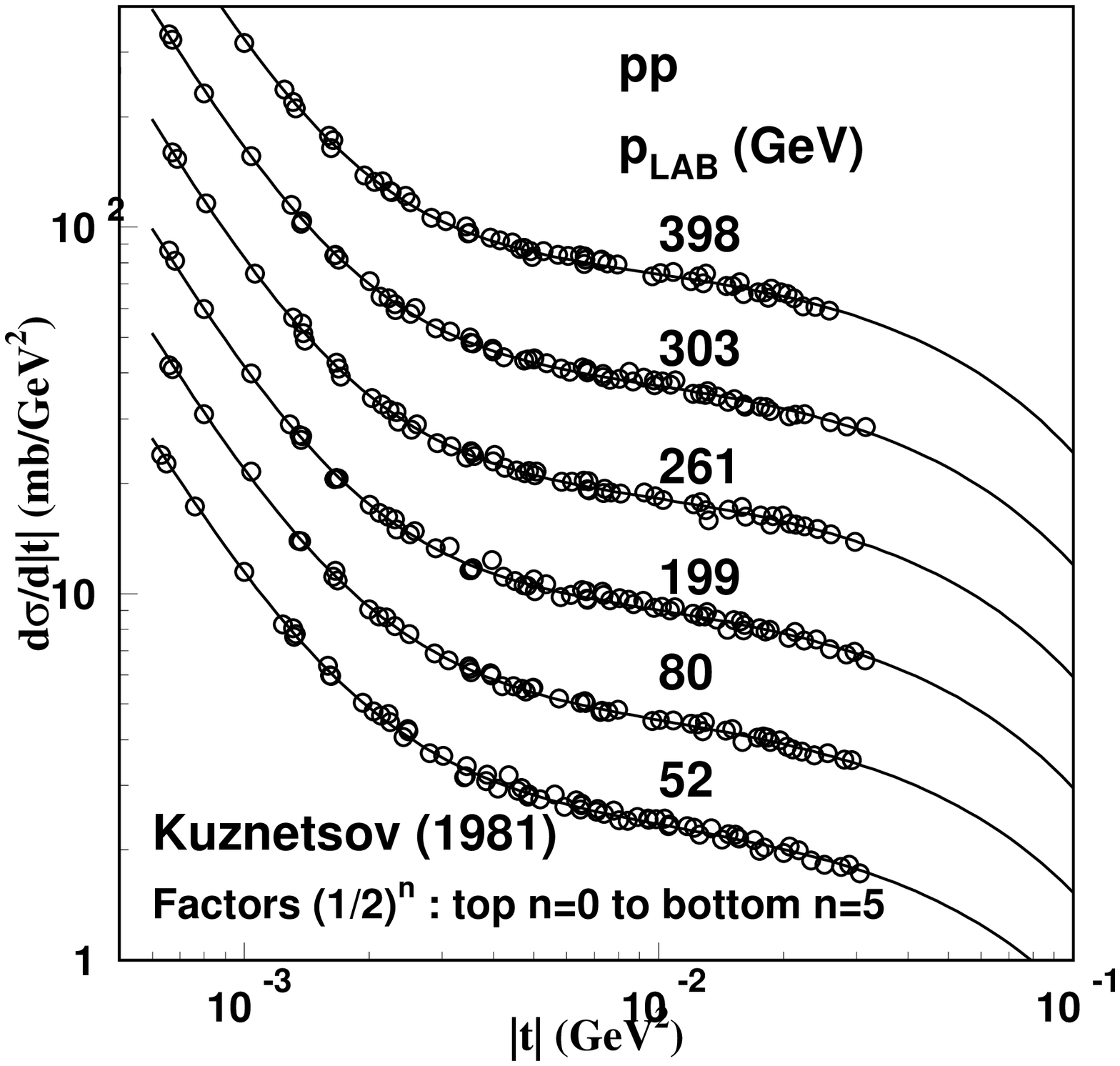} 
   \includegraphics[width=8.0cm]{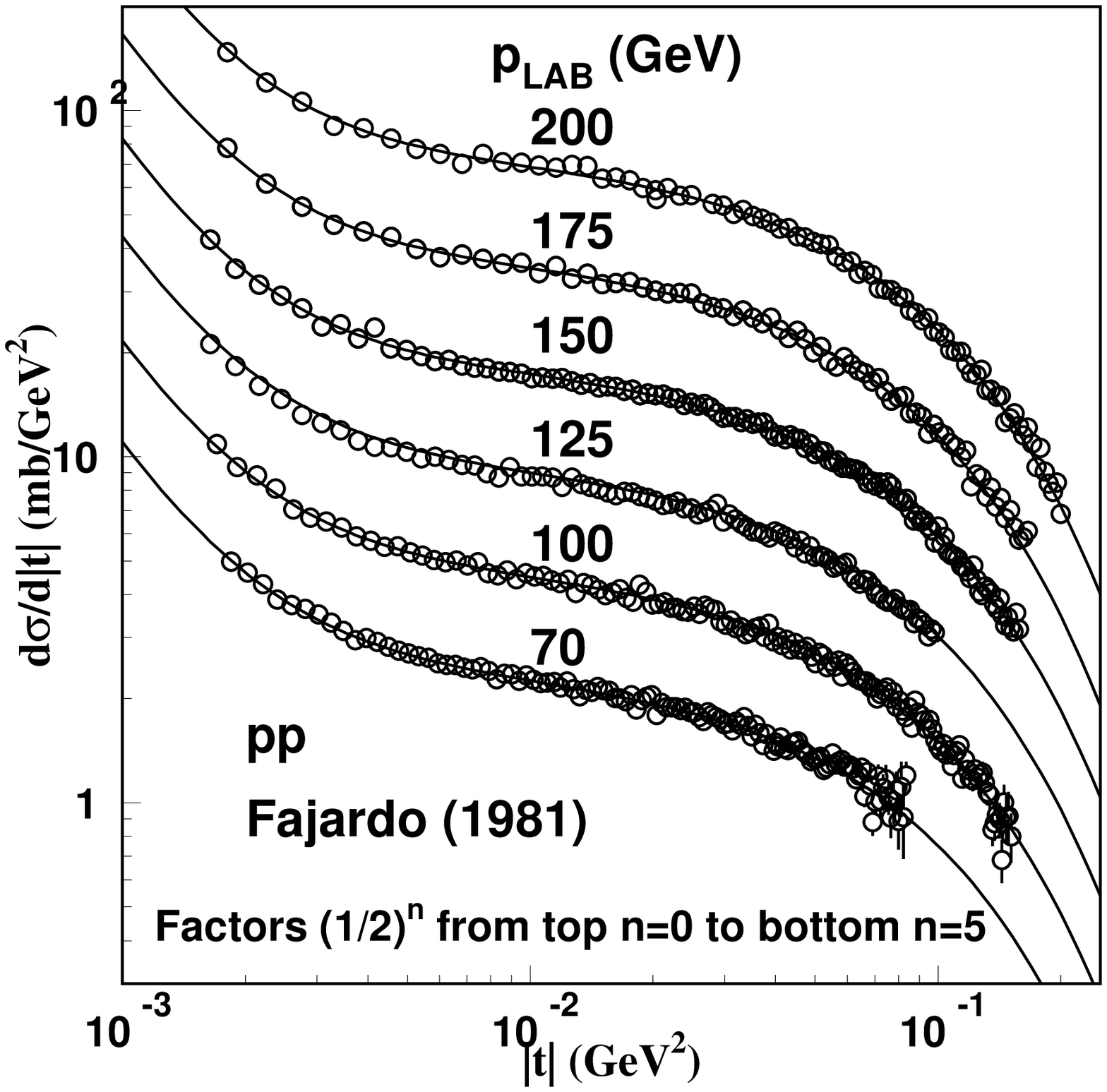}
       \includegraphics[width=8.0cm]{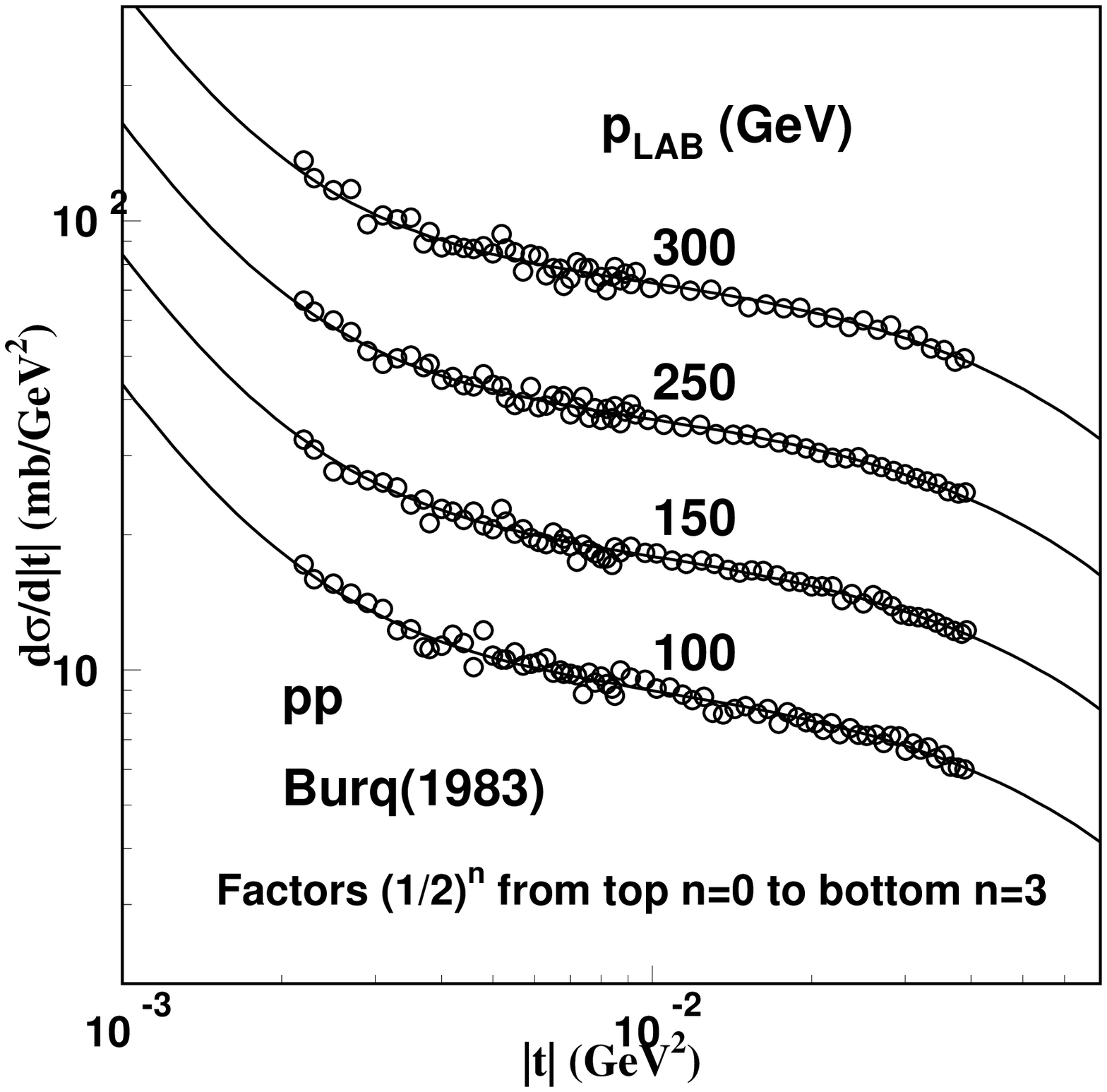}
    \caption{\label{pilhas_pp}   
  Data and suggested representations for pp scattering  in the   energy range 
$p_{\rm LAB}$ = 10  to 400 GeV, as reported in four different series of measurements
  \cite{data_analysed} (h), (i), (n)   and (r).  The data are regular, cover $t$ ranges  
adequate for determination of forward scattering quantities, and the representation  
  with parameters given in Table \ref{table:pp} are extremely precise. 
 The  compatibility among the four experiments is remarkable.  
The parameter $\rho$ and the combination $(\rho B_R/2 ~- \mu_R) $ are in agreement with 
predictions  from dispersion relations for amplitudes DRA and for their derivatives DRS. The
framework is explained in the text. The zero of $\rho$ at $p_{\rm LAB}= 277 \GeV$  that 
occurs in the energy range of these data is well treated by the representations.
 } 
\end{figure*}

\subsubsection{Low energies:  $p_{\rm LAB}= $4 - 10 {\rm GeV} } 
 
In the very low energy end, there are forward data from Jenni et 
al. \cite{data_analysed}(a) at $p_{\rm LAB}$ = 4.2, 7 and 10 GeV. 
At 4.2 and 7 GeV we join Ambats et al. \cite{data_analysed}(q) at 3.65 and 6 GeV , 
with factors 0.99 to  
compensate for the energy dependence. 

At $p_{\rm LAB} = 10 \GeV$, to exhibit  a representative recommendation  
   we join to Jenni et al.  \cite{data_analysed}(a)   the data from 
 Beznogikh et al.  \cite{data_analysed}(h) at 9.43 GeV and from Brandenburg et al. 
\cite{data_analysed}(d) at 10.4 GeV, with energy adjustment  factors, 
forming a large set with 153 data points. 

In all these cases the matching of the experimental sets is very good, and we are then 
able to find representations of very good quality for $d\sigma/dt$ in elastic pp scattering in 
all this energy range. 
The parameters are shown in Table \ref{table:pp} and the data and representative curves 
are shown in Fig. \ref{pp-4-10}.

\begin{figure}[h]
      \includegraphics[width=8.0cm]{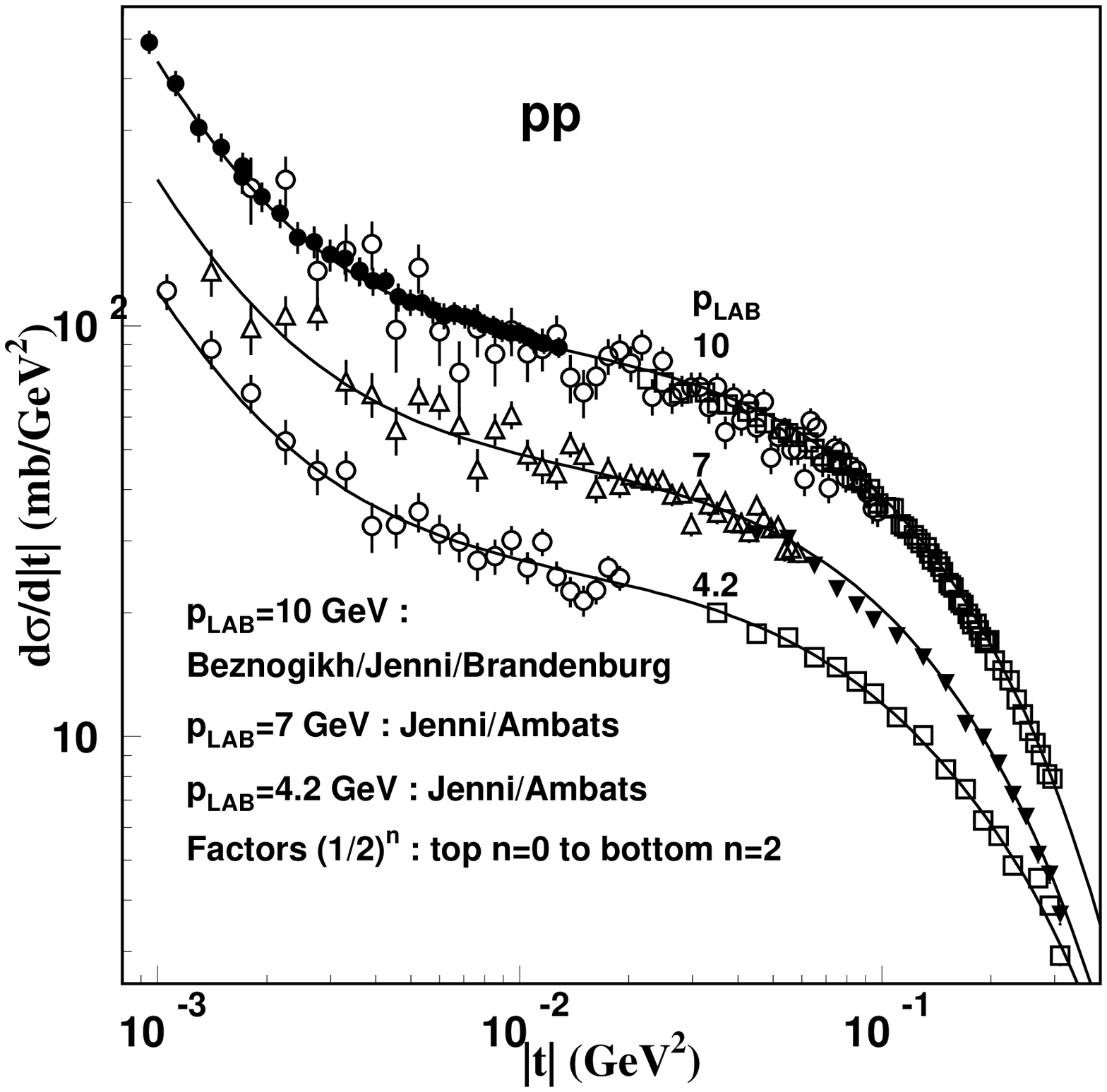} 
       \caption{\label{pp-4-10}  
  Data for pp scattering in the low energy end, with Jenni points \cite{data_analysed}(a) at  
4.2, 7 GeV and 10 GeV in the forward range.  
At 4.2 and 7 GeV, data from Ambats et al.  \cite{data_analysed}(q) at  3.65 GeV and 6 GeV  respectively 
with large $|t|$ are incorporated in the sets, after 
multiplication by the same factor 0.99 accounting for the energy dependence. 
  At $p_{\rm LAB}$ = 10 GeV we plot together in a single set  data from three different experiments 
\cite{data_analysed} (h),(a),(d)  at 9.43  , 10 and 10.4  GeV respectively, introducing conversion 
factor 0.97 at 10.24 GeV to  to account for energy dependence. For 9.43 GeV the factor is nearly 1, 
and is ignored. In all cases the connection of the data sets is remarkable. 
The data at 9.43 GeV are also plotted in a pile of the same experiment \cite{data_analysed}(h)
in Fig. \ref{pilhas_pp}.  The  parameters are given in Table \ref{table:pp}.  } 
\end{figure}

\subsubsection{pp Scattering in  CERN at $p_{\rm LAB}$  =  294.4 , 313.7 and 499.1 {\rm GeV}  } 

The energy $p_{\rm LAB} = 294.4 ~ \GeV$, with $\sqrt{s}= 23.54 ~ \GeV$,  of a CERN measurement 
\cite{data_analysed}(k), shown in Fig. \ref{pilha_AMOS} is close to the energy $\sqrt{s} = 23.88 ~ \GeV $ 
(namely $p_{\rm LAB} = 303 ~\GeV$) 
 of the Fermilab  data \cite{data_analysed}(i) shown in Fig.\ref{pilhas_pp}. The parameters are 
given in Table \ref{table:pp}, showing the  characteristic feature  that $\rho$ crosses zero 
in this region. 
 The data of the CERN  measurements, also  at $\sqrt{s}=24.3 ~\GeV$  (thus $p_{\rm LAB}=313.7 ~ \GeV $) , 
and    $\sqrt{s}=30.63 ~\GeV$  (thus $p_{\rm LAB}=499.1  ~ \GeV $) are shown together in Fig. 
\ref{pilha_AMOS}.   All descriptions are of high precision.  
 Other measurements   in this energy  region, at $p_{\rm LAB}$  = 303  and 398  GeV , 
are included in Fig.\ref{pilhas_pp}. 


\begin{figure}[h] 
  \includegraphics[width=8.5cm]{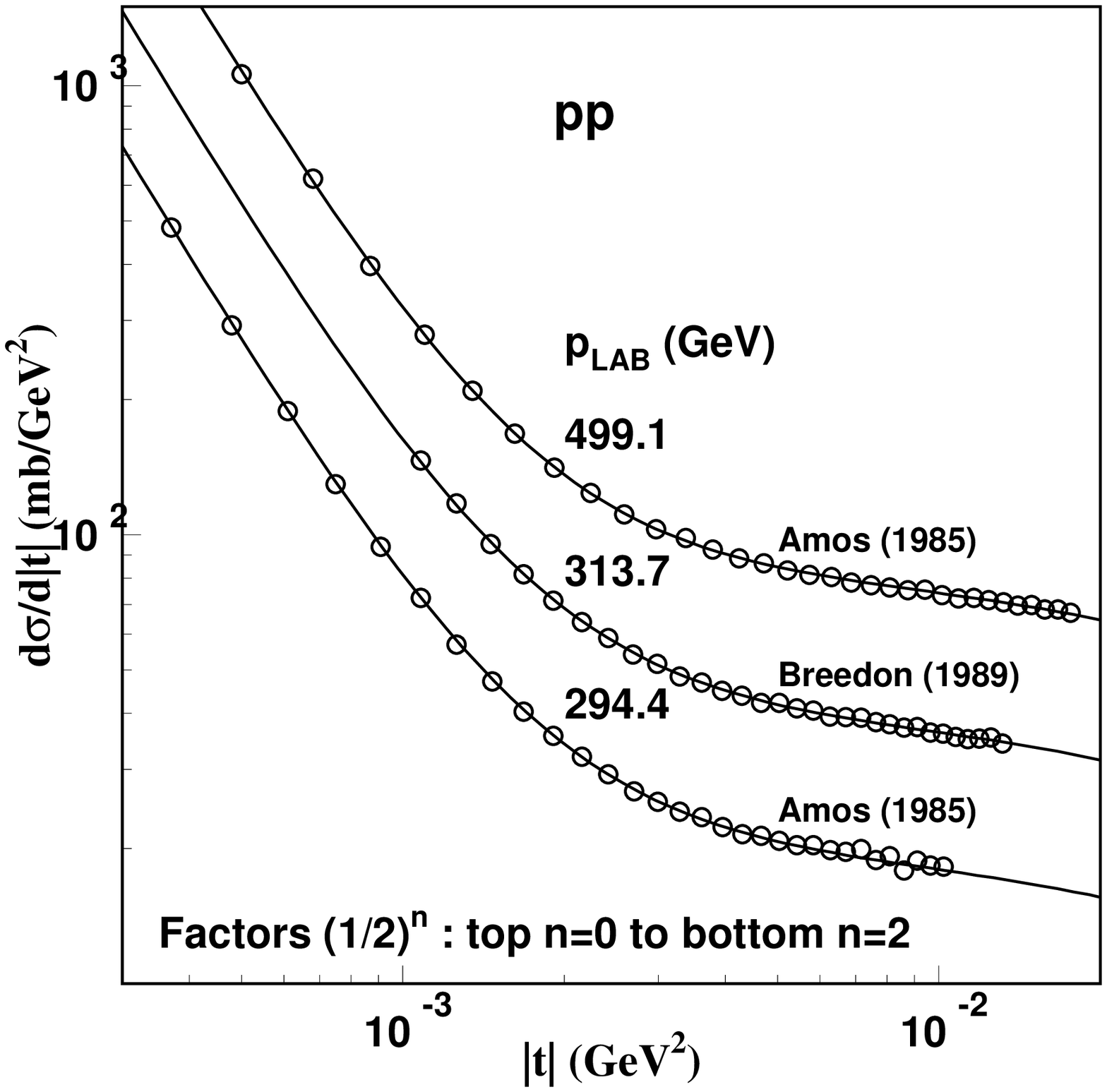}
   \caption { \label{pilha_AMOS}
  CERN data \cite{data_analysed} (k),(l) 
  in the higher energy part  of our analysis   
with parameters given in Table \ref{table:pp}. 
Other measurements in the same region are shown in Fig. \ref{pilhas_pp}.  
 }
   \end{figure}

\subsubsection { Large $|t|$: Behaviour beyond the forward range } 

The situation at large $|t|$ is shown  with  D.S. Ayres et al. data \cite{data_analysed}(f) 
  in Fig.\ref{pilhas_Ayres_Schiz}, where we study the behavior for $|t|$ beyond 0.1 $\GeV^2$. 
These  data do not cover the low-$|t|$ range,  and are plotted  together with the lines
representing the  Beznogikh (1973) and Fajardo (1981) data  at the corresponding energies. 
We see  a remarkable  matching of normalization in the  points with smaller $|t|$ and   
an increasing  deviation as $|t|$ increases. We here confirm that the equations for the 
amplitudes in forward scattering have the  validity  confirmed up to 0.1 GeV, where regular 
deviation of the data upwards  may start.  

Two other plots in the same  Fig.\ref{pilhas_Ayres_Schiz} show the behaviour of data in 
comparison with the analytic representations for $|t|$ beyond  the strict forward region. 
Details are given in the figure caption. 
 \begin{figure*}[h]
    \includegraphics[width=8.0cm]{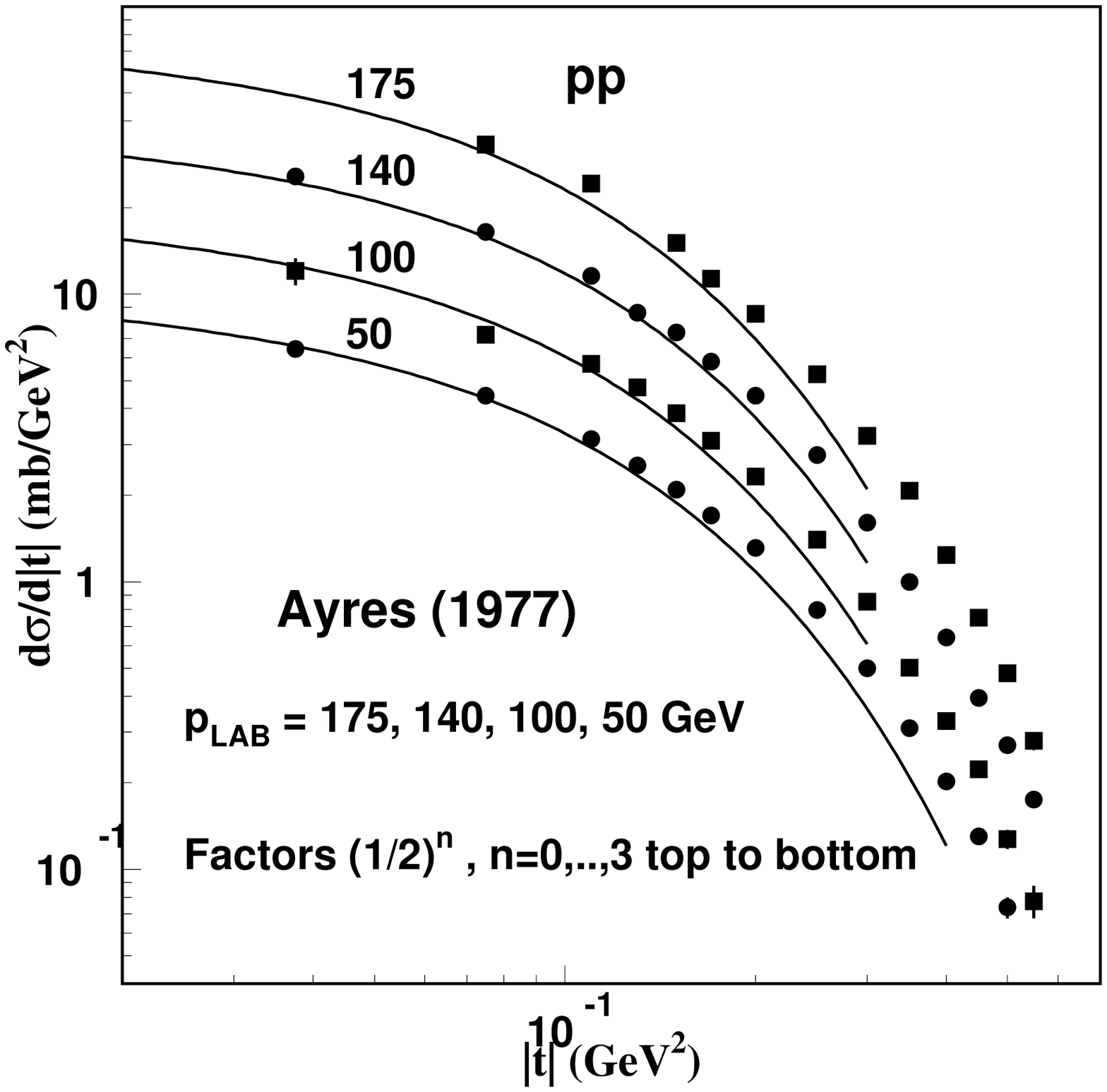}  
    \includegraphics[width=8.0cm]{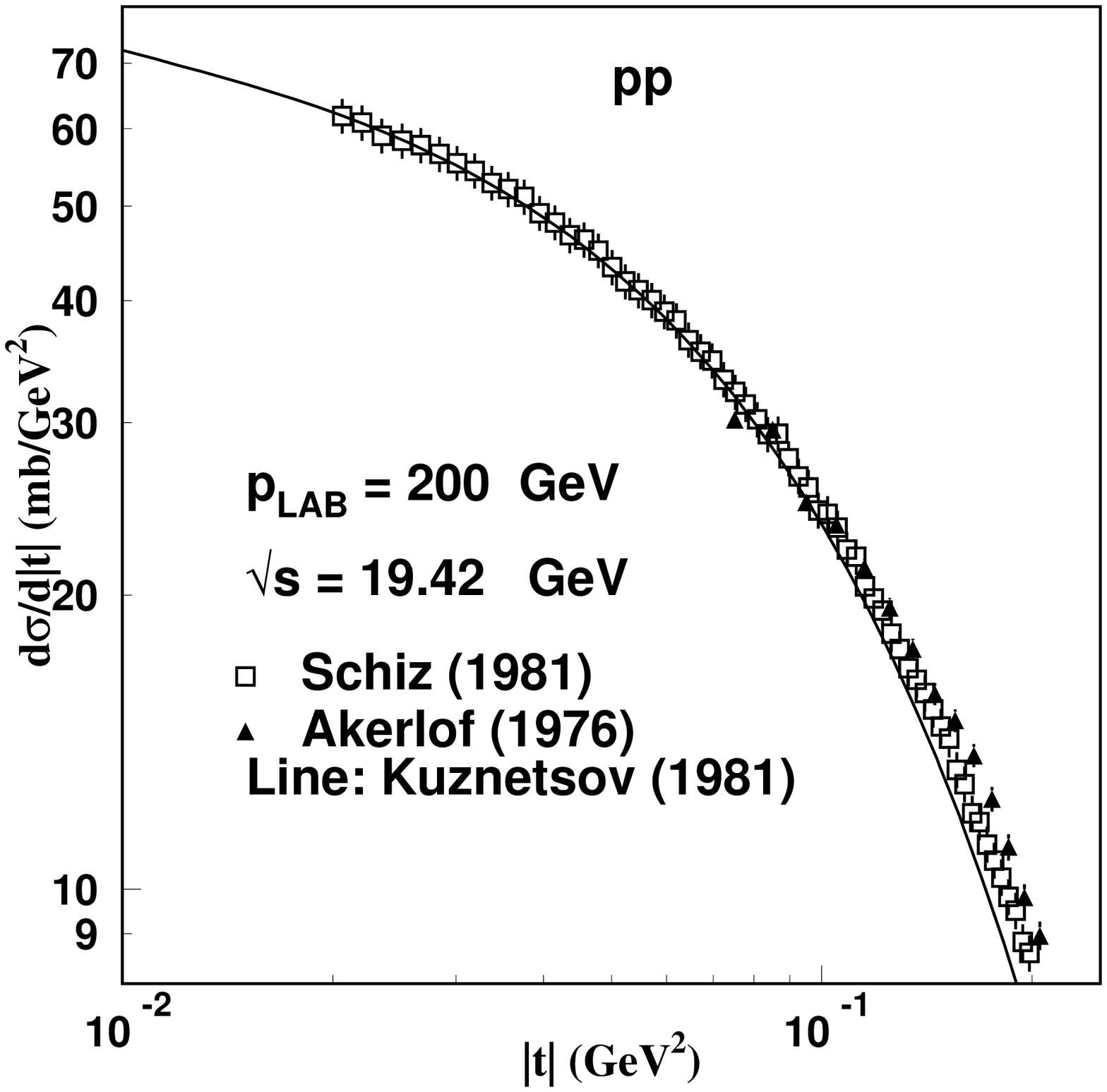} 
    \includegraphics[width=8.0cm]{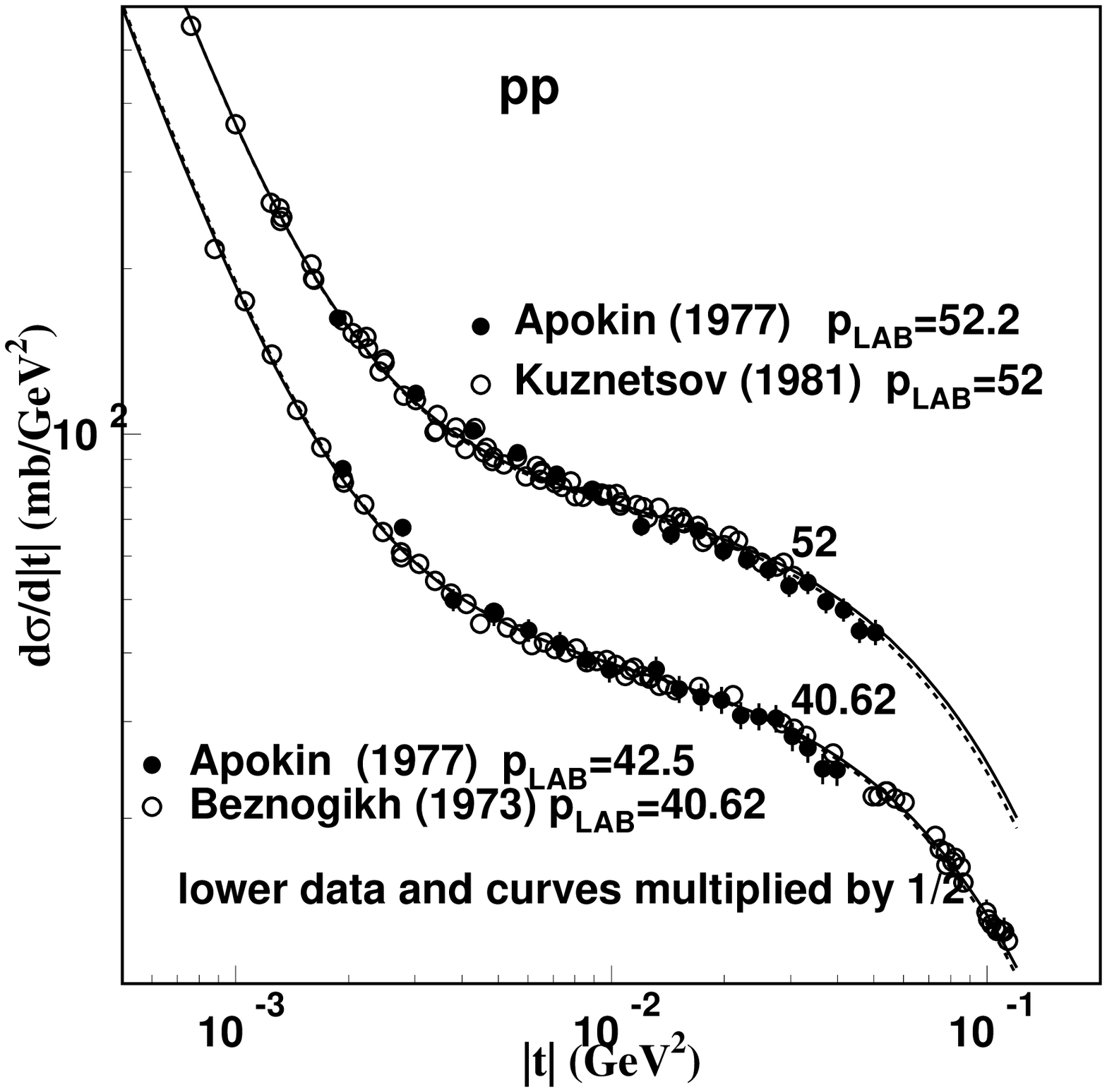}
   \caption{\label{pilhas_Ayres_Schiz}  Deviations for large $|t|$.    
  In the first plot a) (LHS at top) we show data for $|t|$ in the frontier of the forward region from 
Ayres et al.  \cite{data_analysed}(f).   The lines are NOT obtained with these data, but rather are 
taken from  the description of the forward range  in  Fig \ref{pilhas_pp} ,  with the parameters 
as given in Table \ref{table:pp}, at the corresponding energies. For the 140 GeV data 
there is no corresponding low $|t|$ experiment, and we use the line for 150 GeV \cite{data_analysed}(m). 
The matching of    points of lowest $|t|$  at  each energy is impressive, showing that different 
experiments succeed in the normalization of their data.  At all energies there is a progressive 
detachment between data and predicted curves starting at, say,  0.1 GeV , showing clearly that 
these data by themselves cannot be used for determinations.
 In the second  plot b) (RHS at top)  we show the connection between the solution at 199 GeV 
\cite{data_analysed}(i)  and the forward part (up to 0.2 $\GeV^2$)  of data by 
Schiz et al. \cite{data_analysed}(p) that starts at $|t|= 0.02 \GeV^2$. 
  The solid line represents the curve for   Kuznetsov data  at 199 GeV in Fig.\ref{pilhas_pp}
with parameters given in Table \ref{table:pp}.  The agreement of Schiz data with the curve up to 
$|t| \approx 0.1  \GeV^2 $ is remarkable  but it seems that improvement could be obtained with 
normalization.   We observe  that for  higher $|t|$ there appears a displacement between data 
and the curve, exhibiting again the limitation    for the use of the forward   amplitudes. 
For illustrative purpose, we show together   Akerlof et al.  points  \cite{data_analysed}(m) 
added to the 200 GeV plot, demonstrating that they should not be considered in  good agreement 
(notice that the scale is very tight), presenting strong displacement as $|t|$ increases.
 In the third plot c) (bottom)  we show    Apokin et al. data \cite{data_analysed}(j) 
at $p_{\rm LAB} = $ 42.5 and 52.2  GeV , covering smaller $|t|$ ranges, inserted together with  
Beznogikh  \cite{data_analysed}(h)   and Kuznetsov  \cite{data_analysed}(i)  data respectively. 
We observe compatibility (the 42.5 data are corrected 
with factor 1.02 in the plot, to reduce to  energy 40.62 GeV). 
The lines of the solution for Apokin \cite{data_analysed}(j)  are shown in dashed  form, 
while the Beznogikh and Kuznetsov lines are solid. There is reasonable compatibility, but it is clear 
that the  measurements reaching smaller $|t|$ values  are more qualified for parameter 
determination.   }  
\end{figure*}

\clearpage

\subsection{ppbar  data analysed  \label{ppbar_data} }
                           
{\small 
 \begin{center}
\begin{table*}
\setlength{\tabcolsep}{2.0pt}
\caption{ Values of the parameters for the amplitudes of  p\=p elastic scattering } 
\centering 
\begin{tabular}{c c c c c c c c c c c c c c  } 
\hline\hline 
$p_{\rm LAB}$  & $\sqrt{s}$ & N & $t_{\rm min} -  t_{\rm max}$& $f$ & $\sigma$ & $B_I-2\mu_I$ & $\rho$ &  $B_R$ & $\mu_R$ & $\mu_I$ & $\chi^2$  & ${\rm Re}$ \\
 \hline 
(GeV)  & (GeV) &  & (GeV$^{-2}$) &  & (mb) & (GeV$^{-2}$) & & (GeV$^{-2}$) & (GeV$^{-2}$) & (GeV$^{-2}$) &  & \cite{data_analysed}   \\
\hline 
4.2 & 3.14 & 48 &   0.00106-0.54     & 1 & 67.45$\pm$0.32 & 13.4$\pm$0.11 & 0$\pm$0.05 & 25$\pm$2 & -2$\pm$0.5 & 0$\pm$0.1 & 1.33 &   aq  
\\ \hline
6 & 3.63 & 83 &  0.00141-0.42        & 1 & 61.44$\pm$0.26 & 12.8$\pm$0.08 & -0.072$\pm$0.005 & 22$\pm$2 & -1.77$\pm$0.4 & 0$\pm$0.1 & 1.331 &  ab  
\\ \hline
8 & 4.11 & 83 &  0.00181-0.33       & 1.021 & 55.84$\pm$0.05 & 12.5$\pm$0.02 & -0.076$\pm$0.009 & 19$\pm$1 & -1.55$\pm$0.09 & 0.056$\pm$0.009 & 1.567 &  ac  
\\ \hline
10 & 4.54 & 55 &  0.00181-0.355   & 1 & 53.6$\pm$0.1 & 12.35$\pm$0.05 & -0.08$\pm$0.01 & 17$\pm$1 & -1.4$\pm$0.1 & 0.035$\pm$0.2 & 1.106 &  ad  
\\ \hline
16 & 5.64 & 25 &  0.085-0.540     & 1.1084 & 49.71$\pm$0.18 & 11.90$\pm$0.10 & -0.072$\pm$0.006 & 16$\pm$1 & -1.2$\pm$0.2 & -0.05$\pm$0.02 & 2.84 &  c   
\\ \hline
22.4 & 6.62 & 32 &  0.055-0.43     & 1.0242 & 47.53$\pm$0.12 & 12.08$\pm$0.07 & -0.060$\pm$0.020 & 16$\pm$1 & -0.96$\pm$0.04 & -0.08$\pm$0.04 & 0.771 &  g   
\\ \hline
25.2 & 7.01 & 33 &  0.075-0.580     & 1.0914 & 46.87$\pm$0.20 & 12.00$\pm$0.05 & -0.058$\pm$0.041 & 16$\pm$2 & -0.92$\pm$0.24 & -0.10$\pm$0.03 & 1.562 &  e   
\\ \hline
32.1 & 7.88 & 39 & 0.055-0.47   & 1.0950 & 45.66$\pm$0.17 & 12.00$\pm$0.05 & -0.050$\pm$0.010 & 16$\pm$2 & -0.80$\pm$0.26 & -0.14$\pm$0.02 & 3.285 & s
\\ \hline
40.1 & 8.78 & 30 &  0.075-0.520  & 1.1306 & 44.72$\pm$0.30 & 11.90$\pm$0.08 & -0.042$\pm$0.020 & 16$\pm$2 & -0.73$\pm$0.3 & -0.14$\pm$0.03 & 2.362 &  e   
\\ \hline
50 & 9.78 & 11 &  0.0375-0.400   & 1.0136& 43.93$\pm$0.20 & 11.58$\pm$0.08 & -0.036$\pm$0.020& 13$\pm$1.5 & -0.66$\pm$0.10 & -0.18$\pm$0.02 & 2.114 & f   
\\ \hline
70 & 11.54 & 125 &  0.00185-0.08468 & 1.0076 & 42.97$\pm$0.11 & 12.05$\pm$0.15 & -0.02$\pm$0.02 & 16$\pm$2 & -0.56$\pm$0.2 & -0.22$\pm$0.05 & 1.279 &  n  
\\ \hline
70 & 11.54 & 125 &   0.0375-0.500    & 1.0076& 42.97$\pm$0.11 & 12.05$\pm$0.15 & -0.02$\pm$0.02 & 16$\pm$2 & -0.56$\pm$0.2 & -0.22$\pm$0.05 & 1.279 &  f  
\\ \hline
125 & 15.37& 140 &  0.00164-0.20931  & 1.0101 & 41.92$\pm$0.08 & 12.13$\pm$0.1 & 0.003$\pm$0.001 & 16$\pm$2 & -0.43$\pm$0.2 & -0.32$\pm$0.05 & 1.206 &  n  
\\ \hline
150 & 16.83 & 140 &  0.00164-0.20931  & 0.996 & 41.73$\pm$0.11 & 12.14$\pm$0.15 & 0.01$\pm$0.005 & 16$\pm$2 & -0.4$\pm$0.2 & -0.38$\pm$0.05 & 1.205 &  n  
\\ \hline
175 & 18.17 & 86 &  0.00181-0.09766  & 1.0019 & 41.61$\pm$0.14 & 12.19$\pm$0.12 & 0.015$\pm$0.01 & 16$\pm$2 & -0.37$\pm$0.2 & -0.4$\pm$0.05 & 1.051 &  n  
\\ \hline
175 & 18.17 & 86 &  0.0375-0.600     & 1.0019 & 41.61$\pm$0.14 & 12.19$\pm$0.12 & 0.015$\pm$0.01 & 16$\pm$2 & -0.37$\pm$0.2 & -0.4$\pm$0.05 & 1.051 &  f  
\\ \hline
200 & 19.42 & 86 &  0.00181-0.53656  & 1.0125 & 41.54$\pm$0.09 & 12$\pm$0.2 & 0.02$\pm$0.006 & 16$\pm$2 & -0.35$\pm$0.2 & -0.41$\pm$0.07 & 1.359 &  n  
\\ \hline
313.7 & 24.3 & 32 & 0.00108-0.01376 & 0.9911 & 41.51$\pm$0.04 & 12.31$\pm$0.26 & 0.036$\pm$0.002 & 17$\pm$1 & -0.28$\pm$0.01 & -0.46$\pm$0.07 & 0.51& l 
\\ \hline
491.5 & 30.4 & 29 &   0.00067-0.01561  & 0.9859 & 41.76$\pm$0.15 & 12.2$\pm$0.2 & 0.051$\pm$0.003 & 17$\pm$2 & -0.23$\pm$0.18 & -0.53$\pm$0.1 & 1.525 &  k 
\\ \hline  \hline
\end{tabular} 
\label{table:ppbar} 
\end{table*}
      \end{center}
 } 

At low energies, the analysis of the ${\rm p \bar p} $ forward scattering data is 
particularly difficult, because $\rho$ is very small, and changes sign for 
$p_{\rm LAB} = 117 $ GeV ( $\sqrt{s}$ near  12 GeV).

\subsubsection{$\rm{p \bar p}$ in the range $p_{\rm LAB}= $ 4.2-10.0 GeV  }

The range is $\sqrt{s}=3.14-4.54 ~\GeV $ in p\=p scattering is difficult to treat, because $\rho$ 
is very small, possibly passing through zero in this range. 
In Fig.\ref{ppbar_4_10} we show data collected at the 
energies $p_{\rm LAB}$ =   4.2 , 
6.0 , 8.0  and 10.0   GeV   corresponding to   measurements of Jenni et al. \cite{data_analysed}(a) 
with $|t|$ starting at about $10^{-3} \GeV^2$. These data are not smooth and do not cover a 
wide t range, so that they are combined with other data  in the range to build sufficient sets. 
The assemblage is shown in  plots, with the lines representing the results of the analysis, 
with parameters given in Table \ref{table:ppbar}. 

At 4.2 GeV the combination is made with the $p_{\rm LAB}$ = 5 GeV data of Ambats et al. \cite{data_analysed}(q), 
multiplied by 1.06  to account for the energy dependence. 
At 6 GeV the merging is with Braun et al. \cite{data_analysed}(b) at $p_{\rm LAB}$ = 5.7 GeV (multiplied by 
the energy dependence factor 0.98). At 8 GeV the combination is made with Russ
 et al. \cite{data_analysed}(c)  at     the same energy.   
At 10 GeV  the set is completed with data at 10.4 GeV from Brandenburg et al. \cite{data_analysed}(d) with 
a factor 1.0117. The energy factors are based  on the   squared ratio  of the total cross section given by the 
parametrized $\sigma$  input. We thus obtain  values for 
parameters that give reliable  proposals in a region of energies that is very difficult, because 
$\rho$ is very small. Our representations with  Eqs.(\ref{diffcross_eq},\ref{real_TR},\ref{imag_TI}) 
and parameters given in Table \ref{table:ppbar} give correct descriptions of these data. The 
dispersion relations  predictions   for  $\rho$ and for the derivative combination $\rho B_R/2-\mu_R$ 
are satisfied.   

 \begin{figure}[h]
\includegraphics[width=8.0cm]{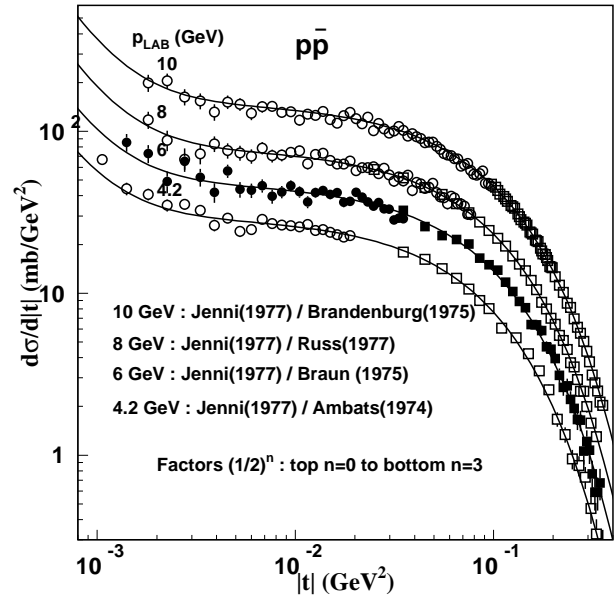} 
   \caption{\label{ppbar_4_10} Low energy end. 
At 4.2 GeV, data from Jenni et al.\cite{data_analysed}(a) together with data from 
Ambats \cite{data_analysed}(q) at 
5 GeV (multiplied by 1.06 to account for energy dependence). 
Data from Jenni et al. \cite{data_analysed}(a) at 6 GeV , 
and from Braun et al. \cite{data_analysed}(b)  for $p_{\rm LAB} = 5.7 \GeV$ (multiplied by 0.98). 
Data from Jenni et al. \cite{data_analysed}(a) and from  Russ et al. \cite{data_analysed}(c) at 
8 GeV , and data from  Jenni et al. \cite{data_analysed}(a) at 10 GeV and from Brandenburg  
et al. \cite{data_analysed}(d) at 10.4 GeV (multiplied by 1.0117). Visually 
(up to $|t| \approx \GeV^2$ ) and in the calculations  the matching of data is impressive. 
Parameter values for each line  are given in Table \ref{table:ppbar}.  }
\end{figure}

\subsubsection{$\rm{p \bar p}$ Data in the mid-$|t|$  range for $p_{\rm LAB}= $ 16 - 50  {\rm GeV}     }

The data in Fig.\ref{ppbar_middle}, with $p_{\rm LAB}$  = 16-50  GeV, cover only the mid-$|t|$ range. 
It is interesting 
observe in general that  the representation of data using the forward scattering 
expression for   $d\sigma/dt$,  in principle is only valid up to  $|t|=0.1 \GeV^2$,  in many cases remain 
good up   to 0.5.     
\begin{figure}[h]
\includegraphics[width=8.0cm]{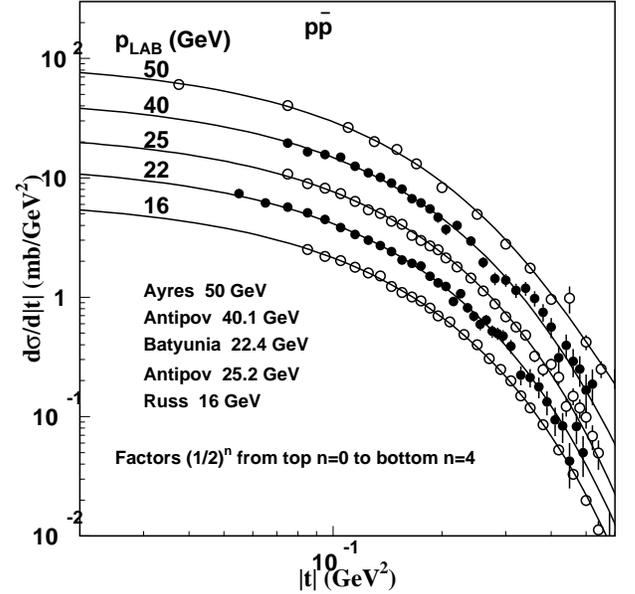} 
   \caption{\label{ppbar_middle} Mid t Range. 
 Data with measurements at mid$|t|$ only.  Data from 
Russ et al. \cite{data_analysed}(c),  Antipov et al. \cite{data_analysed}(e),  
Ayres et al. \cite{data_analysed}(f), 
Batyunia et al. \cite{data_analysed}(g) 
at $p_{\rm LAB}$= 16, 22.4,  25.2, 40.1 and 50 GeV,  described with the parameter values of 
Table \ref{table:ppbar}, presenting  regular energy dependence. 
It is remarkable that the lines reproduce well the data up to 0.5 $\GeV^2$. 
 Notice that the original Antipov data at 40.1 GeV have been remeasured and published with changes in 1977 
\cite{data_analysed}(e), and a factor 1.15 is used to adjust the old to the new published data.  
    }
\end{figure}

According to   Ayres \cite{data_analysed}(a), Akerlof data \cite{data_analysed}(m),
also at mid-$|t|$, have normalization problem,  and then are not included in our report
in Table  \ref{table:ppbar}.

\subsubsection{$\rm{p \bar p}$ in the range $p_{\rm LAB}= $ 70 - 200 {\rm GeV} }

This is a sensitive range for the determination of  $\rho$, due to the change of sign.  
  In $\rm{p \bar p}$ $\rho$ is small and negative at    low energies up to about $p_{\rm LAB}= 116 \GeV$.
 The data by D. S. Ayres \cite{data_analysed}(f) at $p_{\rm LAB} =$   50, 70, 100, 140 and 175 GeV are regular 
and with small error bars but are restricted to 
$|t|$ values that are not small enough for    determination of forward scattering parameters, and we  
then form  combinations with other sets of data, as shown in Fig. \ref{pilha_ppbar}.  
At $p_{\rm LAB}$ = 70 and 175 GeV  these data can be combined with the data from Fajardo et al. \cite{data_analysed}(n) 
to produce sets that cover both small and mid-$|t|$ values. Some parts of Fajardo data \cite{data_analysed}(n)
  in the higher $|t|$ part  are excluded, giving place to Ayres data \cite{data_analysed}(f), which are more 
regular. It  is remarkable that the data from the two experiments match very well, and a common solution 
 seems to be adequate. At $p_{\rm LAB}$ = 70 GeV the data give $\rho$ compatible with zero, and in 
Table \ref{table:ppbar} we choose the  value given by the DRA prediction. This is a choice, and the value  and sign have  the corresponding uncertainty. 
We observe in Table \ref{table:ppbar} that at  $p_{\rm LAB} = 125 ~\GeV $,  $\rho $ is positive, confirming 
the change of sign that occurs at $p_{\rm LAB}$ = 117 GeV.  

 At all energies of the range, the suggested solutions in Table \ref{table:ppbar} give very good
descriptions, as indicated  by the low $\chi^2 $ values. We are able    to follow the DRA  and DRS  
predictions  for the value of the amplitude and for its derivative at the origin. 
It is interesting to observe   that often  the representation of data using the forward scattering 
expression for $d\sigma/dt$, in principle   only valid up to  $|t|=0.1 \GeV^2$,  in many cases remains 
good up to 0.2 $\GeV^2$.  This is seen  also in both   Figs. \ref{ppbar_middle},\ref{pilha_ppbar}.  

We observe that in this range the  numbers reported in  experimental papers  for the $\rho$ parameter  
are very irregular and with contradictions.

 \begin{figure}[h]
\includegraphics[width=8.0cm]{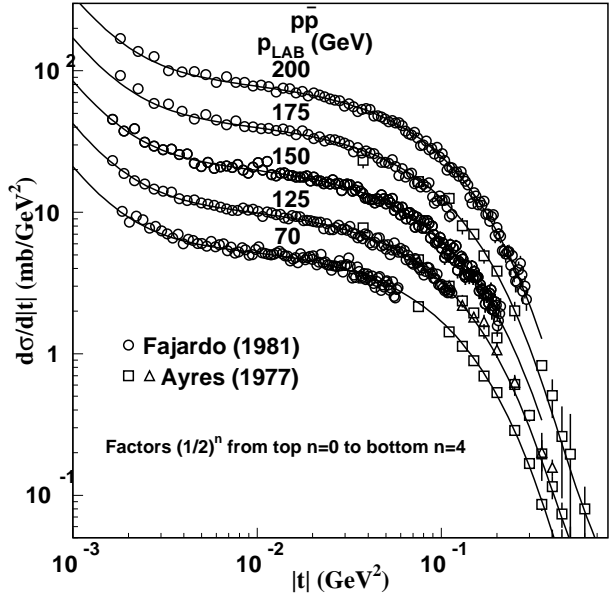} 
   \caption{\label{pilha_ppbar}  Data from  Fajardo et al.  \cite{data_analysed}(m) , 
and from D. S. Ayres et al.  \cite{data_analysed}(f) described by the   parameters  
given in Table \ref{table:ppbar}. At 125 GeV the solution refers only to Fajardo; the Ayres points at 100 
and 140 GeV are not fitted, and only marked  together  with the   125 GeV  line of Fajardo. 
 The  Jenni data are restricted to a range of large
 $|t|$ values, with difficulty for determination of  parameters, but are important to confirm the 
data of Fajardo. It is interesting to observe that $d\sigma/dt$  for large  $|t|$ is not very 
sensitive to the energy. At 70 and 175 GeV, there are measurements for both Jenni and Fajard, 
and then the respective data are merged to form   joint sets for the unified determination of parameters 
that are given in Table  \ref{table:ppbar}.  
At $p_{\rm LAB}$ = 150 and 200 GeV the   points are only  from Fajardo. } 
\end{figure}

\subsubsection{$\rm{p \bar p}$ in the range $p_{\rm LAB}= $ 300 - 500 {\rm GeV}}

   At the high energies  $\sqrt{s}$ = 24.3 and 30.4 GeV there are data from CERN ISR, with good  
precision in forward scattering measurements. 
The treatment of these data with our inputs is reproduced in  Fig.\ref{ppbar-250-500}.

\begin{figure}[h]
 \includegraphics[width=8.0cm]{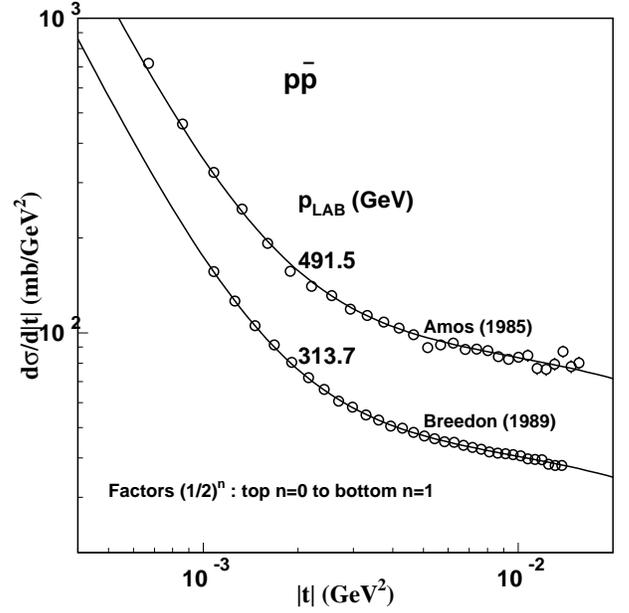} 
    \caption{\label{ppbar-250-500}  
Data from   R. E. Breedon et al. \cite{data_analysed}(l) and from N. Amos et al. \cite{data_analysed}(k) described by the  parameters  
given in Table \ref{table:ppbar}.  The numerical values  of $d\sigma/dt$ for $p_{\rm LAB}$ = 313.7 GeV ( $\sqrt{s}$ = 24.3 GeV)
are available from  the thesis of the author  Breedon   \cite{data_analysed}(l), and also in CERN
 }
\end{figure}
 
\clearpage
 
\section{Concluding Remarks \label{conclusions-sec}  }   
  
In this paper we propose a   revision of the existing information on forward  
parameters in pp and ${\rm p \bar p} $   scattering, giving coherent and precise description 
of all elastic  data in the energy range from  ${\rm p_{LAB}} \approx $ 4 to 500 GeV, 
namely $ \sqrt{s} \approx $ = 3 to 30 GeV.    

We treat the fundamental problem of the identification of the real and imaginary parts 
of the complex amplitude in $d\sigma/dt$, using  Eqs.(\ref{real_TR},\ref{imag_TI})
and   exact solutions of dispersion relations for amplitudes (DRA) and for slopes (DRS). 
The Coulomb interference is included with correct expression for the phase in 
Eq.(\ref{diffcross_eq}), which depends on the actual analytical   form 
of the real amplitude, and  is studied  in   Appendix {\ref{coulomb_phase}}.   
We  do not perform open fittings, but rather rationalize descriptions,
in an interplay between data   and mathematical representations,  
building  a bridge     for the  construction of full dynamical models.

The analysis of the $d\sigma/dt$  data leads to the parameter values presented 
 in   Tables \ref{table:pp}, \ref{table:ppbar}, given in  Sec.\ref{data_analysis}. 
Determination of parameters is guided by dispersíon relations and criteria of 
regularity in the energy dependence, in general achieving low $\chi^2$ values, with 
   visual performances  exhibited in plots in Sec.\ref{data_analysis}.
 The description is coherent and theoretically respectable and formally simple.  

A global analysis in  the low energy range must be restricted to the forward range, 
because there are not enough data for a full-$t$ study of the amplitudes. 
Since the imaginary amplitudes dominate at small $|t|$, our precision  work 
has particular importance as an investigation of the structure of the real parts. 
At low energies  the use of dispersion relations  requires the use of exact solutions 
for principal value  integrals, as we prove in Appendix \ref{DR_expanded}}. 

For a precise treatment of the data, the $t$ dependence of the amplitudes cannot 
be limited to exponential forms. It is known that the imaginary part  must point to a 
 zero because there is a dip in $d\sigma/dt$ shown when data have 
more extended $|t|$. Also the real part must show a zero when $\rho$ is positive, 
determined by Martin's theorem \cite{Martin}. We are thus led to the amplitudes 
written in Eqs.(\ref{real_TR},\ref{imag_TI}). To deal with this complexity without
ambiguities in the analysis of data, we make use of the dispersion relations for slopes
(DRS), which determine the derivatives of the real part when the derivatives of the 
imaginary parts are given as function of the energy. 

 The  proposals for $\sigma$, $B_I$ and $\mu_I$ (for pp and for p\=p) in 
Eqs.(\ref{sigma_x}), (\ref{BI_x}) and (\ref{MUI_x}), obtained from the data,  
  are basic in our work as inputs for DRA and DRS. Fig.\ref{inputs_fig} shows that 
the analytical forms  have correct extensions up to the very distant LHC energies. 

  Starting with the unitarized form  for $\sigma(x)$ originated in 
Pomeron/Reggeon phenomenology \cite{Cudell}, and complemented with   the $|t|$ 
structure of the imaginary part, the description of the $d\sigma/dt$  data  shows 
that all information in elastic scattering  at low energies  can be represented  by 
 simple expressions for the amplitudes.   We believe  that the simple  forms of 
energy dependence are a consequence of the very complex gluonic dynamics of QCD. Local 
effects that may be meaningful microscopically are absorbed in the complexity 
and do not become visible in elastic data.  

In Appendix \ref{Landshoff} it is shown that the log-squared form of the total
 cross section  in Eq.(\ref{sigma_x}) is  equivalent in chosen ranges 
to   simple-pole-Pomeron/Reggeons forms, with powers instead of logarithms, and thus
without limitation for the analytical solutions of  PV integrals.

 As an important test for the amplitudes, Fig.\ref{sig_elastic} in Sec.\ref{formalism} 
shows that the   integrated elastic cross section $\sigma^{\rm elastic}$  very well 
reproduces the data at higher energies (ISR, Fermilab, LHC) \cite{data_HE,data_LHC}. 
 The ratio $\sigma^{\rm elastic}/\sigma^{\rm total}$  is small, 
corresponding to large fraction of diffractive and inelastic  contributions, which
have not been measured in the low energy range and cannot enter  in our scope. 
The   single  and double diffractive processes,  measured at ISR/CERN, Fermilab 
and LHC   \cite{data_HE,data_LHC}, are theoretically studied in microscopic terms based 
on multi-Pomeron/Reggeon ideas  \cite{Tel-Aviv}   and on  gluonic dynamics of the 
color glass condensate \cite{CGC}.

The analysis of the ${\rm p \bar p} $   data at low energies is particularly 
delicate, because $\rho$ is very small and has change of sign. 
Most  p\=p  data do not cover  $|t|$ ranges   sufficient for  determination 
of the amplitude parameters  without support of DRA, DRS and search for regular energy 
dependence.  We include as much p\=p information as possible, forming combinations 
of different experiments at nearby energies   to have  sets suitable for analysis, 
  arriving at reliable solutions.   

Figure \ref{DR_results} shows that  $\rho$ and  the derivative 
$D_R=\rho B_R/2-\mu_R$   follow DRA and DRS predictions. Fig.\ref{muR_results} 
shows the proposed decomposition of $D_R$, with $\mu_R$  given by 
Eqs.(\ref{muRpp}) and (\ref{muRppbar}), and shows the calculated $B_R$, compared 
to values obtained  from the data. Everything seems satisfactory, including 
the behaviour as high energies, where these quantities appear \cite{us_LHC,LHC_2017}. 
Negative  $\mu_R$ predicts the zero of the 
real amplitude in a region where $\rho$ is positive, as at high energies \cite{Martin}, 
with the real zero approaching $|t|=0$ with increasing energy.
 Of course alternative solutions  can be obtained for $\mu_R$ and $B_R$, maintaining the 
 combination  Eq.(\ref{coefficient_R}) as fixed by DRS. 
 
 This extensive  work  is an effort to place order in the multitude of determinations 
of pp and p\=p forward scattering parameters at low energies. The treatment is 
as simple as possible,  without relation  to any specific model based on
microscopic ingredients, and identifies the real and imaginary parts of the 
complex elastic amplitude. We hope it can be considered as necessary and useful. 

The experiments at very high energies in the TeV range at LHC offer data of good  precision in 
the forward range with very small $|t|$ reaching near 0.0001 $\GeV^2$, providing material for the 
study of the properties of forward elastic amplitudes. Precise determination of parameters will 
provide material for investigation of asymptotic propeties in QCD, tests of validity of dispersion relations
and of  possibilities of new physics, mainly through  properties of the real part \cite{REAL_PART}. Also the 
structure of proton electromagnetic form factor and of the  theory for  the interference phase of Coulomb and 
strong forces  influence these measurements, affecting the extraction of fundamental information. 
In the forward range the real part is very small, with ratio of about 1 percent between the real and imaginary 
contributions to the cross sections. This requires   high statistics and regularity in the observations 
of $d\sigma/dt$  so that the subtraction of the 99\%  of the  imaginary part still leaves the real contribution 
with observable properties. Up to now, the analysis of LHC data \cite{LHC_2017} has faced difficulties and 
ambiguities for the determination of  the amplitude parameters.  The technology and experience of analysis of 
forward scattering learned in the present paper dealing with the difficult low energy data will be important 
in the high energy domain. 

\clearpage
 
\appendix

\section{ \label{coulomb_phase} Coulomb Phase}

We here present the Coulomb interference phase $\phi$ used in the phenomenology.
This phase  was
studied   by  several authors  \cite{phase}.

The expression for the phase depends on   
the analytical structure of the real and imaginary parts of the nuclear amplitude and also on 
the Coulomb amplitude and electromagnetic form factor. The real and imaginary nuclear amplitudes 
are given by Eqs. (\ref{real_TR}) and (\ref{imag_TI}) respectively and the Coulomb amplitude 
is given by Eq.(\ref{coulomb}) with the proton form factor (\ref{ff_proton}).
To simplify calculations,  we may alternatively use the     exponential representation for  the form factor
 \begin{equation}
f(t)=e^{2t/\Lambda^2}~, 
\label{form_factor_1} 
 \end{equation}     
with $\Lambda^2=0.71$ GeV$^2$, trusting that  differences in results for the phases with different form factors  are unimportant, as confirmed by 
Cahn \cite{phase}.

The Coulomb-nuclear-interference amplitude is given by
\begin{equation}
F^{N+C}(s,t)=F^{C}(s,t) ~ e^{i\alpha \phi}  + F^N(s,t) 
\label{Coulomb_ampl_2}
\end{equation}
with normalization defined by 
\begin{eqnarray}
&&  \sigma=\frac{4 \pi}{s} {\rm Im}~ F^N(t=0)    ~ , \nonumber \\
&& \frac{d \sigma}{dt}= \frac{\pi}{s^2} |F|^2 ~ . 
\label{normalization}
\end{eqnarray}
 The correspondence between the dimensionless nuclear amplitude $F^N$ and the phenomenological $T^N$ is given by
 \begin{equation}
 T^N(s,t)=\frac{\sqrt{\pi}}{(\hbar c)^2}\frac{F^N(s,t)}{s}~.
 \label{FN_TN}
 \end{equation}



 We start from the  expression for the phase
\begin{eqnarray}
\phi (s,t)=\mp \int_0^{\infty}dt'\ln\Big(\frac{t'}{t}\Big)\frac{d}{dt'}
\Big[f^2(t')\frac{F_N(s, t')}{F_N(s,0)}\Big]~, \nonumber \\
\label{Cahn_phase_app}
\end{eqnarray}
with the signs  $\mp$ corresponding to the  $ {\rm pp/p\bar p } $ systems.

As a generalization with respect to Cahn's calculation, we take for the nuclear amplitude 
the same expressions in Eqs.(\ref{real_TR}) and (\ref{imag_TI}). 
Then we  need to evaluate integrals 
\begin{eqnarray} 
&& H_k(t,b_\beta)=  \int_0^{-\infty}dt'\ln\Big(\frac{t'}{t}\Big)~
\frac{d}{dt'}\big[t'^k ~ e^{4t'/\Lambda^2} e^{B_\beta t'/2}  \big] ~ \nonumber \\
&& = \int_0^{-\infty}dt'\ln\Big(\frac{t'}{t}\Big)~\frac{d}{dt'}\big[t'^k ~ e^{b_\beta t'}\big]~ ,  
\label{integral_N} 
\end{eqnarray} 
where we have used the definition
\begin{eqnarray}
b_\beta= \frac{4}{\Lambda^2} +\frac{B_\beta}{2} 
\label{generic_slope}
\end{eqnarray}
with $\beta= R, I $.  

The results of the integrations (k=0,1) are 
\begin{eqnarray}
&&  H_0=\gamma ~+ ~\log(- b_\beta t) ~ , \nonumber \\
&&  H_1= \frac{1}{b_\beta} ~ , 
\end{eqnarray} 
where $\gamma=0.5772$  is the Euler Gamma constant.
The  phase is then  written 
\begin{eqnarray}
&& \phi (s,t)=\mp \frac{1}{\rho + i} \Big\{ \Big[- \frac{\mu_R}{b_R}  
 + \rho \big(\gamma + \log(- b_R t)\big) \Big]  \nonumber  \\
&& + i ~  \Big[- \frac{\mu_I}{b_I}+ \gamma + \log(- b_I t) \Big]   \Big\} ~ ,  
\label{our_phase_1}
\end{eqnarray}
with real and imaginary parts respectively 
 \begin{eqnarray}
&& \phi_{R} (t)=\mp\bigg\{\frac{1}{1+\rho^2}  
  \bigg[  \Big(- \frac{\mu_I}{b_I}+\log(b_I) \Big) \nonumber \\
&& + \rho \Big( - \frac{\mu_R}{b_R} 
       +\rho ~ \log(b_R)\Big) \bigg]+ \gamma + \log(-t)  \bigg\}
\label{Cahn_phase_real_2}
\end{eqnarray} 
and  
\begin{eqnarray}
&& \phi_{I} (t)=   \mp \frac{1}{1+\rho^2} 
   \bigg\{ \rho  \Big(- \frac{\mu_I}{b_I}+\log( b_I) \Big) \nonumber \\
&& - \Big(- \frac{\mu_R}{b_R}   + \rho \log( b_R ) \Big) \bigg\} ~. 
\label{Cahn_phase_imag_2}
\end{eqnarray} 
 Equations (\ref{Cahn_phase_real_2}) and (\ref{Cahn_phase_imag_2})   are our final results for the 
phase calculated with form factors,  in a generalization of the work by Cahn \cite{phase},  
assuming more complete structures for the real and imaginary parts of the scattering amplitude. 
   It may be of practical usefulness  to define  
\begin{equation}
  C_R =   - \frac{\mu_R}{b_R}   + \rho ~ \log( b_R ) 
\label{C_real}
\end{equation}
and 
\begin{equation}
  C_I =   -\frac{\mu_I}{b_I}  +\log( b_I )  ~.     
  \label{C_imag}
\end{equation}
  and then write     
\begin{equation}
 \phi_{R} (t)= \mp  \bigg[     \frac{1}{1+\rho^2}     
 \big[ C_I+ \rho ~ C_R  \big]   +  \gamma + \log(-t) \bigg]  
\label{final_R}
\end{equation}
and 
\begin{equation}
 \phi_{I} (t)= \mp \frac{1}{1+\rho^2}\big[\rho ~ C_I-C_R \big]    ~. 
\label{final_I}
\end{equation}

It must be observed that in these expressions  $b_R$ , $b_I$   and $-t$  have 
compatible units, as $\GeV^{-2}$  and $\GeV^2$. 
The result is simple: in the real part the $t$ dependence is purely linear in $\log(-t)$, 
the imaginary part is very small  constant, and there is no explicit energy dependence. 

In the simplified case  
$\mu_R=\mu_I=0,   ~ B_R=B_I=B, $
$$ b_R=b_I=b=\frac{4}{\Lambda^2}+\frac{B}{2} $$ 
we obtain  Cahn's original form.
 


\clearpage 

 \section{ \label{DR_expanded}  Dispersion Relations in Expanded Forms}

\subsection { Dispersion Relations for Amplitudes   \label{DRA-section}    }

Introduction of the inputs of $\sigma(x)$,  $B_I(x)$, $\mu_I(x)$  given in 
Eqs.(\ref{sigma_x}), (\ref{BI_x}) and (\ref{MUI_x}), DRA and DRS become sum of terms with 
principal value integrals, all of the  general  form 
\begin{eqnarray}
I(n,\lambda,x)=
{\bf P}\int_{1}^{+\infty}\frac{x^{\prime\lambda} \, \log^n (x^\prime)}{ x^{\prime 2}-x^2}\,dx^\prime  ~,
\label{PV_int_2}
\end{eqnarray} 
belonging to a family of integrals   that can be solved 
analytically  in terms of Lerch's  transcendents  $\Phi$ \cite{Exact} with the form 
 \begin{eqnarray}
I(n,\lambda ,x) &=&\frac{\pi}{2x}\frac{\partial ^n}{\partial \lambda^n}\left[x^{\lambda} \tan \left(\frac{\pi}{2}\lambda\right)\right]  \nonumber  \\
& & +\,\frac{(-1)^n\, n!}{ 2^{n+1}\,x^2} \,\Phi\left(\frac{1}{x^2},n\! +\! 1,\frac{1\! +\! \lambda}{2}\right) ~ .
\label{PV_integrals_app}
\end{eqnarray}
 
Collecting terms, we have for the even and odd parts of DRA    
\begin{eqnarray}
&&{\rm Re}\,F_+(x,0) = K + \frac{4\,m^2\,x^2}{\pi} \,\Big[  I(0,0,x)\left(P+H \log^2 x_0\right) \nonumber \\
&&  +I(1,0,x) \left(-2 H \log x_0\right) \nonumber \\
&&+\,I(2,0,x)\, H   + I(0,-\eta_1,x)\, R_1 \,x_0^{\eta_1} \Big] ~  
\label{even_even}
\end{eqnarray}
and
 \begin{eqnarray}
 {\rm Re}\,F_-(x,0) =   \frac{4\,m^2\,x}{\pi}\, I(0,1-\eta_2,x)  \, R_2 \,x_0^{\eta_2}  ~.
\label{odd_odd}
 \end{eqnarray}
Equations (\ref{even_even}) and (\ref{odd_odd}) lead to the real parts of the complex
amplitude for   pp and   p\=p elastic scattering, and then we have predictions for the parameters $\rho$ (pp, p\=p).   
  
Using  practical truncated expression for the transcendents (up to first order 
in $1/x$) we have  for the even combination   
 \begin{eqnarray}
&& \frac{1}{2}\Big[(\sigma\rho) ({\rm p \bar p})+(\sigma\rho) ({\rm pp})\Big]  
  = \frac{1}{2 m^2 x}\,{\rm Re}~ F_+(x,0) \nonumber \\   
&& \approx  T_1(x)+T_2(x)+T_3(x) ,
\label{rhosigma_terms}
\end{eqnarray}
with
\begin{eqnarray}
&& T_1(x)= H \pi \log\Big(\frac{x}{x_0}\Big) ~ ,
\label{T1-eq}  \\
&& T_2(x) = \frac{K}{2 m^2 \,x}   \nonumber  \\ 
&&  +\frac{2}{ \pi\,x } \Big( P+H \left[  \log ^2\left(x_0\right)
+2 \log\left(x_0 \right) +2   \right] \Big) \, ,
 \label{T2-eq}  \\
&& T_3(x)=   R_1 x_0^{\eta_1} \Big[  -x^{-\eta_1}\, \tan \Big(\frac{\pi \eta_1}{2}\Big)
 +    \frac{1}{x}  \frac{2/\pi}{1-\eta_1} \Big] ~,
\label{T3-eq}
\end{eqnarray}
and for the odd part
\begin{eqnarray}
&& \frac{1}{2}\left[(\sigma\rho) ({\rm p\bar p})-(\sigma\rho) ({\rm p  p}) \right]  
  =\frac{1}{2 m^2x}{\rm Re}\,F_-(x,0) \nonumber \\
&& \approx   R_2\,x_0^{\eta_2} \Big[x^{-\eta_2}\, \cot\Big(\frac{\pi\, \eta_2}{2}\Big)
 +     \frac{1}{x^2} \,  \frac{2/\pi}{2-\eta_2}   \Big].
\label{T_odd-eq}
\end{eqnarray}
Additional  terms are of order ${\cal O}(x^{-4})$.

Using the parameter values for $P, H, R_1, R_2$ in Table \ref{table:inputs} and with  $K=-310$,  
the calculated $\rho$ values  are  plotted in Fig. \ref{DR_calculations}. To demonstrate the 
importance of  the calculation with the exact solutions for the  PV integrals, 
   we plot in the figure  the $\rho$ values obtained with the above 
expressions (namely first order in the transcendents)  and  
in zero order (very large $x$, meaning the above expressions without the last terms with  ${2/\pi}/{1-\eta_1}$  and  $ {2/\pi}/{2-\eta_2}$). 
From the figure we observe the 
importance of the improved  solutions at low energies. 

 \subsection { Dispersion relations for slopes   \label{DRS-section}    }

In the same way , the DRS terms are collected in terms of the standard integrals.
We obtain for the even part
\begin{eqnarray}
&& \frac{\partial{\rm Re}\,F_+(x,t)}{\partial t}\Big|_{t=0} =
 \frac{2\,m^2\,x^2}{\pi}\, \Big\{ I(0,0,x)~  \Big(P+H \log^2{x_0}\Big)~  b_0^\prime    \nonumber \\
&& +\,I(1,0,x) \Big[ \Big(-2 H \log{x_0}\Big)~ b_0^\prime  + \Big(P+H \log^2{x_0}\Big)~ b_1^\prime \Big]  \nonumber \\
&&  +\,I(2,0,x)\Big[  H ~ b_0^\prime
 -2 H \log{x_0} ~  b_1^\prime + \Big(P+H \log^2 {x_0} \Big)~ b_2  \Big]   \nonumber  \\
&&  +\,I(3,0,x)  ~ \Big[ -2 H \log{x_0} ~   b_2    +  H ~ b_1^\prime \Big] +I(4,0,x) ~     H ~  b_2 \nonumber \\
&&  + \,  R_1 ~ x_0^{\eta_1}  \Big(I(0,-\eta_1,x)~b_0^\prime+I(1,-\eta_1,x)~b_1^\prime \nonumber \\
&& +I(2,-\eta_1,x)~ b_2+I(0,-\eta_1-\eta_3,x)~b_3\Big)  \nonumber \\
&& + \,R_2~x_0^{\eta_2}~I(0,-\eta_2-\eta_4,x)~ b_4       \nonumber \\
&&+\, \big[(P+H\log^2 {x_0})\,I(0,-\eta_3,x)\nonumber \\
&&  -2H\log{x_0}\,I(1,-\eta_3,x)+H~I(2,-\eta_3,x)\big] ~b_3 \Big\} ,
 \label{drdtplus_expanded}
\end{eqnarray}
and for the odd part
\begin{eqnarray}
&& \frac{\partial{\rm Re}\,F_-(x,t)}{\partial t}\Big|_{t=0} =
 \frac{2\,m^2\,x}{\pi}\, \Big\{ R_2 ~ x_0^{\eta_2}\Big(I(0,1-\eta_2,x)~ b_0^\prime  \nonumber \\
 && + I(1,1-\eta_2,x)~ b_1^\prime
   +\,I(2,1-\eta_2,x) ~ b_2  \nonumber \\
&& + I(0,1-\eta_2-\eta_3,x)~b_3 \Big)  \nonumber \\
&& +\,\Big( (P+H\log^2{x_0})\,I(0,1-\eta_4,x) \nonumber \\
&&  -2H\,\log{x_0}~I(1,1-\eta_4,x)  \nonumber \\
&& +\,H~I(2,1-\eta_4,x) \nonumber \\
&& +R_1~x_0^{\eta_1}~I(0,1-\eta_1-\eta_4,x)\Big)  ~ b_4~\Big\}~.
\label{drdtminus_expanded}
\end{eqnarray}
We recall Eq.(\ref{change}) with the definitions of $b_0^\prime$ and
$b_1^\prime$.

    Explicit expressions for the derivative DR   including 
only the first  terms of the expansions of the transcendents, are    
\begin{eqnarray}
&& \frac{1}{2m^2 x}\frac{\partial{\rm Re}\,F_+(x,t)}{\partial t}\Big|_{t=0} \nonumber \\ 
&& =  \frac{1}{2} \big\{\sigma_{\rm p\bar p} [\rho B_R/2 -\mu_{R}]({\rm p\bar p}) + \sigma_{\rm p p}[\rho B_R/2 -\mu_{R}]({\rm p p})    \big\} \nonumber \\
&& = \frac{ 1}{\pi} ~ \Big[(P+H\log^2 x_0)~G_1(x)+H~G_2(x)  \nonumber \\
&& +R_1~G_3(x)+R_2~G_4(x)\Big]~,
 \label{drdtplus_expanded_1}
\end{eqnarray}
 where
\begin{eqnarray}
&& G_1(x)\equiv \frac{b_0^\prime-b_1^\prime+2b_2}{x}+\frac{b_1^\prime \pi^2}{4}+\frac{b_2 \pi^2}{2}\log x  \nonumber \\
&&  +b_3\bigg(-\frac{\pi}{2}x^{-\eta_3}\tan\left(\frac{\pi\eta_3}{2}\right)+\frac{1}{x}\,\frac{1}{1-\eta_3}\bigg)~,
\label{G_1}  \\
&& G_2(x)\equiv \bigg[\frac{\pi^2}{4}\Big(3\log^2 x+\frac{\pi^2}{2}\Big)-\frac{6}{x}\bigg](b_1^\prime-2b_2\log x_0)\nonumber  \\
&& -2b_0^\prime\log x_0\Big(\frac{\pi^2}{4}-\frac{1}{x}\Big)  
 +\,(b_0^\prime-2b_1^\prime\log x_0)\Big(\frac{\pi^2}{2}\log x+\frac{2}{x}\Big) \nonumber \\
&&  +b_2\bigg[\pi^2\log x\Big(\log^2x+\frac{\pi^2}{2}\Big)+\frac{24}{x}\bigg] \nonumber \\
&&-\,\pi\, b_3 \,x^{-\eta_3}\bigg[\log x~\tan\Big(\frac{\pi\eta_3}{2}\Big)\Big(-\log x_0+\frac{1}{2}\log x\Big) \nonumber \\
&& -\frac{\pi}{2}\sec^2 \Big(\frac{\pi\eta_3}{2}\Big)\bigg(\log \Big(\frac{x}{x_0}\Big)  
    -\,\frac{\pi}{2}\tan\left(\frac{\pi\eta_3}{2}\right)\bigg)\bigg]  \nonumber  \\
&&  +\frac{2b_3}{x(1-\eta_3)^2}\Big(\log x_0+\frac{1}{1-\eta_3}\Big)~,
\label{G_2}  \\
&& G_3(x)\equiv x_0^{\eta_1}\bigg\{
b_0^\prime\bigg[\frac{\pi}{2}x^{-\eta_1}\tan\left(-\frac{\pi\eta_1}{2}\right)+\frac{1}{x}\,\frac{1}{1-\eta_1}\bigg] \nonumber   \\
&&+\,b_1^\prime\,\frac{\pi}{2}\,x^{-\eta_1}\bigg[\frac{\pi}{2}\sec^2\left(\frac{\pi\eta_1}{2}\right)
-\tan\left(\frac{\pi\eta_1}{2}\right)\log x \bigg] \nonumber \\
&&-\,b_2\,\frac{\pi^2}{2}\,x^{-\eta_1}\bigg[\sec^2\left(\frac{\pi\eta_1}{2}\right)\bigg(\frac{\pi}{2}\tan\left(\frac{\pi\eta_1}{2}\right) -\log x\bigg) \nonumber \\
&&+\frac{1}{\pi}\tan\left(\frac{\pi\eta_1}{2}\right)\log^2 x \bigg]  \nonumber \\
&&+\,b_3\bigg[-\frac{\pi}{2}x^{-\eta_1-\eta_3}\tan\Big(\frac{\pi(\eta_1+\eta_3)}{2}\Big)
+\frac{1}{x}\,\frac{1}{1-\eta_1-\eta_3}\bigg]  \nonumber  \\
&&+\,\frac{1}{x}\,\frac{1}{(1-\eta_1)^2}\bigg(-b_1^\prime+\frac{2 b_2}{(1-\eta_1)}\bigg)\bigg\}  ~,
\label{G_3} 
\end{eqnarray} 
\begin{eqnarray} 
  && G_4(x)\equiv x_0^{\eta_2}\,b_4\,\Big[-\frac{\pi}{2}x^{-\eta_2-\eta_4}\tan\Big(\frac{\pi(\eta_2+\eta_4)}{2}\Big)
          \nonumber \\
  &&  +\frac{1}{x}\,\frac{1}{1-\eta_2-\eta_4} \Big]~.  
    \label{G_4}
   \end{eqnarray}

For the odd combination we have
 \begin{eqnarray}
&& \frac{1}{2m^2 x}\frac{\partial{\rm Re}\,F_-(x,t)}{\partial t}\Big|_{t=0} \nonumber \\
&& = \frac{1}{2} \big\{\sigma_{\rm p\bar p}[\rho B_R/2 -\mu_{R}]({\rm p\bar p})-\sigma_{\rm p p}[\rho B_R/2 -\mu_{R}]({\rm p p})    \big\} \nonumber \\
&& = \frac{1}{\pi}~ \Big[(P+H\log^2 x_0)~F_1(x)+H~F_2(x) \nonumber \\
&&  +R_1~F_3(x)+R_2~F_4(x)\Big],
\label{drdtminus_expanded_1}
\end{eqnarray}
where
\begin{eqnarray}
&& F_1(x) \equiv b_4\bigg(\frac{\pi}{2}x^{-\eta_4}\cot\left(\frac{\pi\eta_2}{2}\right)+\frac{1}{x^2}\,\frac{1}{2-\eta_4}\bigg)~,
\label{F_1}  \\
&&  F_2(x)\equiv b_4\bigg\{\frac{\pi}{2}\,x^{-\eta_4}\bigg[\pi\csc^2\Big(\frac{\pi\eta_4}{2}\Big)\bigg(\log \Big(\frac{x}{x_0}\Big)
            \nonumber \\
&& + \frac{\pi}{2}\cot\Big(\frac{\pi\eta_4}{2}\Big)  \bigg)    \nonumber \\
&&+\,\cot\Big(\frac{\pi\eta_4}{2}\Big)\log x\,(-2\log x_0+\log x)  \bigg]  \nonumber \\
&& +\frac{2}{x^2}\,\frac{1}{(2-\eta_4)^2}\Big(\log x_0+\frac{1}{2-\eta_4}\Big)  \bigg\}~,    
\label{F_2}   
\end{eqnarray}
\begin{eqnarray}
&& F_3(x)\equiv b_4\,x_0^{\eta_1}\bigg[\frac{\pi}{2}x^{-\eta_1-\eta_4}\cot\Big(\frac{\pi}{2}(\eta_1+\eta_4)\Big) \nonumber \\
&& +\frac{1}{x^2}\frac{1}{2-\eta_1-\eta_3} \bigg]~,  
\label{F_3}  \\
&& F_4(x)\equiv x_0^{\eta_2}\bigg\{\frac{\pi}{2}x^{-\eta_2}\bigg[ \cot\left(\frac{\pi\eta_2}{2}\right)\left(b_0^\prime+b_1^\prime\log x+b_2\log^2 x\right) \nonumber \\
&&+\,\frac{\pi}{2}\csc^2\left(\frac{\pi\eta_2}{2}\right)\bigg(b_1^\prime+\pi\cot\left(\frac{\pi\eta_2}{2}\right)b_2+2\log x~b_2 \bigg)
    \nonumber \\
&& +x^{-\eta_3}~\cot\Big(\frac{\pi}{2}(\eta_2+\eta_3)\Big)b_3 \bigg]  +\,\frac{1}{x^2}\frac{1}{2-\eta_3}\bigg[b_0^\prime-\frac{b_1^\prime}{2-\eta_2} \nonumber \\
&& +\frac{2 b_2}{(2-\eta_2)^2}+\frac{(2-\eta_2)b_3}{2-\eta_2-\eta_3}  \bigg]\bigg\} ~  . 
\label{F_4}
\end{eqnarray}

In Fig.{\ref{DR_calculations}} the quantities  $D_R$ for pp and p\=p are plotted 
using the above expressions (first order in the Transcendents) and the 
calculations where $\Phi$ is ignored. 

\begin{figure*}[h]
            \includegraphics[width=8.0cm]{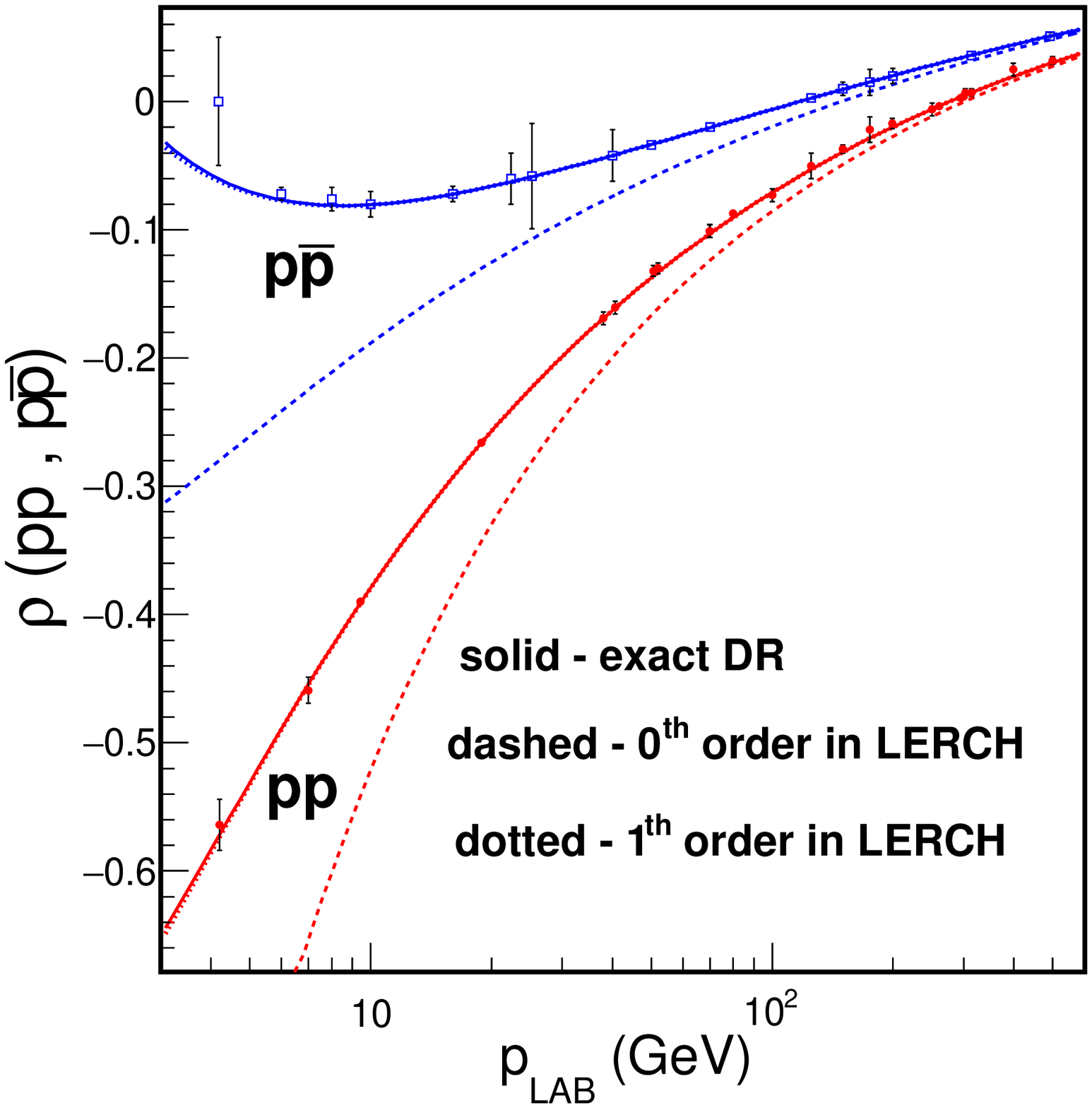}
   \includegraphics[width=8.0cm]{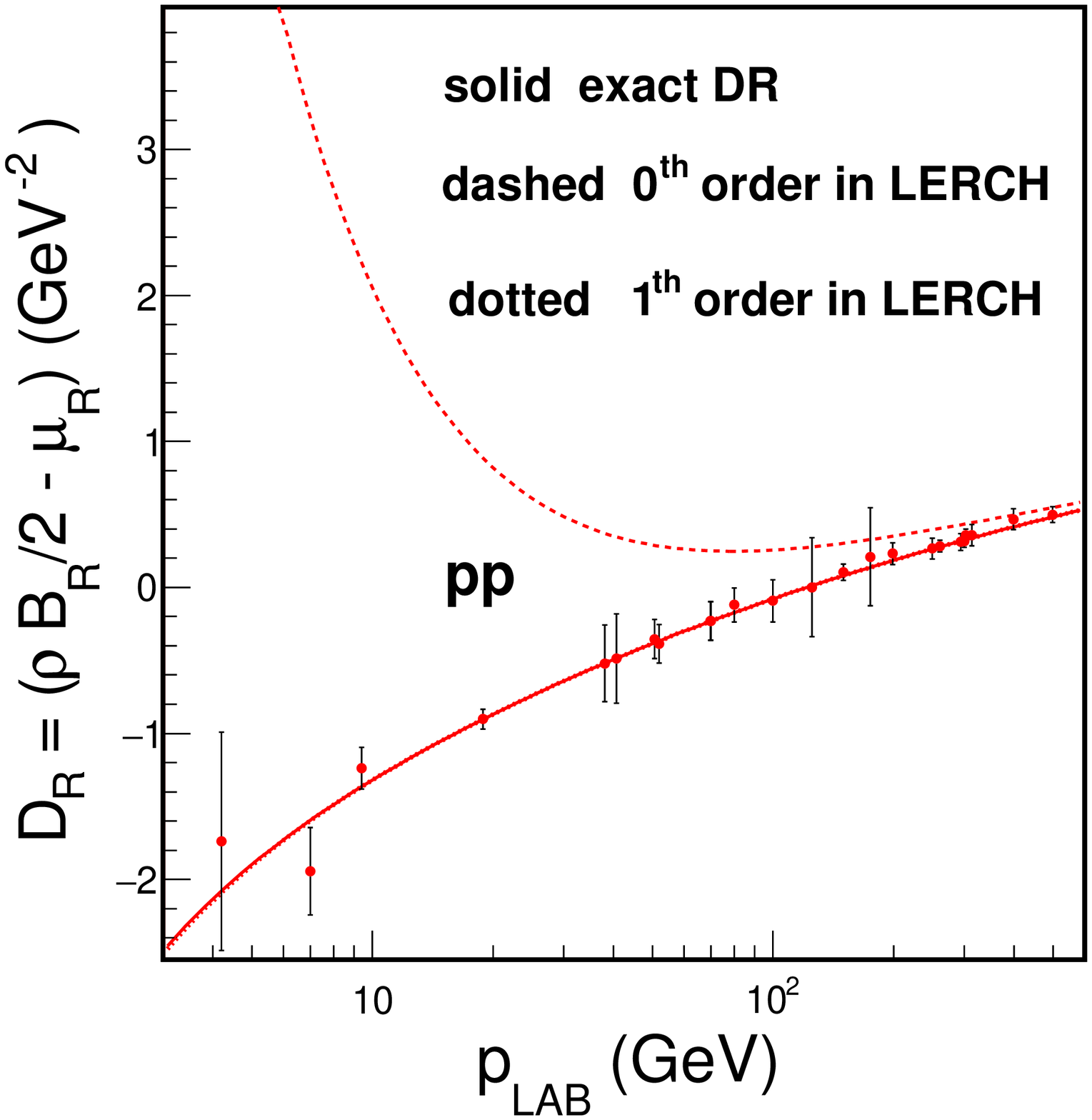}
   \includegraphics[width=8.0cm]{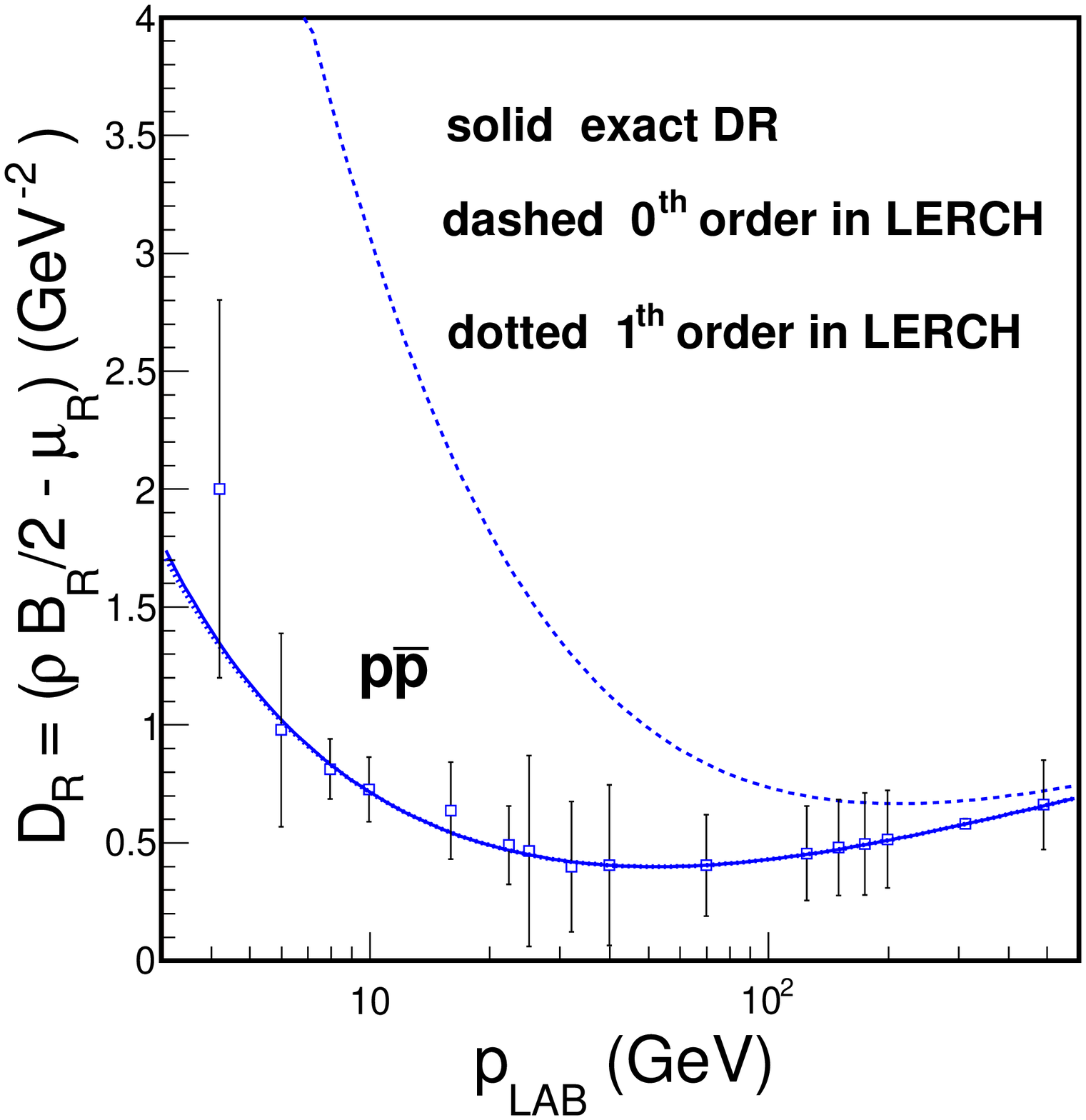}
        \caption{\label{DR_calculations}
  Demonstration of the importance of calculations with improved solutions for the PV integrals. 
(a) $\rho$ parameters ; (b) and (c)   Energy dependence of the derivatives of 
the real amplitudes at $t=0$. The lines are fully predicted by the dispersion relations for the amplitudes 
(that give $\rho$ for pp and p\=p) and by dispersion relations for slopes DRS
(that give $D_R=\rho B_R/2-\mu_R$ for pp and p\=p).  }
   \end{figure*}

These equations, here called Dispersion Relations for Amplitudes (DRA) 
and Dispersion Relations for  Slopes (DRS ), control quantities observed in
forward scattering and should be used as basic information for
phenomenological and  theoretical description of
forward pp and $\rm{ p \bar p}$ scattering. Since their introduction, \cite{EF2007}
they were shown to be important for the
analysis of forward scattering, determining structure and parameters of the real amplitude.

\clearpage

\section{ \label{Landshoff}  Alternative Single Pole  Pomeron  Approach}

In the  $\log^2 s$ description of total cross section (high energies) the Pomeron intercept can be
 recovered when we take a convenient energy range  of the total cross section in Eq.(\ref{sigma_s}). 
 Comparative analysis between the two main alternatives for the total cross section was
 performed \cite{Cudell} before the LHC era. In the present work we based our calculation 
of dispersion relations in the $\log^2s$ approach (unitarized model). On the other hand   
in the  $\log^2 s$ description of total cross section (high energies) the Pomeron intercept 
(single pole description) can be  recovered when we take a convenient energy range  of the total 
cross section.  
First we rewrite the dominant terms at high energies  in the form  
\begin{eqnarray}
\sigma_{HE}(s)&=& P + H\log^2\left(\frac{s_0}{ s_1}\right)-4H\log\left(\frac{s_0}{s_1}\right)\log\left(\sqrt{\frac{s}{s_1}}\right)  \nonumber \\
&&+4H\log^2\left(\sqrt{\frac{s}{s_1}}\right)~,
\label{Pomeron}
\end{eqnarray}
 where we have introduced  a new scale $s_1$ (in GeV$^{2}$) chosen appropriately in order obtain the 
Pomeron intercept.  After some algebra we can rewrite  the above equation as
\begin{eqnarray}
\sigma_{HE}(s)&=& P-H\log^2\left(\frac{s_0}{s_1}\right)+2H\log^2\left(\frac{s_0}{s_1}\right)\left(1+y+\frac{1}{2}y^2\right)~, \nonumber \\
\label{Pomeron1}
\end{eqnarray}
where 
\begin{eqnarray}
y\equiv \log\left[\left(\frac{s}{s_1}\right)^{-1/\log\left(\frac{s_0}{s_1}\right)}\right]~.
\label{variable}
\end{eqnarray}
 For values  $y<<1$, corresponding to $\sqrt{s}<<10^6$ GeV, we can exponentiate the terms in 
parenthesis of Eq.(\ref{Pomeron1}), writing   
\begin{eqnarray}
&&\sigma_{HE}(s)\simeq  P-H\log^2\left(\frac{s_0}{s_1}\right)+2H\log^2\left(\frac{s_0}{s_1}\right)e^{\left(y\right)} \nonumber \\
&&=P+H\log^2\left(\frac{s_0}{s_1}\right)\left[2\left(\frac{s}{s_1}\right)^{\epsilon_0}-1\right]
\label{Pomeron2}
\end{eqnarray}
 where 
\begin{equation}
 \epsilon_0=-1/\log\Big(\frac{s_0}{s_1}\Big) , 
\end{equation} 
which is the Pomeron intercept coefficient similar to the one in Donnachie and Landshoff description. 
From the definition of $\epsilon_0$ in terms of the scales, $s_0$ and $s_1$, we can rewrite 
 Eq.(\ref{Pomeron2}) as
\begin{eqnarray}
\sigma_{HE} (s)\simeq P+\frac{H}{\epsilon_0}\left[2~e^{-1}\left(\frac{s}{s_0}\right)^{\epsilon_0}-1\right]~.
\label{Pomeron3}
\end{eqnarray}
  Of course Eq.(\ref{Pomeron3}) is compatible with Eq.(\ref{Pomeron1}) only in a limited range 
of energy, and if one wants a better description of low energy data using Eq.(\ref{Pomeron1}), 
the parameters $P$, $H$ and also the parameters for the low energies $R_1$ and $R_2$, here not 
included explicitly,  should be reobtained. The complete formula for total cross section in this 
power-like representation is}
\begin{eqnarray}
&&\sigma^\mp(s)= P'+H'\left(\frac{s}{s_0}\right)^{\epsilon_0} +R_1' \left(\frac{s}{s_0}\right)^{-\eta_1'} \pm\, R_2' \left(\frac{s}{s_0}\right)^{-\eta_2' }~,  \nonumber \\
\label{Pomeron4}
\end{eqnarray}
 where we re-write the parameter with prime in order  to distinguish from Eq.(\ref{sigma_s}). 
In Table \ref{table:Pomeron} we   give values for this representation.  It is important to observe 
that the term $P'$ in Eq.(\ref{Pomeron4}) corresponds to a Regge-trajectory 
with zero power coefficient.

  \begin{table}
 \setlength{\tabcolsep}{3pt}
\caption{Suggested values for the parameters of the total cross section in the power-like representation. }  
\centering 
\begin{tabular}{c c c c c c c} 
\hline\hline 
$ P' $(mb )  & $H'$  (mb) & $ R_1'$(mb) &$R_2'$(mb) & $\eta_1'$ &  $\eta_2'$ & $\epsilon_0$  \\
\hline 
 -49.1 &  59.4 &37.1 &  8.143 &0.225  & 0.6144 &0.059   \\
\hline \\
\end{tabular} 
\label{table:Pomeron} 
\end{table} 

Obviously this alternative representation can be written in terms of the variable $x$ 
for use in dispersion relations.

\clearpage 

  \begin{acknowledgments}
The authors wish to thank the Brazilian agencies CNPq and CAPES  and Gobierno de Arag\'on (Spain)
for financial support.
\end{acknowledgments}

\end{document}